\documentclass[pdflatex,sn-mathphys-num]{sn-jnl}

\usepackage{graphicx}%
\usepackage{multirow}%
\usepackage{amsmath,amssymb,amsfonts}%
\usepackage{amsthm}%
\usepackage{mathrsfs}%
\usepackage[title]{appendix}%
\usepackage{xcolor}%
\usepackage{textcomp}%
\usepackage{manyfoot}%
\usepackage{booktabs}%
\usepackage{algorithm}%
\usepackage{algorithmicx}%
\usepackage{algpseudocode}%
\usepackage{listings}%

\usepackage{mathtools}
\usepackage{braket}
\usepackage{physics}
\usepackage{bm}
\usepackage{multirow}

\theoremstyle{thmstyleone}%
\newtheorem{theorem}{Theorem}
\theoremstyle{thmstyletwo}%

\theoremstyle{thmstylethree}%

\newcommand{\ci}{\mathrm{i}}
\newcommand{\ee}{\mathrm{e}}

\newcommand{\tv}{\mathrm{v}}
\newcommand{\vtv}{\mathbf{v}}
\newcommand{\vx}{\mathbf{x}}
\newcommand{\vy}{\mathbf{y}}
\newcommand{\tx}{\mathrm{x}}
\newcommand{\ty}{\mathrm{y}}
\newcommand{\vk}{\mathbf{k}}
\newcommand{\tk}{\mathrm{k}}
\newcommand{\vq}{\mathbf{q}}

\newcommand{\cs}{\mathrm{c_s}}
\newcommand{\bdg}{\mathcal{L}_\mathrm{BdG}}

\newcommand{\omax}{\omega_\mathrm{max}}
\newcommand{\omh}{\Omega_\mathrm{h}}

\newcommand{\dif}{\mathrm{d}}
\newcommand{\vna}{\bm{\nabla}}
\newcommand{\rh}{r_\textrm{h}}
\newcommand{\rerg}{r_\textrm{erg}}

\newcommand{\absv}[1]{|#1|}
\renewcommand{\expval}[1]{\langle#1\rangle}

\renewcommand{\dd}{\text{d}}
\newcommand{\bs}[1]{\boldsymbol{#1}}

\newcommand{\lr}[1]{\left(#1\right)}
\newcommand{\lrsq}[1]{\left[#1\right]}

\newcommand{\w}{\omega}
\newcommand{\hsp}{\hspace{0.5cm}}

\newtheorem{problem}{Problem}

\begin{document}

\title[Article Title]{Analogue gravity with Bose-Einstein condensates}


\author*[1,2]{\fnm{Adri\`a} \sur{Delhom}}\email{adria.delhom@gmail.com}

\author*[3]{\fnm{Luca} \sur{Giacomelli}}\email{lucagiacomelli92@gmail.com}

\affil[1]{Departamento de F\'isica Te\'orica and IPARCOS, Facultad de Ciencias F\'isicas, Universidad Complutense de Madrid, Plaza de Ciencias 1, 28040 Madrid, Spain}

\affil[2]{Department of Physics and Astronomy, Louisiana State University, Baton Rouge, LA 70803, U.S.A.}

\affil[3]{Université Paris Cité, CNRS, Matériaux et Phénomènes Quantiques, 75013 Paris, France}


\abstract{
Analogue gravity explores how collective excitations in condensed matter systems can reproduce the behavior of fields in curved spacetimes. An important example is the \textit{acoustic} black holes that can occur for sound in a moving fluid. In these lecture notes, we focus on atomic Bose–Einstein condensates (BECs) - quantum fluids that provide an interesting platform for analogue gravity studies thanks to their accurate theoretical description, remarkable experimental control, and ultralow temperatures that allow the quantum nature of sound to emerge. We give a pedagogical introduction to analogue black holes and the theoretical description of BECs and their elementary excitations, which behave as quantum fields in curved spacetimes. We then apply these tools to survey the current understanding of black-hole superradiance and analogue Hawking radiation, including explicit examples and numerical methods.
}





\maketitle

\tableofcontents

\section{Introduction}
These lecture notes, based on a course by Luca Giacomelli at the school \textit{Analogue Gravity in 2023} and expanded by Adrià Delhom, intend to provide an introduction on the use of atomic Bose Einstein condensate in analogue gravity. This field of research, which started as a clever idea to mimic the black hole radiance predicted by Hawking \cite{Hawking:1975vcx} in moving fluids \cite{Unruh:1980cg}, has matured into a complex interdisciplinary field where theoretical insights and cutting edge experiments intertwine. The accumulated literature (both theoretical and experimental) on the topic is vast, and many experiments were and are being performed. Here, we do not attempt a complete review, and instead refer to the existing ones \cite{barcelo2011analogue,novello2002artificial,faccio2013analogue,braunstein2023analogue,almeida2023analogue,schutzhold2025ultra}. We instead aim at providing a pedagogical account of the foundations and some aspects of the field, introducing in detail the essential concepts and techniques in gravitational physics, field theory, and many body physics needed to understand the topic at the level of current research.

The original analogy to investigate the Hawking effect in classical fluids has extended in two directions. On the one hand, other platforms such as Helium 3 \cite{kopnin1998critical}, Bose Einstein condensates, non-linear optics \cite{philbin2008fiber,belgiorno2010hawking}, gravity waves in water \cite{schutzhold2002gravity,rousseaux2008observation,weinfurtner2011measurement}, fluids of light \cite{marino2008acoustic,nguyen2015acoustic},  superconducting circuits \cite{nation2009analogue} have been proposed and experimentally investigated as laboratory analogues of the Hawking effect. On the other hand, the analogy has transcended the Hawking effect, and analogue simulators for other phenomena in (relativistic) field theory have been put forward as well. Examples include superradiance, particle creation in an expanding universe, the Unruh effect, vacuum decay, quasi-normal modes, to name a few.

Most of these phenomena were originally predicted within the framework of Quantum Field Theory (QFT) in curved spacetime, i.e. in the presence of external gravitational fields. However, we must keep in mind that the interest in these phenomena goes beyond the analogy with quantum fields in gravitational scenarios. In general, they are realizations of QFTs in presence of external classical fields. Having laboratory realizations of QFTs interacting with classical background fields allows to probe a regime of QFT in which experimental data is scarce. In particular, analogue quantum simulators of QFTs have the potential to probe quantum aspects of these phenomena, which are at the heart of key fundamental predictions such as the expected quantum origin of the Cosmic Microwave Background inhomogeneities that evolved into the rich structure that we observe today. Needless to say, experimentally accessing non-perturbative quantum aspects of QFT can have a profound impact on our understanding of our most fundamental framework to describe nature, potentially shedding light into some of the most basic questions about our universe.

For what concerns Bose-Einstein condensates, after they were achieved experimentally in 1995, their fine experimental control was quickly developed. This tunability, together with their low temperature, made them one of the first platforms to be considered for the study of the analogue Hawking effect \cite{garay2001sonic,barcelo2001analogue}. Theoretical work on the Hawking effect in BECs has been extensive, with some of the earliest works being \cite{leonhardt2003theory,macher_blackwhite_2009,balbinot2008nonlocal,carusotto2008numerical,recati2009bogoliubov,larre2012quantum}. The first experimental realization of a 1+1 dimensional analogue black hole was reported in \cite{lahav2010realization}. The same group then performed a series of experiments investigating Hawking emission and its spectrum \cite{steinhauer2014observation,steinhauer2016observation,munoz2019observation,kolobov2021observation}. Besides Hawking radiation, other experiments have also investigated the physics of expanding universes \cite{eckel2018rapidly,viermann2022quantum}. Theoretically, other phenomena have also been considered, such as cosmological pair production \cite{prain2010analogue}, black hole superradiance \cite{Giacomelli:2019tvr,giacomelli2021superradiant,giacomelli2021understanding,giacomelli2021spontaneous} and the Unruh effect \cite{gooding2020interferometric}.

These lecture notes aim at providing a conceptual and technical introduction to the use of atomic Bose-Einstein condensates as tools to explore effects in quantum and classical field theories. As far as the effects go, we will focus on superradiance and the Hawking effect. 
The presentation is structured as follows. After a brief introduction of some general concepts in Analogue Gravity in Section \ref{sec:Basics}, we will give an account of the theory of weakly interacting BECs in Section \ref{sec:BECBasics}. In Section \ref{sec:LowEnergyExcitations}, we devoted particular attention to the description and properties of linear excitations and to their quantization in Section \ref{sec:QuantumSound}. The field describing these excitations is the central object of Analogue Gravity and we put a specific focus on the relation with the treatment of quantization in relativistic field theories, since we found that the standard treatment in the cold-atom community can be confusing for someone more accustomed to that perspective. In Section \ref{sec:SR} we then proceed to describe the phenomenon of superradiant scattering and its quantum counterpart, spontaneous pair production. Although superradiance occurs in more structured spacetimes than the Hawking effect, from the analogue point of view superradiance is simpler, and also the analogue Hawking effect can be seen as a type of superradiance. Finally, in Section \ref{sec:Hawking} we discuss the analogue Hawking effect and its relation to the astrophysical Hawking emission. In Appendix \ref{app:ComputationScattCoef} we provide an algorithm to compute scattering coefficients in stationary setups, which can be applied to superradiance and analogue Hawking emission, and in Appendix \ref{app:Problems} we provide a list of solved problems that complement the notes.

\section{Basic concepts in analogue gravity}
\label{sec:Basics}
Wave equations are hyperbolic PDEs which have an in-built causal structure that can be obtained from their characteristic curves. For example, the characteristic curves of Maxwell equations provide the light cone structure which describes the propagation of electromagnetic waves. These are one of the most universal types of equations in physical phenomena, and they can be thought generally as describing perturbations around a background solution in generic physical systems \cite{Barcelo:2007mga}. Remarkably, even in systems without a definite causal structure, the evolution of perturbations usually displays a hyperbolic structure dictated by the background solution. Physically, this reflects into a finite propagation speed via an in-built lightcone, leading to an emergent causal structure \cite{Barcelo:2007mga,Barcelo:2001ah,Barcelo:2001cp}.

For waves that propagate in media, such as fluids, the properties of the media can modify this causal structure in nontrivial ways. Homogeneous and isotropic media will lead to a trivial causal structure with the corresponding associated global symmetry group. This is parallel to what occurs to relativistic fields in Minkowski spacetime, which can be encapsulated by a global Lorentz symmetry of the propagation equations. Interestingly, departures from homogeneity will modify this in-built causal structure in a similar manner in which a non-vanishing gravitational field deforms the Minkowskian lightcones. These modifications inply that the in-built lightcones can tilt, and the global causal structure felt by the waves can develop interesting features such as causal horizons or ergoregions. 

For sound waves in fluids, these structures can be reproduced by sub-to-supersonic transitions in different directions in which the fluid propagates. For instance, an analogue ergoregion occurs in the supersonic region of a system with a sub-to-supersonic transition. There sound waves cannot propagate against the local flow. An analogue of a single event horizon requires a sub-tosupersonic transition whose supersonic region is also trapping. This is usually achieved by having the local flow velocity perpendicular to the surface defining the sub-to-supersonic transition, and pointing towards the supersonic region. As a result, sound waves emitted in the supersonic region cannot escape, in full analogy to their trapping by a black hole\footnote{Strictly, this is only true for idealized fluids which are inviscid, irrotational and barotropic. In more realistic fluids, which usually violate some of these conditions, sound propagation is relativistic only at large wavelenghts. In some cases, short wavelength perturbations can travel against the flow in supersonic regions.}. 

Naively, we can derive an acoustic metric controlling the behavior of sound waves in fluids as follows. If the soundspeed is $\cs$, the position of a sound wave propagating in direction $\hat{\vk}$ emitted at a point where the local flow has velocity $\vtv$ will change as
\begin{equation}
    \frac{\dd \vx}{\dd t}= \cs \hat{\vk} + \vtv.
\end{equation}
One can then write $(\dd \vx-\vtv \dd t)^2= \cs^2 \dd t^2$. Postulating the \textit{sound rays}\footnote{These are actually , the characteristic curves of the corresponding wave equation.} to be null directions of our acoustic spacetime, so that the line element has to vanish on these curves, leads to the acoustic line element
\begin{equation}
    \dd s^2=-(\cs^2-\tv^2)\dd t^2 - 2\vtv \cdot \dd \vx \dd t + \dd \tx^2,
\end{equation}
where bold letters are vectors and their light versions are their modulus with the euclidean norm. This results in the acoustic metric with components in $(\vx,t)$ coordinates
\begin{equation}
    [g_{\mu\nu}]\propto \begin{pmatrix}
        -(\cs^2-\tv^2) & -\vtv\\
        -\vtv^\top & \mathbb{I}_d
    \end{pmatrix},
\end{equation}
where the square brackets stems for the matrix representation of the (0,2) tensor $g_{\mu\nu}\dd \tx^\mu \dd \tx^\nu$, $\tx^0=t$ and $\tx^{i}$ are the components of the possition vector $\vx$. Here $d$ is the spatial dimension, greek indices run from $0$ to $d$ and latin indices from $1$ to $d$. Note that, in general, $\cs$ and $\vtv$ can be functions of space and time. 

From the way we arrived at this metric, we conclude that these null geodesics are the characteristic curves of the PDE describing sound wave propagation. Furthermore, we did not assume anything in particular about the system, except from the existence of a set of characteristic curves that define an absolute lightcone for the sound waves, which is true for linear 2nd order hyperbolic PDEs. In fact, this argument above can be formalized by a theorem that applies to some kinds of fluids as follows. For a proof of the theorem see section 2.3 of \cite{barcelo2011analogue}.

\begin{theorem}{\textbf{--- Acoustic metric for sound waves}}
\label{thm:AcousticMetric}

    Let a fluid with density $\rho$, pressure $p$ and velocity $\vtv(\vx,t)$ be an isolated, inviscid, barotropic $\rho(\vx,t)=f(p(\vx,t))$, and irrotational $\vna\wedge\vtv=0$ fluid. The irrotational condition ensures that there is a velocity potential $\theta$ such that $\vtv=\vna\theta$. 
    
    The theorem states that linear perturbations $\theta_1$ of the velocity potential $\theta=\theta_0+\theta_1$ are described by a curved spacetime Klein-Gordon equation
    \begin{equation}
        \Box_g \theta_1=\frac{1}{\sqrt{-|g|}}\partial_\mu \sqrt{-|g|}g^{\mu\nu}\partial_\nu\theta_1=0 
    \end{equation}
    whith a d'Alambertian operator associated to a Lorentzian acoustic metric which in cartesian components is of the form
    \begin{align}
        [g_{\mu\nu}]=\frac{\rho}{\cs^{\frac{2}{d-1}}}
        \begin{pmatrix}
            \tv^2-{\cs^2} & -\tv^i\\[.3cm]
            -\tv^j &\hspace{.2cm} \delta^{ij}
        \end{pmatrix},
        \label{eq:GeneralAcousticMetric}
\end{align}
where $d$ is the spatial dimension and $\cs^2=\frac{\dd p}{\dd\rho}$ is the square of the speed of sound. 
\end{theorem}

This metric will give a correct description of the propagation of small perturbations in the fluid if the speed of sound is independent on the wavevector $\vk$ of the perturbations, i.e. if their dispersion relation is linear $\omega=c_s {\tk}$. In general, {the microscopic physics of a real fluid} will make the speed of the perturbations depend on their wavevector by introducing {higher spatial derivatives in the wave equation (corresponding to nonlinearities in their dispersion relation).} The modifications are classified as \textit{subluminal} if $\cs$ is an upper limit for the velocity of the perturbations, and \textit{superluminal} if high wavevectors $\vk$ can travel faster than $\cs$. In the later case, the nontrivial causal structure generated by the background flow is only seen by low $\tk$ modes. In these lectures we will deal with Bose Einstein condensates which, as we will see, have a superluminal dispersion relation.

\subsection{Ergoregions and horizons: an acoustic black hole through a draining vortex}
\label{sec:vortex-geometry}

\begin{figure}
    \center
    \includegraphics[width=0.4\textwidth]{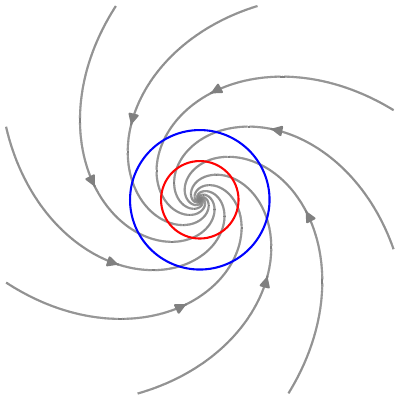}
    \caption{Streamlines of the vortex geometry. The location of the ergosurface and of the horizon are indicated respectively with a blue and a red circle.}
    \label{fig:vortex-geometry}
\end{figure}

Ergoregions and horizons are familiar concepts in gravitational physics which play a central role in black hole physics and, therefore, in analogue gravity. We will now explore those concepts, being as general as possible, but providing specific examples through draining vortex flows for illustrative purposes.

Starting general, from the theorem in the previous section \eqref{thm:AcousticMetric}, we know that sound waves in an inviscid, barotropic and irrotational fluid propagate causally according to the acoustic metric \eqref{eq:GeneralAcousticMetric}. In 2+1 D and using cartesian spatial coordinates and laboratory time, the corresponding line element is
\begin{equation}
    \dif s^2=\frac{\rho}{\cs^2}\lrsq{ (\tv^2-\cs^2)\dif t^2-2\vtv\cdot \dif\mathbf{x}\,\dif t+\dif\mathbf{x}^2},
\end{equation}
where $\cs$ is the speed of sound in the fluid. Given this metric, we can discuss some generic features of acoustic spacetimes, without having to specify the density or velocity profiles. 

Let us start with symmetries. Symmetries of a metric, a.k.a.  isometries, are geometrically realized by the presence of Killing vector fields\footnote{A vector field $X$ is a Killing vector field of a metric $g$ if the Lie derivative of the metric along $X$ vanishes $\mathcal{L}_X g=0$. The integral curves of a Killing vector field are the orbits of the 1-parameter group of isometries it generates.}. In stationary fluid profiles where either the flow velocity, its density, or its sound speed do not depend on laboratory time $t$, the vector $\partial_t$ is a Killing vector field of the acoustic metric. Its causal character is obtained from
\begin{equation}
    g(\partial_t,\partial_t)=\frac{\rho}{\cs^2}(\tv^2-\cs^2).
\end{equation}
Thus, we see that $\partial_t$ is timelike in regions where $\tv<\cs$, null where $\tv=\cs$, and spacelike in regions where $\tv>\cs$. A spacetime is said to be stationary\footnote{A stronger definition of stationary requires that the Killing vector is everywhere timelike \cite{Wald:1995yp}. We will use stationary in this weaker sense so that rotating black holes are within the category of stationary spacetimes.} if there is a Killing vector field which is timelike in an open neighborhood of infinity. This definition of stationary is equivalent to the requirement that there is a Killing vector field which is asymptotically timelike at infinity.

Now, imagine we have an acoustic spacetime where the flow velocity vanishes asymptotically, but it becomes larger than the speed of sound in a given region. Assymptotically, $\partial_t$ is tangent to the worldlines of observers who are at rest with respect to the laboratory frame, and is a timelike Killing vector. However, in the region where $\tv>\cs$, $\partial_t$ becomes spacelike.  Such region is called an (acoustic) ergoregion, and the locations at which $\tv=\cs$ constitute its ergosurface \cite{barcelo2011analogue}. An example of acoustic metric with an ergoregion is the 2 dimensional (2+1 D) vortex flow. This acoustic metric is realized by a fluid which rotates around the center of the vortex. To describe it, it is easier to use polar coorindates $(r,\theta)$ adapted to the vortex. In those coordinates, where the velocity of a general flow is $\vtv=\tv_r\hat{\mathbf{r}}+\tv_\theta\hat{\boldsymbol{\theta}}$, the acoustic metric reads
\begin{equation}
    \dif s^2=\frac{\rho}{\cs^2}\lrsq{(\tv^2-\cs^2)\dif t^2-2(\tv_r \dif r\,+r\tv_\theta \dif \theta)\dif t+\dif r^2+r^2\dif\theta^2},
\end{equation}
The mentioned vortex geometry is typically realized through the velocity profile
\begin{equation}
    \vtv=\frac{B}{r}\hat{\boldsymbol{\theta}},
\end{equation}
so that the corresponding acoustic metric is
\begin{equation}
    \dif s^2_{\rm vortex}=\frac{\rho}{\cs^2}\lrsq{\lr{\frac{B^2}{r^2}-\cs^2}\dif t^2-2B \,\dif \theta\dif t+\dif r^2+r^2\dif\theta^2},
\end{equation}
where we have used $\tv_r=0$ and $\tv_\theta=B/\cs$.
In these coordinates, it is explicit that $\partial_\theta$ is a Killing vector field of the vortex geometry, so that in addition to being stationary, the vortex flow is also invariant under rotations. Furthermore, note that there is a radius 
\begin{equation}
 \rerg=\frac{B}{\cs}   
\end{equation}
at which $\tv=\cs$, with $\tv>\cs$ at smaller radii. This is a paradigmatic example of an ergoregion, enclosed by an ergosurface. Within the ergoregion, where the flow rotates faster than the speed of sound, acoustic observers cannot remain static in the laboratory frame and must co-rotate with the flow. We will later see that this feature is tied to superradiant amplification of waves which scatter with the vortex (see section \ref{sec:SR-vortex}).

In adition to the ergoregion, this 2D vortex geometry can easily be generalized to accommodate a black hole. To that end, one adds radial flow towards the center of the vortex, making a 2D draining vortex flow with a velocity profile required by the fluid equations to be of the form
\begin{equation}
    \vtv=-\frac{A}{r}\hat{\boldsymbol{r}}+\frac{B}{r}\hat{\boldsymbol{\theta}},
    \label{eq:vortex-geometry-flow}
\end{equation}
where $A$ controls the radial velocity and $B$ the angular one. A representation of the streamlines of this flow is shown in Figure \ref{fig:vortex-geometry}. The acoustic metric resulting from this velocity profile reads
\begin{equation}
    \dif s^2=g_{\mu\nu}\dd x^\mu\dd x^\nu=\frac{\rho}{\cs^2}\lrsq{\lr{\frac{A^2+B^2}{r^2}-\cs^2}\dif t^2+2\lr{\frac{A}{r} \dif r- B\, \dif \theta}\dif t+\dif r^2+r^2\dif\theta^2},
    \label{eq:AcousticMetricPolar}
\end{equation}
This metric also contains an ergoregion, enclosed by the ergosurface which is now at
\begin{equation}
 \rerg=\frac{\sqrt{A^2+B^2}}{\cs}.  
\end{equation}
Furthermore, this metric also contains a trapped surface inside the ergoregion at
\begin{equation}
 \rh=\frac{A}{\cs}.
\end{equation}
The presence of this trapped surface is independent of the rotation of the flow, and encloses the interior of an acoustic black hole. This trapped surface is the black hole horizon, which forms where the radial velocity overcomes the speed of sound of the fluid. The interior of the acoustic black hole is part of the acoustic ergoregion. In fact, in the case of pure drain without rotation ($B=0$), we find $\rerg=\rh$, and the interior of the black hole is the full ergoregion. This is a generic feature of black holes: their interior is an ergoregion, where there can be no acoustic observers which are static with respect to geodesic observers at infinity (i.e. with respect to laboratory observers). 

The physical difference between the \textit{rotational ergoregion} of a vortex geometry and the ergoregion in the black hole interior is that acoustic observers can escape rotational ergoregions but cannot escape from the interior of the black hole. This is due to the horizon being a (marginally) trapped surface, i.e. a closed surface such that the flow velocity in the direction of the inward normal to that surface is equal to the speed of sound of the fluid \cite{barcelo2011analogue}. This is not true for an ergoregion formed due to rotational velocity, since the flow velocity at the corresponding ergosurface has a component tangent to the surface, and hence the normal component is necessarily smaller than the speed of sound. Thus, while the ergosurfaces formed due to inward radial flow are also black hole horizons, those formed due to rotational flow (or a combination of rotation + radial flow) are not trapped surfaces. In the draining vortex geometry we have the ergosurface at $\rerg$, which is not trapped, and the trapped surface $\rh$ which is an acoustic black hole horizon. This structure is parallel to the one found in rotating gravitational black holes. In fact, the draining vortex metric can be seen as the equatorial slice of a rotating Kerr black hole \cite{Visser:2004zs}.

As a last remark that will be useful later, let us discuss the surface gravity of the rotating acoustic black hole we presented. To that end, we introduce the vector 
\begin{equation}
    \xi=\partial_t+\omh\partial_\theta, 
\end{equation}
where $\omh=B/\rh^2=B \cs^2/A^2$ is the angular velocity of the horizon. This is a Killing vector, since it is a linear combination of Killing vectors with constant coefficients. Its causal character is given by the sign of
\begin{equation}
    g(\xi,\xi)=\rho\lr{\frac{\rh^2}{r^2}-1}\lrsq{\frac{\rerg^2+r^2}{\rh^2}-\frac{r^2\rerg^2}{\rh^4}}
\end{equation}
where we have used several relations between the parameters characterizing the draining vortex. Note that this Killing vector field is null at the horizon, timelike in the region between the horizon and $\rerg\lr{\rerg^2/\rh^2-1}^{-1/2}$, and spacelike at larger radii. It can be seen that this Killing vector field is the generator of the horizon, and therefore the surface gravity $\kappa$ of the acoustic black hole, which is defined by $\xi^a\nabla_a\xi^b=\kappa\xi^b$, is given by \cite{barcelo2011analogue}
\begin{equation}
\label{eq:SurfGravDV}
    \kappa=\left.\cs\frac{\partial(\cs-\tv_r)}{\partial r}\right|_{\rh}.
\end{equation}
\section{Basics of Bose-Einstein condensates}
\label{sec:BECBasics}
Here we introduce the notions about Bose Einstein condensates (BECs) that we will need to discuss analogue gravity experiments. This is far from a complete account of their physics and we will not talk about the experimental realization. For complete references we redirect to \cite{pitaevskii2016bose,pethick2008bose,castin2001bose}.

Bose-Einstein condensation occurs when, in a system of many bosonic microscopic constituents\footnote{These bosonic components are typically atoms, and we will call them microscopic or atomic degrees of freedom (dof). However, bear in mind that these microscopic dof can be other more exotic states of matter.}, a macroscopic fraction of the particles occupies the same single-particle state. These particles are called the condensate. For free bosonic constituents at zero temperature, all of them would be in the condensed state, but finite temperature effects and particle-particle interactions result in what are known as thermal and quantum depletion of the condensate. In these notes we will focus on the zero-temperature case. 
As we will see, excitations above the many-body ground state can be described as collective excitations of the gas that can be understood as quantum sound waves, or phonons. The general idea of analogue gravity in these systems is that the condensate will form the \textit{classical curved spacetime}, while the quantum field describing the phonons will be the quantum field propagating in this spacetime.

\subsection{Weakly interacting dilute Bose gas}\label{sec:WeakBoseGas}

To understand under which conditions an atomic BEC typically forms let us start by considering a \textit{dilute} gas of bosons\footnote{This discussion is based on the one of Section 4.1 of \cite{pitaevskii2016bose}.}. Diluteness is defined in terms of the mean inter-atomic distance $d_0$ and the range of inter-atomic forces $r_0$ by the condition $d_0\gg r_0$. This condition has some important consequences for the description of the system
\begin{enumerate}
    \item Only interaction events involving two particles are important, hence one can neglect configurations in which three or more particles interact simultaneously. This implies that the microscopic dof can be modeled by a quartic Hamiltonian
\begin{equation}
\begin{split}
    H=\int_\Sigma \dd^d\tx &\lrsq{\frac{\hbar^2}{2m}\vna\hat{\Psi}^\dagger(\vx)\cdot\vna\hat{\Psi}(\vx)+\hat{\Psi}^\dagger(\vx)V_\textrm{ext}\hat{\Psi}(\vx)}
    \\
    &+\frac{1}{2}\int_{\Sigma^2} \dd^d\tx \,
\dd^d \mathrm{y}\, \hat{\Psi}^\dagger(\vx)\hat{\Psi}^\dagger(\mathbf{y})V(\vx-\mathbf{y})\hat{\Psi}(\vx)\hat{\Psi}(\mathbf{y}),
\end{split}
\label{eq:AtomicHamiltonian}
\end{equation}
where the field operator $\hat{\Psi}(\vx)$ destroys an atom at position $\vx$ and $\hat{\Psi}^\dagger(\vx)$ creates it, so that their commutation relations are $[\hat{\Psi}(\vx),\hat{\Psi}^\dagger(\vy)]=\delta(\vx-\vy)$, and $d$ are the number of spatial dimensions on which the condensate effectively lives in. We have included an external potential $V_\textrm{ext}$, as typically used in experiments to trap the system.
    
\item Since the particles are in average far apart one can describe the scattering between two of them with the asymptotic expression of the wavefunction which. To leading order in the partial wave expansion, the interaction depends only on the scattering angle and the (absolute value of) the momentum of the incoming wave $\hbar\vk$ through the scattering amplitude $f(\theta,\tk)$.
    
\item The typical momentum of the atoms $\hbar\vk_0$ will always be small compared to the range of interaction, i.e. $\tk_0 r_0\ll 1$. In this regime, the scattering amplitude becomes independent of the angle and momentum and is hence a constant: the s-wave scattering length $f(\theta,\tk) a$. More in general, low momenta guarantee that we can replace the $2$ body potential $V(\vx-\mathbf{y})$, typically diverging at vanishing distance, by an effective soft version of it\footnote{Basically one can think of this soft version of the potential as the infrared Fourier component of it $\displaystyle V_\mathrm{soft}(\vx)\coloneqq\lim_{\vk\rightarrow0}\int_\Sigma \dd^d\mathrm{y} \, \ee^{\ci \vk\cdot\mathbf{y}}V(|\mathbf{y}-\vx|)$.} that reproduces the infrared scattering.  Low momenta futher imply that we can consider only the zero-momentum component of $V_{\mathrm{soft}}$. This constant potential in momentum space corresponds in real space to a contact potential $V_\mathrm{soft}(\vx-\mathbf{y})\sim g \delta(\vx-\mathbf{y})$, with the interaction constant
\begin{equation}
    g\coloneqq\int_\Sigma \dd^d \vx V_{\mathrm{soft}}(\vx)=\frac{\Omega_d\hbar^2 a}{m},
\end{equation}
where the second equality establishes the relation with the s-wave scattering length and $\Omega_d$ is the $d$-dimensional solid angle.
\end{enumerate}

Notice that the diluteness condition does not imply that the interaction is weak, that is the scattering length could be large with respect to the interparticle distance. However, for a BEC to form, the dilute Bose gas needs to be also \textit{weakly interacting}, that is $|a|\ll d_0$. Hence, from now on we will focus on the \textit{weakly interacting dilute Bose gas}.

Under these assumptions, we can rewrite Hamiltonian \eqref{eq:AtomicHamiltonian}
\begin{equation}
    \hat{H}=\int_\Sigma \dd^d\tx\, \lrsq{\frac{\hbar^2}{2m}\vna\hat{\Psi}^\dagger(\vx)\cdot\vna\hat{\Psi}(\vx)+\hat{\Psi}^\dagger(\vx)V_\textrm{ext}\hat{\Psi}(\vx)+\frac{g}{2}\hat{\Psi}^\dagger(\vx)\hat{\Psi}^\dagger(\vx)\hat{\Psi}(\vx)\hat{\Psi}(\vx)}.
    \label{eq:GPHamiltonian}
\end{equation}
Using the standard properties of Dirac deltas, the Heisenberg equations of motion for the above Hamiltonian yield the evolution equation\footnote{$\hat\Psi(t,\vx)$ is here the Heisenberg picture atomic annihilation operator.}  (see Problem \ref{prob:QuantumGPE})
\begin{equation}\label{eq:heisenberg-eom}
\ci\hbar\partial_t \hat{\Psi}(t,\vx)=\lrsq{-\frac{\hbar^2}{2m}\nabla^2+V_{\mathrm{ext}}+g\hat{\Psi}(t,\vx)^\dagger\hat{\Psi}(t,\vx)}\hat{\Psi}(t,\vx).
\end{equation}

\subsection{Classicalization: the Bogoliubov approximation and the Gross-Pitaevskii equation}

As explained before, at vanishing temperature (almost) all the atoms in the system will be in the single-particle ground state, characterized by the single-particle wavefunction $\Psi_0(t,\vx)$. It is then convenient to write the bosonic field operator as
\begin{equation}
    \hat\Psi(t,\vx)=\Psi_0(t,\vx)\hat a_0 +\sum_{i\neq 0}\Psi_i(t,\vx)\hat a_i,
\end{equation}
where the operators $\hat a_i$ destroy particles in the $i$-th single-particle state of wavefunction $\Psi_i(t,\vx)$. For a $N$-particle system, the (approximate) condensed state is
\begin{equation}
    \ket{\mathrm{con}(N)}=\frac{(\hat a_0)^N}{\sqrt{N!}}\ket{0}.
\end{equation}
The main point of the classical mean-field treatment of BECs lays in the fact that this condensed state can, for a large number of particles, be approximated by the following coherent state
\begin{equation}
    \ket{\mathrm{coh}(\Phi)}=e^{\Phi\hat a_0^\dag -|{\Phi}|^2/2}\ket{0}=e^{-\absv{\Phi}^2/2}\sum_{M=0}^\infty \frac{\Phi^M(\hat a_0^\dag)^M}{M!}\ket{0},
\end{equation}
    with the requirement $\bra{\mathrm{coh}(\Phi)}\hat a_0^\dag\hat a_0\ket{\mathrm{coh}(\Phi)}=\absv{\Phi}^2=N$. This is a state that does not have a well defined number of particles but that becomes indistinguishable from $\ket{\mathrm{con}(N)}$ in the thermodynamic limit (i.e. $\bra{\mathrm{con}(M)}\ket{\mathrm{coh}(\Phi)}\to \delta{(N-M)}$ for $N\to\infty$)\footnote{The use of states with a non-fixed number of states is related to working with the grand-canonical ensemble.}.

    An interesting mathematical property of the coherent state is the fact that it is an eigenstate of the annihilation operator, i.e. $\hat a_0\ket{\mathrm{coh(\Phi)}}=\Phi\ket{\mathrm{coh(\Phi)}}$. One can hence think of the condensed state as being characterized by $\langle\hat a_0\rangle\neq 0$. This essentially allows to treat it as a classical field, meaning that the quantum fluctuations on the number of particles in the condensate are much smaller than their number $\langle\hat a_0\rangle\sim\sqrt{N}\gg \langle[\hat a_0,\hat a_0^\dag]\rangle$. This is often referred to as \textit{Bogoliubov approximation} and, in practice, corresponds to \textit{substituting} the quantum field $\hat\Psi(\vx)$ in the Heisenberg equation \eqref{eq:heisenberg-eom} with a classical field $\Psi(\vx)$, usually referred to as \textit{macroscopic wave function} or \textit{order parameter}, since it is non-zero in the condensed state\footnote{Crudely speaking one can think of this field as $\Psi(\vx)=\psi_0(\vx)\Phi$.}.

This leads to the Gross-Pitaevskii equation (GPE)
\begin{equation}\label{eq:TimeDepGPE}
\ci\hbar\frac{\partial {\Psi}}{\partial t}=\lrsq{-\frac{\hbar^2}{2m}\nabla^2+V_{\mathrm{ext}}+g n}{\Psi}
\end{equation}
where $V_{\mathrm{ext}}$ is a potential used to trap the system and $\Psi(t,\vx)$ is normalized so that $\int_\Sigma \dd^d x \absv{\Psi(t,\vx)}^2=N$, so that $n(\vx)=|\Psi(t,\vx)|^2$ is the local density of condensed particles. While the coherent state we are using is not an eigenstate of the number of particles, it is an eigenstate of the \textit{phase}. Hence, though the original Hamiltonian \eqref{eq:GPHamiltonian} is invariant under a change of phase of the quantum field, the condensation will spontaneously break this symmetry.

An important class of solutions to the GPE are stationary ones. We say that a solution to the GPE (or a condensate) is stationary if there is a frame of reference in which $g$ and $V_\mathrm{ext}$ are time-independent. In that case, there are solutions where the time dependence of the wavefunction can be trivialized in a global phase that evolves linearly with time. In other words, in that case, we have a time independent GPE Hamiltonian. In that frame of reference, an eigenstate of the Hamiltonian with eigenvalue $\mu$ is of the form $\Psi(t,\vx)=\Psi_0(\vx) \ee^{-\ci {\mu t}/{\hbar}}$, where $\Psi_0(\vx)$ satisfies the stationary GPE
\begin{equation}
\lrsq{\frac{\hbar^2}{2m}\nabla^2+(\mu-V_\mathrm{ext})-gn}\Psi_0=0.
    \label{eq:StaionaryGPE}
\end{equation}
 These eigenstates are also called stationary solutions to the GPE with chemical potential $\mu$. In a homogeneous stationary condensate of density $n$ which flows with a homogeneous velocity\footnote{The velocity of the condensate is the gradient of its phase multiplied by $\hbar/m$, as will be explained later.} $\vtv$, the above equation turns into an equation of state which takes the form
\begin{equation}
    \mu=\frac{m}{2}\tv^2+V_\mathrm{ext}+gn.
    \label{eq:ChemPotHomog}
\end{equation}
This tells us that the chemical potential $\mu$ is the sum of the kinetic and potential energies of a particle in the condensate (i.e. the energy needed to add another atom to the condensate). The total energy of a stationary homogeneous condensate can be then be obtained as
\begin{equation}
    E=\int \mu dN = \frac{g N^2}{2V}+N\lr{\frac{m}{2}\tv^2+V_\mathrm{ext}}.
\end{equation}
One can see that this result is consistent if one computes the energy of a grand canonical ensemble with chemical potential $\mu$ described by the (classical version of the) GPE Hamiltonian.  For completeness, let us also note that the GPE can be equivalently derived from the Lagrangian density
\begin{equation}
    \label{eq:GPELag}
    \mathcal{L}_{\rm GPE}=\frac{\ci\hbar}{2}\big(\Psi^*\partial_t\Psi-(\partial_t\Psi^*)\Psi\big)-\frac{\hbar^2}{2m}\vna\Psi^*\cdot\vna\Psi-V_{\rm ext}|\Psi|^2-\frac{g}{2}|\Psi|^4,
\end{equation}
which can be useful to analyze the physics of small excitations on top of the condensed state in a systematic way, akin to standard analysis in field theory. \eqref{eq:GPHamiltonian}.

\subsection{Density-phase variables and the hydrodynamical approimation}\label{sec:HydroBckg}

There is a way to parametrize the complex scalar field in terms of variables that connect directly to macroscopic properties of the system, which are its density and phase. This representation is actually general for any quantum fluid, and it is usually called the Madelung or hydrodynamical representation for $\Psi$ (the reason will become apparent in a moment).

Te hydrodynamical representation is a field redefinition (you can see it as a change of coordinates in phase space) that rewrites the real variables $\mathrm{Re}[\Psi]$ and $\mathrm{Im}[\Psi]$ in terms of the density and phase of the condensate. This field redefinition can be found from $\Psi=\sqrt{n}\ee^{\ci S}$ and is defined by
\begin{equation}
\begin{split}
        &n=|\Psi|\\
        &S=\text{Arg}(\Psi)
\end{split}
\quad\Longleftrightarrow\quad
\begin{split}
        &\mathrm{Re}[\Psi]=\sqrt{n}\cos S\\
        &\mathrm{Im}[\Psi]=\sqrt{n}\sin S
\end{split}
 \label{eq:DPtoAtomic}
\end{equation}
which consists on passing from cartesian to polar coordinates in the complex plane. In this representation, the real and imaginary parts of the GPE for the complex field $\Psi$ provide two equations for the real fields $n$ and $S$ which read
\begin{align}
\begin{split}
&\partial_t n+\frac{\hbar}{m}\vna\cdot(n \vna S)=0\,,
\\
&\hbar\partial_t S+\frac{\hbar^2}{2m} (\vna S)^2+V_\mathrm{ext}+gn\lr{1-\frac{\xi^2}{2}\frac{\nabla^2\sqrt{n}}{\sqrt{n}}}=0\,,
\end{split}
\label{eq:DensityPhaseGPE}
\end{align}
which, by defining $\vtv\coloneqq\frac{\hbar}{m}\vna S$, can be rewritten as a continuity equation plus an Euler equation for an inviscid irotational fluid
\begin{align}
&\partial_t n+\vna\cdot(n \vtv)=0\\
&\partial_t m\vtv+\vna\lrsq{\frac{1}{2}m \tv^2+V_\mathrm{ext}+gn\lr{1-\frac{\xi^2}{2}\frac{\nabla^2\sqrt{n}}{\sqrt{n}}}}=0,\label{eq:HydrodynamicEqs}
\end{align}
where
\begin{equation}
    \xi=\frac{\hbar}{\sqrt{mgn}}
    \label{eq:HealingLength}
\end{equation} 
is called the healing length of the condensate (see below). If we interpret it as the velocity of a fluid. The Euler equation can also be written as
\begin{align}
m n\left(\partial_t\vtv+(\mathbf{v} \cdot \vna) \mathbf{v} + \frac{\vna V_{\mathrm{ext}}}{m}\right)=-\vna\lrsq{\frac{g n^2}{2}\lr{1+\frac{\xi^2}{2}\frac{\nabla^2\sqrt{n}}{\sqrt{n}}}},
\end{align}
where the $\xi-$dependent term provides a quantum correction to the pressure due to interactions\footnote{Note this contribution to the equations is independent of the interaction coupling since $gn^2\xi^2=\hbar^2n/m$. This quantum contribution to the dynamics is also known by quantum or Bohm potential, as it was identified by D. Bohm when he presented his hidden-variable theory as a potential interpretation of quantum mechanics \cite{Bohm:1951xw,Bohm:1951xx}.}. Particularizing for stationary condensates defined by a time-independent density and velocity profiles, requires $\partial_t V_\mathrm{ext}=0$ and $S(t,\vx)=-\mu t/\hbar+S_\mu(\vx)$, which leads again to the continuity equation and a generalized equations of state of the form
\begin{align}
&\vec{\nabla}\cdot(n \vtv)=0\\
&\mu=\frac{1}{2}m \vtv^2+V_\mathrm{ext}+gn\lr{1-\frac{\xi^2}{2}\frac{\nabla^2\sqrt{n}}{\sqrt{n}}}.
\label{eq:StationaryDensPhase}
\end{align}
Having this general equation of state, we notice that, strictly speaking, the theorem \eqref{thm:AcousticMetric} does not apply for BECs when seen as fluids. This is because the quantum pressure precludes BECs from being barotropic fluids. Indeed, the quantum contribution to pressure implies that the value of the density at a point is not enough to compute the pressure. One also needs to know the density in a neighbourhood of that point (to compute the Laplacian of the density). Interestingly, the relevance of this contribution to the total pressure is controlled by the healing length. If the density varies on a characteristic length scale $\lambda$, the quantum contribution to pressure will be of order $(\xi/\lambda)^2$. Hence, we can safely neglect it provided that variations in the background density occur within length scales much greater than the healing length $\lambda\gg\xi$. In these case, we say that the we are in the hydrodynamical regime of the condensate, or that the hydrodynamic approximation holds. On the contrary, the quantum contribution to pressure dominates when these variations in the condensate density occur at length scales shorter than the healing length $\lambda\ll\xi$, and the hydrodynamic approximation breaks down. 

We can then see the healing length as a scale below which there is a transition from a barotropic behavior to a non-barotropic behavior, dominated by quantum contributions pressure. Well below the healing length, we can apply the theorem \eqref{thm:AcousticMetric}, and we will be able to model the condensate as an analog spacetime with an acoustic metric describing the propagation of sound waves. Note, however, that the analogy is not an analogy of the dynamics of the gravitational field by the condensate, since Einstein's equations are different from the GPE, but it will rather be an analogy of the dynamics of a massless scalar field propagating in a given curved spacetime, with geometry dictated by the specific condensate profile. The object that will be analogue to the scalar field will not be the condensate itself, but its low-energy excitations, as we will see later.

\subsection{Canonical structure of the GPE and a Noether current}

In order to better understand the physics behind the GPE, let us  discuss its canonical structure, and understand how many degrees of freedom it describes. Naively, one would think that we have two scalar degrees of freedom corresponding to a complex field. However, this would imply that the phase space corresponding to the GPE is 4-dimensional (or alternatively, that we need 4 initial conditions to fully specify time evolution). This is not exactly what occurs. Due to the fact that the GPE equation is of first order in time, we only need to specify 2 real initial conditions, the imaginary and real parts fo $\Psi$ (or the density and the phase). Therefore, we should only have one scalar degree of freedom. In the density-phase picture, this manifests in the fact that density and phase are conjugated variables. To see this, we take the classical version of the Hamiltonian \eqref{eq:GPHamiltonian} written in terms of $(n,S)$, which reads
\begin{equation}
H=\int_\Sigma dV \lrsq{\frac{\hbar^2}{2m}\lr{\frac{\lr{\vna{n}}^2}{4 n}+n (\vna S)^2}+V_{\mathrm{ext}}n+\frac{g}{2}n^2}.
\label{eq:GPEHamDP}
\end{equation}
Then we can check that the density-phase equations \eqref{eq:DensityPhaseGPE} are reproduced by Hamilton's equations
\begin{equation}
   \partial_tS=\frac{\delta H}{\delta n}
   \qquad\qquad
   \partial_tn=-\frac{\delta H}{\delta S} 
\end{equation}
so that $S$ plays the role of the configuration space coordinate and $n$ that of its conjugate momentum. This proves that we actually have only one real scalar degree of freedom in the system, which will be relevant for the quantization of the perturbations later. Again, another way to see this is by resorting to the GPE Lagrangian in density-phase variables
\begin{equation}
    \label{eq:GPELagDP}
    \mathcal{L}_{\rm GPE}=-\hbar n \partial_t S-\frac{\hbar^2}{2m}\lr{\frac{(\vna n)^2}{4n}+n(\vna S)^2}-V_{\mathrm{ext}} n - \frac{g}{2}n^2,
\end{equation}
which recovers the field equations for density and phase as its Euler-Lagrange equations. As usual, from the lagrangian, one can compute the canonical momentum conjugated to $S$ as
\begin{equation}
\pi_S=\frac{\partial\mathcal{L}_{\rm GPE}}{\partial \partial_tS}=-\hbar n.
\end{equation}
which confirms that $n$ does not describe an independent dynamical degree of freedom, but only the conjugate pair $(S,n)$ does. Indeed, we can recover the hamiltonian density \eqref{eq:GPEHamDP} as $\pi_S \partial_t S-\mathcal{L}_{\rm GPE}$. Hence, we are now convinced that the GPE describes a single real scalar degree of freedom. Of course, to describe such degree of freedom, we have the freedom to choose canonically conjugate real field variables $(S,n)$, or a complex field variable $\Psi$ (with its conjugate). 

To square the two pictures, note that the conjugate momentum of $\Psi$ is $\pi_\Psi=\ci \hbar\Psi^*/2$ and that of $\Psi^*$ is $\pi_{\Psi^*}=-\ci\hbar \Psi/2$. From the point of view of a canonical analysis, these are two primary constraints. After Dirac's algorithm for constrained theories is enforced \cite{DiracConstraints}, the two constraints eliminate one of the two degrees of freedom usually described by the complex scalar field, leaving only a single degree of freedom. We note that this is not special about the GPE, as it also occurs in standard Schr\"odinger theory, or Dirac theory, and it is due to having only first-order time derivatives in the kinetic term.

As a last remark, we note that the continuity equation for the density can also be recovered from the original phase symmetry of the GPE lagrangian \eqref{eq:GPELag}, which here is manifested as a global shift symmetry for the phase $S\mapsto S+c$, with $c$ constant. Noether's theorem associates a spatiotemporal current $j^\mu$ to this symmetry such that $\partial_\mu j^\mu=0$ for solutions to the field equations. From the standard Noether recipe, this current takes the form
\begin{equation}
    j^\mu\coloneqq(Q,\boldsymbol{j})=-\hbar\lr{n,\frac{\hbar}{m}{n\vna S}},
\end{equation}
where $Q$ is called the charge and $\boldsymbol{j}$ the spatial current. Note that $Q$ coincides with the conjugate momentum to $S$ as expected for shift-symmetric theories. In general, Noether's result that $\partial_\mu j^\mu=0$ implies that $Q$ is conserved in time, in the sense that its temporal variation in any closed spatial region is accounted for by the flux of $\boldsymbol{j}$ through its boundary. Thus, $\boldsymbol{j}$ can be seen as the flux of $Q$. 
\section{Low energy excitations of the condensate}
\label{sec:LowEnergyExcitations}

We have described equilibrium states of the BEC as stationary solutions to the GPE. The next step to take revolves around the question: What happens if we slightly perturb a BEC in a stationary state? As we shall see, the answer is that low-energy perturbations do not excite single atoms but rather correspond to excitations of the system as a whole, described by collective degrees of freedom known as phonons. These are quantum waves of density/pressure, analogous to sound waves occurring in classical fluid systems. As the energy of the perturbations increases instead, they start to look more and more as the excitation of a single atom. This transition is beautifully encoded in the Bogoliubov dispersion relation, which governs the collective excitations of BECs, and the scale at which this occurs is controlled by the healing length $\xi$. In the following, we will give a detailed description of these results through several procedures, all of which converge in their results, but emphasize different physical aspects of the phenomenon. First we will present the Bogoliubov approximation to the full quantum GPE Hamiltonian, which provides clear insight on how the atomic degrees of freedom combine into collective excitations. Second, we will arrive to equivalent results considering linear perturbations the atomic field variables, a procedure more akin to standard field-theoretic methods. Then we will also discuss the problem as linear perturbations of density-phase variables, which leads to a more intuitive picture of the collective phenomena and from the acoustic metric which governs low-energy BEC perturbations arises more clearly.

\subsection{Collective excitations from the quantum Hamiltonian: the Bogoliubov approximation}\label{sec:CollectiveBogoApp}

To discuss the Bogoliubov approximation and how it leads to collective excitations as combinations of atomic degrees of freedom, we start by considering a stationary and homogeneous condensate with chemical potential $\mu$ in the frame where it has vanishing velocity (or comoving frame). Following \cite{pitaevskii2016bose} we expand the atomic field in atomic creation and annihilation of momentum eigenstates as
\begin{equation}
    \hat\Psi=\frac{1}{2\pi \sqrt{V}}\int\dd\vk  \ee^{\ci\vk\cdot\vx}\hat{b}_{\vk},
\end{equation}
where $V$ is the volume of our system. Stationarity and homogeneity ensures respectively that modes with different $\omega$ and $\vk$ will not mix as time evolves. We can then write the Hamiltonian \eqref{eq:GPHamiltonian} in terms of atomic creation and annihilation operators. In the Bogoliubov approximation $\hat{b}_{0}$ and $\hat{b}^\dagger_{0}$ are approximated by $\sqrt{N}$, and we keep terms up to quadratic in $\hat b_{\vk}$ for non-vanishing $\vk$, which leads to (see Problem \ref{prob:FluctHam})
\begin{equation}
    \hat{H}=\frac{g n N}{2}\hat{\mathbf{I}}+\int \dd\vk \lrsq{\frac{\hbar^2\tk^2}{2m}\hat{b}^\dagger_{\vk}\hat{b}_{\vk}+\frac{g
n}{2}\lr{2 \hat{b}_{\vk}^\dagger \hat{b}_{\vk} + \hat{b}_{\vk}^\dagger\hat{b}_{-\vk}^\dagger + \hat{b}_{\vk}\hat{b}_{-\vk}}}
    \label{eq:QuantumLowEnergyHam}
\end{equation}
This result tells us that the Hamiltonian describing small perturbations from the ground state is not diagonal in the atomic degrees of freedom due to the \textit{anomalous} terms
\begin{equation}
    \hat{b}_{{\vk}}^\dagger\hat{b}^\dagger_{-{\vk}}\qquad\text{and}\qquad\hat{b}_{{\vk}}\hat{b}_{-{\vk}}.
\end{equation}
This kind of quadratic Hamiltonian would also be obtained for a system of coupled springs in terms of the ladder operators of the single oscillators. Because of the couplings, the normal modes of the system, those that would diagonalize the Hamiltonian, are not the displacements of each of the springs, but rather linear combinations of them. To diagonalize the Hamiltonian, we can look for new degrees of freedom constructed as linear combinations of the atomic ones, as was first done by Bogoliubov in \cite{bogoliubov1947theory}. To do that, he defined new degrees of freedom described by $\hat{a}_{\vk}$ and $\hat{a}_{\vk}^\dagger$ through the relations
\begin{equation}
    \hat{b}_{{\vk}}=u_\tk\hat{a}_{{\vk}}+v_{\tk}^*\hat{a}^{\dagger}_{{-\vk}}\qquad\text{and}\qquad \hat{b}^\dagger_{{\vk}}=u_k^*\hat{a}^\dagger_{{\vk}}+v_{\tk}\hat{a}_{{-\vk}},
    \label{eq:CondBogDiag}
\end{equation}
where, in order for the transformation to be canonical (i.e. preserve the commutation relations), the coefficients of the combination must satisfy
\begin{equation}
    |u_{{\tk}}|^2-|v_{{\tk}}|^2=1.
    \label{eq:CondBogCoeff}
\end{equation}
Such a transformation is now in general known as a Bogoliubov transformation, and can be used to diagonalize any quadratic bosonic Hamiltonian, i.e. to write it as a sum of terms of the form $\hat{a}_{\vk}^\dagger\hat{a}_{\vk}$ (when normal ordered). Bogoliubov found that, in order to make Hamiltonian \eqref{eq:QuantumLowEnergyHam} diagonal in this sense, the coefficients of the linear combination must also satisfy
\begin{equation}
    \frac{g n}{2}\lr{|u_{{\tk}}|^2+|v_{{\tk}}|^2}+\lr{\frac{\hbar^2 \tk^2}{2m}+g n}u_{{\tk}}v_{{\tk}}=0,
    \label{eq:UVEquation}
\end{equation}
where $n=N/V$ is the atomic density. The coefficients $u_\tk$ and $v_\tk$ depend only on the modulus of the momentum $\tk$ and can be found to be\footnote{The phase of the coefficients is free, but can be reabsorbed into a re-definition of the modes by a global phase. Also, there are infinitely many solutions to \eqref{eq:UVEquation} of the form $Au_\tk$ and $Av_\tk$ where $A\in\mathbf{R}$, but only the $A=1$ solution is normalized, in the sense that it satisfies \eqref{eq:CondBogCoeff}.}
\begin{equation}
    |u_\tk|=\sqrt{\frac{1}{2\hbar\w_\tk}\left(\frac{\hbar^2\tk^2}{2m}+gn\right)+\frac{1}{2}}\qquad\text{and}\qquad |v_\tk|=\sqrt{\frac{1}{2\hbar\w_\tk}\left(\frac{\hbar^2\tk^2}{2m}+gn\right)-\frac{1}{2}},
    \label{eq:BogoCoeffBECs}
\end{equation}
where we defined the frequency
\begin{equation}\label{eq:bogo-transf-eigenvalues}
    \omega_\tk\coloneqq\sqrt{\frac{\hbar^2\tk^4}{4m^2}+\frac{gn \tk^2}{m}}.
\end{equation}
This is the celebrated Bogoliubov dispersion relation, relating the frequency of collective excitations to their momentum. We will encounter this dispersion relation multiple times in the rest of these notes.

The above results tell us that the low-energy eigenstates of the Hamiltonian do not describe the dynamics of individual atoms but rather collective excitations of the whole system, just as the normal modes in a system of coupled springs are not excitations of the individual springs. In particular, in the limit the limit $\tk\to0$, $u_\tk\simeq v_\tk$, so that low-energy excitations are a symmetric superposition of creation and annihilation of the atoms.
In fact, one can see that the presence of collective excitations oscillating at frequency $\omega_\tk$ involves the presence of out-of-condensate atoms oscillating frequencies $\mu/\hbar+\omega_\tk$ and $\mu/\hbar-\omega_\tk$. Heuristically, we can imagine how, locally, a collective excitation excites atoms in a way in which they approach each other and, due to their interactions, then bounce back, transmitting a density/pressure wave along the condensate, i.e., producing sound waves. The collective excitations of low momenta are called phonons, and are understood as quanta of sound. At large momenta instead, $|u_\tk|\to 1$ and $|v_\tk|\to 0$, and one recovers simple single-atom excitations, in which one atom is pushed out of the condensate.

Note that transformation \eqref{eq:CondBogDiag} is a 2-mode squeezing transformation involving atoms with opposite momenta. This implies that the vacuum of low energy excitations ($\ket{0_a}$ such that $\hat a_{\tk}\ket{0_a}=0$) does not correspond to the absence of atoms outside of the condensate in the $\vk=0$ mode. Instead, it is populated by couples of atoms with opposite momenta that are pushed outside of the condensate. In fact, by substituting the Bogoliubov transformation in the expectation value of the number of atoms at $\vk\neq0$, one obtains
\begin{equation}\label{eq:quantum-depletion}
    \bra{0_a}\hat b_\vk^\dagger \hat b_\vk \ket{0_a}=|v_\tk|^2.
\end{equation}
This phenomenon is called quantum depletion and is due to quantum fluctuations and the presence of interactions.
\subsection{Collective excitations from the linearization of the GPE}
\label{sec:linearization-gpe}

While the fully quantum treatment of fluctuations of the previous section can be done for a uniform condensate, for general non-uniform configurations this proves difficult. It is instead easier to approach the study of excitations of the condensate from the classical field perspective, by studying perturbations around a solution of the GPE \cite{castin2001bose}. To treat them at the quantum level one can then re-quantize the excitation field, as we are going to discuss in the next section.

To treat excitations at the classical level we can introduce a small amplitude fluctuation around a stationary solution of the GPE
\begin{equation}
    \Psi(\vx,t)=\ee^{-\ci \mu t/\hbar}\lrsq{\Psi_0(\vx)+{\psi}(t,\vx)},
    \label{eq:DefPertGPE}
\end{equation}
where $\Psi_0(\vx)$ solves  \eqref{eq:StaionaryGPE}. Here, the overall temporal exponential means that we are going to measure the frequency of the excitations with respect to the chemical potential of the stationary solution. Inserting this expression in the time-dependent GPE \eqref{eq:TimeDepGPE}, and keeping only terms that are linear in $\psi$ and $\psi^*$ we obtin (see Problem \ref{prob:BdG})
\begin{equation}
    \ci\hbar\partial_t\psi=\lrsq{-\frac{\hbar^2}{2m}\nabla^2+V_{\mathrm{ext}}-\mu + 2 g n_0}\psi + g\Psi_0^2 \psi^*.
\end{equation}

This equation is a differential equation of 1st order in time for a complex field. Hence, the evolution is determined by a complex initial condition $\psi(t_0)$, or two real ones $\mathrm{Re}[\psi(t_0)]$ and $\mathrm{Im}[\psi(t_0)]$. The phase space associated to this equation can then be seen as a 2 dimensional space, so it describes one real scalar degree of freedom as expected. However, this equation is not strictly linear in $\psi$, because of the coupling to $\psi^*$. As explained in \cite{castin2001bose}, a possible way out to end with a linear problem is to consider the real and imaginary parts of $\psi$ as independent variables, though it is more common to work with $\psi$ and $\psi^*$ as independent variables which are components of a two-dimensional complex vector
\begin{equation}
     \ket{\psi}=
    \begin{pmatrix}
       \psi\\
       \psi^*
    \end{pmatrix}.
    \label{eq:BdGvector}
\end{equation}
By using the conjugate to \eqref{eq:DefPertGPE} we can write the dynamical problem as the system
\begin{equation}
     \ci\hbar\partial_t\ket{\psi}=\mathcal{L}_\mathrm{BdG}\ket{\psi}\qquad\text{with}\qquad \mathcal{L}_\mathrm{BdG}=\begin{pmatrix}
       A & B\\
       -B^* & -A^*
    \end{pmatrix},
    \label{eq:BdGProblem}
\end{equation}
where 
\begin{equation}
    A=-\frac{\hbar^2}{2m}\nabla^2+V_{\mathrm{ext}}+2gn_0-\mu\qquad\text{and}\qquad B=g \Psi_0^2.
\end{equation}
This can be seen as an eigenvalue problem for $\bdg$, known as the Bogoliubov-de Gennes (BdG) operator.  This can be thought of as the evolution operator of the system since, formally,
\begin{equation}
    \ket{\psi(t)}=\ee^{-\ci \bdg \frac{t-t_0}{\hbar}}\ket{\psi(t_0)}.
\end{equation}
However, given that $\bdg$ is in general non-hermitian, this evolution operator is not unitary. This non-hermiticity of $\bdg$ implies that we can obtain complex eigenvalues (frequencies), and that the standard inner product in not conserved. We will leave for Section \ref{sec:BdGEigenvalue} the properties and existence of eigenvectors of $\bdg$. 

\subsubsection{Bogoliubov dispersion in homogeneous condensates: comoving vs. non-comoving frames}

While the linearization performed above is a fully general way of treating fluctuations, it is instructive to first focus on a homogeneous condensate. Since in analogue gravity we are interested in moving fluids, let us consider a condensate that is moving with a velocity $\vtv=\hbar\vk_0/m$. The corresponding wavefunction is $\Psi_0(\vx)=\sqrt{n_0}\ee^{\ci \vk_0\cdot\vx}$.

\begin{figure}
    \center
    \includegraphics[width=0.9\textwidth]{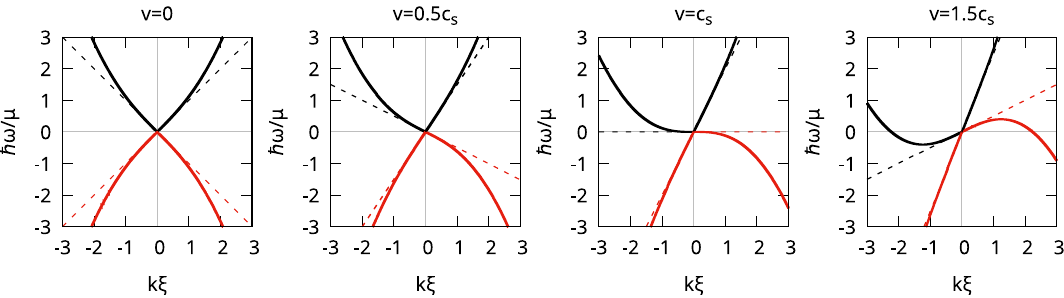}
    \caption{Plots of the Bogoliubov dispersion relation in a homogenous condensate for different values of the background flow velocity. Black and red lines are the plus and minus branches of the eigenvalues. Dashed lines are the corresponding hydrodynamic linear approximation of the dispersion relation.}
    \label{fig:bogoliubov-dispersion}
\end{figure}

We can consider two different reference frames in which to describe excitations. One is the comoving frame, that is obtained with the expression \eqref{eq:DefPertGPE}. The invariance under space translations ensures that perturbations with different wavevectors will not mix. It is then convenient to expand perturbations in plane waves $\ket{\psi}=\sum_{\vk}\ee^{-\ci(\omega_\tk t- \vk\cdot \vx)}\ket{\psi_\tk}$. When acting on a plane wave, the differential operator $\bdg$ becomes a block diagonal matrix, so that the different blocks can be diagonalized independently. By taking $V_{\rm ext}=0$ and the chemical potential $\mu=gn_0$, if we write the background condensate as $\Psi_0=\sqrt{n_0}\ee^{\ci\theta_0}$, these blocks have the form \eqref{eq:BdGProblem} with

\begin{equation}
    A=\frac{\hbar^2 \tk^2}{2m}+gn_0\qquad\text{and}\qquad B= g n_0\ee^{2\ci\theta_0}.
\end{equation}
The two eigenvalues of this matrix for each $\vk$ are $\hbar\omega_\tk^\pm$, with
\begin{equation}
    \omega^{\pm}_{\tk}=\pm\Omega(\tk)
    \coloneqq\pm\sqrt{\frac{\hbar^2}{4m^2}\tk^4+\frac{gn_0}{m}\tk^2}
    =\pm \cs\tk\sqrt{1+\frac{\xi^2}{4}\tk^2}
    \label{eq:BogoliubovDispersionComoving}
\end{equation}
where $\xi$ is the healing length of the condensate defined in \eqref{eq:HealingLength} and we defined the speed of sound in the condensate
\begin{equation}
    \cs\coloneqq\sqrt{\frac{gn_0}{m}}.
    \label{eq:SpeedOfSoundBogolons}
\end{equation}
This is indeed the Bogoliubov dispersion relation found in \eqref{eq:bogo-transf-eigenvalues}, showing that linear perturbations to exact solutions of the GPE describe the collective low-energy excitations of a condensate discussed in \ref{sec:CollectiveBogoApp}. To finish, we compute also the corresponding eigenvectors, which are the normal modes of our dynamical system. These are
\begin{equation}
   \ket{\psi_{\pm}(\vk)}=
   \begin{pmatrix}
       e^{2 i \theta_0 }\\[5pt]
       D_\pm(\tk)
   \end{pmatrix}
   \qquad\text{where}\qquad 
   D_\pm(k)=-1-\frac{\xi^2k^2}{2}\pm\xi k\sqrt{1+\frac{\xi^2k^2}{4}}.
   \label{eq:EigenstatesHomogBdG}
\end{equation}
Thus, plane waves that solve the BdG problem under these assumptions will satisfy the celebrated Bogoliubov dispersion relation, that we already obtained in equation \eqref{eq:bogo-transf-eigenvalues}, and that describes the relation between the frequency and momentum of the fluctuations of a homogeneous stationary condensate as measured in its comoving frame, where it does not flow. A plot of this dispersion relation is shown in Figure \ref{fig:bogoliubov-dispersion}, where the black and red curves correspond to the plus and minus signs in \eqref{eq:BogoliubovDispersionComoving}. 

The Bogoliubov dispersion is linear in the momentum at small momenta, and becomes quadratic at high momenta, reflecting the fact that excitations become the ones of a single atom. The small momenta condition $\xi\tk\ll 1$ identifies the \textit{hydrodynamice regime}, in which the non-linear terms of the Bogoliubov dispersion can be neglected, and it becomes a relativistic dispersion for a massless field, with a lightcone (soundcone) defined by the speed of sound $\cs$. Recall that neglecting $\xi$ dependent terms amounts to neglect the quantum pressure in the hydrodynamical representation of the BEC (see section \ref{sec:HydroBckg}). If we do not neglect it, on the other hand, we find a nonlinear dispersion relation which is superluminal/supersonic in the sense that the group velocity of the perturbations is greater than $\cs$ due to the positivity of the $\tk^4$ coefficient. This implies that there is no universal limiting speed for the perturbations, since high $\tk$ perturbations can propagate outside the soundcone defined by the soundspeed. These effects become relevant for perturbations with wavelength comparable to the healing length, namely, those that become sensitive to the microstructure of the condensate. In that sense, forcing the analogy, one could think of the healing length as a sort of Planck length, below which the microstructure of the acoustic spacetime cannot be neglected anymore, and the emergent Lorentz symmetry is broken. Remarkably, this could provide insights on how some quantum gravity theories work. For instance, emergent spacetime proposals expect that at such scales the continuum spacetime is replaced by discrete building blocks playing a parallel role as atoms in a BEC (see \cite{Margoni:2024ixx} and references within).

We have found the dispersion relation for perturbations in the comoving frame of the condensate, where it is at rest. But we can also describe the perturbations from a frame where the condensate moves at a uniform velocity $\vtv=\hbar\vk_0/m$. This is straightforwardly achieved by defining the perturbation field of the form
\begin{equation}
    \Psi(\vx,t)=\ee^{-\ci(\mu t/\hbar-\vk_0\cdot\vx)}\lrsq{\sqrt{n_0(\vx)}+\chi(t,\vx)}.
    \label{eq:DefPertGPELab}
\end{equation}
Again, the different momenta will not mix, and repeating the linearization procedure as above, one obtains the eigenvalue problems
\begin{equation}\label{eq:bogo-problem-labframe}
	\begin{bmatrix}
		 D_+ & gn_0\\
		-gn_0 & -D_-
	\end{bmatrix}
	\begin{pmatrix}
		\chi\\\chi^*
	\end{pmatrix}
    =
    \hbar\omega_{\tk}
	\begin{pmatrix}
		\chi\\\chi^*
	\end{pmatrix},
\end{equation}
where
\begin{equation}
	D_\pm=\frac{\hbar^2 \tk^2}{2m}\pm\hbar\vk\cdot\vtv+gn_0.
\end{equation}
Notice that now the diagonal blocks are not one minus the other, because of the sign on the term $\hbar\vk\cdot\vtv$. This is clear if one conjugates before applying the derivative operators.
The eigenvalues of this modified BdG problem are now
\begin{equation}\label{eq:doppler-shift}
    \omega_{\vk}^\pm=\vtv\cdot\vk\pm\Omega(\tk),
\end{equation}
where $\Omega(\tk)$ is defined in \eqref{eq:BogoliubovDispersionComoving}. This is related to the Bogoliubov dispersion in the comoving by a Doppler shift of the frequencies, that can be seen as the result of a change of reference frame, from one in which the condensate is moving, to one in which the condensate is still, and is expected from the Galilean invariance of the GPE (see Problem \ref{prob:BdG}).

Plots of this Doppler-shifted dispersion relation are shown in Figure \ref{fig:bogoliubov-dispersion} for diffrent values of the condensate velocity. The corresponding linear dispersion in the hydrodynamic approximation is also shown. The effect of the Doppler shift is to \textit{tilt} the dispersion relation. For supersonic velocities this tilt is such that the minus branch of the eigenvalues raises to positive frequencies. As it will be clearer in the following, this feature of the supersoni regime is what makes it interesting for analogue gravity.

\subsection{The Bogoliubov-de Gennes inner product}\label{sec:BdGEigenvalue}

In this section we will show the existence of an inner product for the BdG problem which is conserved in laboratory time. This allows for a physically meaningful definition of a (pseudo-) norm for its solutions. To that end, let us recall that for a system in which the time evolution is given by a hermitian Hamiltonian, the usual $L^2$ inner product is conserved. However, the BdG matrix --i.e. the operator determining the time evolution of the BdG problem-- is non-hermitian. There is a way of generalizing the inner product as a linear deformation of the $L^2$ inner product, thanks to the pseudo-hermiticity of the BdG  matrix with respect to the Pauli matrix $\sigma_3$. Let us show how to construct this product for general pseduo-hermitian Hamiltonians.

A matrix $\mathcal{L}$ is pseudo-hermitian with respect to the matrix $\sigma$ (or $\sigma$-pseudo-hermitian) if it satisfies $\mathcal{L}^\dagger=\sigma \mathcal{L}\sigma^{-1}$. The solution space of any linear system of PDEs of the form $\ci\partial_t\ket{\psi}=\mathcal{L}\ket{\psi}$ where $\mathcal{L}$ is $\sigma$-pseudo-hermitian admits a conserved bilinear form $\expval{\cdot|\cdot}_\sigma$ defined as 
\begin{equation}
    \expval{\psi_1|\psi_2}_\sigma\coloneqq\bra{\psi_1}\sigma\ket{\psi_2},
\end{equation}
where the right hand side is the standard $L^2$ inner product and $|\psi_{1,2}\rangle$ are any two solutions of the PDE system. One can check that for a matrix of the form $\bdg$, the equation does not admit solutions unless $A^*=A$. In that case one can verify that $\mathcal{L}_{\rm BdG}$ is $\sigma_3$-pseudo-hermitian, where
\begin{equation}
    \sigma_3=
    \begin{pmatrix}
        1 & 0\\
        0 & -1
    \end{pmatrix},
\end{equation}
So that the conserved inner product for the BdG problem is $\langle\psi_1|\psi_2\rangle_{\rm BdG}\coloneqq\langle\psi_1|\psi_2\rangle_{\sigma_3}$. In terms of the components of the eigenvectors $|\psi_i\rangle=(u_i(\vx),v_i(\vx))^\top$ of the BdG problem, the BdG inner product reads
\begin{equation}
\expval{\psi_1|\psi_2}_{\rm BdG}= \int_{\Sigma} \dd^d\tx \lr{u_1^*(\vx) u_2(\vx)-v_1^*(\vx)v_2(\vx)},
    \label{eq:BdGInnerProd}
\end{equation}
where $\Sigma$ is any spatial region where the condensate is supported.

Note that this inner product is reminiscent of the standard Klein-Gordon symplectic product (see equation \eqref{eq:KGNorm} further down). This is suggesting the fact that the perturbation equations are a constrained system for a complex field which propagates only one real scalar degree of freedom, as the Klein-Gordon field does. In fact, we will see in the next section that the perturbations can be in general described by a single PDE with 2nd order time derivatives for a real scalar field, a sort of a generalized Klein-Gordon equation with higher-order spatial derivatives, which can be neglected in for perturbations with large-enough wavelength. As the Klein-Gordon product, the Bogoliubov inner product is not positive definite, but can be used to define a (pseudo-)norm in solution space that allows for quantization of the perturbations as $||\psi||\coloneqq\expval{\psi|\psi}_{\rm BdG}$. With this definition of inner product, we can now discuss some properties of the eigenvectors and the spectrum of the BdG problem related with their orthogonality relations. 

Hermitian operators, the ones we usually encounter in quantum mechanical setups, have the same right and left eigenvalues and eigenvectors, and the existence of an orthonormal basis of eigenvectors is guaranteed by the spectral theorem. Pseudo-hermitian operators instead do not generally have these properties, although they satisfy a useful generalization of them. For example, there is a distinction between left- and right-eigenvectors and eigenvalues of an operator: given a linear operator $\mathcal{L}$ on a vector space, $\ket{R_i}$ is a right-eigenvector with right-eigenvalue $\rho_i$ if $\mathcal{L}\ket{R_i}=\rho_i\ket{R_i}$. Similarly, a left-eigenvector $\bra{L_i}$ with left-eigenvalue $\lambda_i$ is such that $\bra{L_i}\mathcal{L}=\bra{L_i}\lambda_i$. A crucial property of pseudo-hermitian operators is that their sets of right- and left- eigenvalues are related by complex conjugation $\{\rho_i\}_i=\{\lambda^*_i\}_i$. Hence, if $\{\bra{L_i}\}_i$ and $\{\ket{R_i}\}_i$ are the sets of left and right eigenvectors, we can order them such that if $\lambda_i$ is the right eigenvalue corresponding to $\ket{R_i}$, then $\lambda_i^*$ is the left eigenvalue corresponding to $\bra{L_i}$. Moreover, one can check that, if the right eigenvector with eigenvalue $\lambda_i$ is $\ket{R_i}=(u_i,v_i)^T$, then the corresponding left eigenvector with eigenvalue $\lambda_i^*$ is $\bra{L_i}=(u_i^*,-v_i^*)$. With these considerations, the so called biorthogonality relations between left and right eigenvectors of pseudo-hermitian operators holds
\begin{equation}
    (\lambda_i-\lambda_j^*)\expval{L_j|R_i}=(\lambda_i-\lambda_j^*)\expval{\psi_i|\psi_j}_{\rm BdG}=0.
    \label{eq:BiOrtthogonality}
\end{equation}
Since the eigenvalues of the BdG operator are frequencies, this means that eigenvectors with different frequencies are orthonormal with respect to the BdG inner product, and those with complex frequencies have vanishing norm, and are hence non-normalizable. Sticking to real-frequency modes, the biorthogonality property leads to the orthonormality components of normalized eigenstates
\begin{equation}
    \expval{\psi_i|\psi_j}_{\rm BdG}=\int_\Sigma\dd V (u^*_i u_j-v^*_i v_j)=\pm\delta_{ij},
\end{equation}
where the plus or minus sign are necessary because the eigenstates can have negative norm.

Another relevant property of the BdG operator is \textit{particle-hole symmetry}, an important property of collective excitations of many condensed matter systems. This property stands from the identity $\sigma_1\bdg\sigma_1=-\bdg^*$, where $\sigma_1$ is the first Pauli matrix, and it implies that if $|\psi\rangle=(u,v)^\top$ is a right eigenvector with eigenvalue $\lambda$, then $\sigma_1|\psi\rangle=(v^*,u^*)^\top$ is an eigenvector with eigenvalue $-\lambda^*$. Furthermore, biorthogonality among them implies the relation
\begin{equation}
    \int_\Sigma\dd V (u^*_i v^*_j-v^*_i u^*_j)=0
    \label{eq:BdGBiOrtho}
\end{equation}
Hence, if there is a solution mode with a real positive frequency, there is a solution with the same negative frequency, and they have opposite sign of their norm. 
Indeed, physically, they both describe the same atomic degree of freedom. Then, we find that an atomic degree of freedom with energy $\hbar\omega$ is described by the most general combination of Boglyubov modes
\begin{equation}
\label{eq:field-eigenmode}
    \begin{pmatrix}
        u_\omega\\
        v_\omega
    \end{pmatrix}
    \ee^{-\ci\omega t}
    \qquad\text{and}\qquad
    \begin{pmatrix}
        v_\omega^*\\
        u_\omega^*
    \end{pmatrix}
   \ee^{\ci\omega^* t}\,.
\end{equation}
This is known in the BEC literature as \textit{doubling of degrees of freedom}. In order to describe the modes without redundancies, one can either choose to work with positive norms only and both signs of the frequency, or with positive frequencies only and both signs of the norm. We will make use of this later.

Finally, we present the normalization for eigenvectors of the BdG problem with definite frequency and wavevector. Using \eqref{eq:EigenstatesHomogBdG} we find
\begin{equation}
\begin{split}
    \expval{\psi_\pm(\vk)|\psi_\pm(\vk')}_\textrm{BdG}&=\int_\Sigma \dd V \big(1-D_\pm(\tk)D_\pm(\tk')\big)\ee^{\ci(\omega_\vk-\omega_{\vk'})t}\ee^{\ci \lr{\vk'-\vk}\cdot\vx}
    \\&=(2\pi)^d  \lr{1-D^2_\pm(k)}\delta^d(\vk-\vk')\,,
\end{split}
\end{equation}
where $d$ is the spatial dimension. Inverting the relation between $\omega$ and $\vk$ and using properties of the Dirac delta, we find the normalization
\begin{equation}
    \expval{\psi_\pm(\vk(\omega))|\psi_\pm(\vk(\omega')}_\textrm{BdG}=(2\pi)^d  \lr{1-D_{k}^2}\tv_{\rm g}\delta(\omega'-\omega)\,,
    \label{eq:PlaneWaveNormBdG}
\end{equation}
where $\vtv_{\rm g}\coloneqq\vna_{\vk}\omega$ is the group velocity of the mode.

\subsection{The energy of the Bogoliubov 
eigenvectors}

To finish the general characterization of the BdG eigenvectors, we compute their energies. Starting from the classical-field version of \eqref{eq:AtomicHamiltonian}, giving the energy functional $E(\Psi)$, we consider the grand canonical energy functional $E(\Psi)-\mu\int_\Sigma \dd V |\Psi(\vx)|^2$. By inserting the perturbation around a stationary state \eqref{eq:DefPertGPE} in this functional and expanding to second order in the fluctuations $\psi$, one obtains\footnote{The first-order term vanishes because the functional is zero on a stationary state.}
\begin{equation}\label{eq:bogo-energy-functional}
	E^{(2)}(\psi)=\int_\Sigma\mathrm{d}V\;\left[\frac{\hbar^2}{2m}\abs{\nabla\psi}^2+(V+2g\abs{\Psi_0}^2-\mu)\abs{\psi}^2 + \frac{g}{2}(\Psi_0^*)^2\psi^2 +\frac{g}{2}\Psi_0^2(\psi^*)^2\right],
\end{equation}
that is the (quadratic) Hamiltonian of the fluctuations. This can be written in terms of the two dimensional vector \eqref{eq:BdGvector} as
\begin{equation}
	E^{(2)}(\psi)=\frac{1}{2}\bra{\psi}\bdg\ket{\psi}_{\rm BdG},
\end{equation}
that is in terms of the BdG operator without the minuses in the second row, that is a hermitian operator. By expanding $\ket{\psi}$ in terms of eigenmodes $\ket{\psi_k}$ of frequency $\omega_k$ of the Bogoliubov matrix one obtains
\begin{equation}\label{eq:modes-energy} 
	E^{(2)}=\sum_k \expval{\psi_k|\psi_k}_\mathrm{BdG}\hbar\w_k.
\end{equation}
One can see that the energy of an eigenmode is not simply given by its frequency, but is also proportional to its norm. That is, negative-norm modes with a positive frequency have a negative energy. The presence of such modes is called \textit{energetic instability}, and it means that the considered stationary solution around which we study the fluctuations does not correspond to a minimum of the energy functional, and hence the energy of the condensate can be decreased by populating this mode. An important example of the occurrence of energetic instabilities is the Landau criterion for superfluidity, that sets an upper bound on the flow velocity beyond which negative-energy excitations exist. For the weakly interacting BECs we are considering, this occurs for velocities larger than the speed of sound. The presence of these energetic instabilities in supersonic flows is crucial to have analogue gravity phenomena like superradiance and the Hawking effect, as we are going to see in the following.

Equation \eqref{eq:modes-energy} also tells us that complex-frequency eigenvectors have zero energy, since they have zero norm. From the properties listed above we can see that complex-frequencies come in complex conjugates couples, corresponding to an eigenvector that  grows exponentially in time and one that decays. The presence of such modes is called \textit{dynamical instability}, and can be thought as the continuous production of excitations with opposite energies. Even if the linear theory predicts an exponential growth in time of the dynamically unstable modes, this does not of course occur indefinitely. At some point the mode amplitude will enter a regime in which the perturbation theory is no longer a good description and nonlinear effects start to be important, requiring the full GPE to be captured.

\subsection{The acoustic metric from density-phase perturbations}
\label{sec:AcousticMetric}
We will finish this section by explicitly showing how low momentum collective excitations of a BEC are described by a relativistic wave equation with coefficients that depend on the background fluid configuration. This leads to the well known analogy between acoustic waves in a BEC and a massless scalar field in a curved spacetime described by a Lorentzian metric. This metric, called acoustic metric, depends locally on the background fluid, which can be engineered to mimic spacetimes that occur in gravitational settings. We will see later how quantization of these perturbations leads to phenomena that are analogue to those occurring in Quantum Field Theory in Curved Spacetimes, thus allowing to make indirect tests of the latter, and providing laboratory tests to the predictions of QFT in the presence of non-trivial backgrounds.

Let us start with the GPE Lagrangian in density-phase variables \eqref{eq:GPELagDP}, and factorize density and phase as
\begin{equation}
    n=n_0(1+\eta)
    \qquad \text{and} \qquad
    S=S_0+ \varphi
\end{equation}
where we assume that $\Psi_0(\vx,t)=\ee^{i\mu t/\hbar}\sqrt{n_0(\vx)}\ee^{iS_0(\vx)}$ are a stationary solution of the GPE. 
Differently from equation \eqref{eq:DefPertGPELab}, this corresponds to factorizing also the density when defining the fluctuations field
\begin{equation}
    \begin{split}
   \Psi(\vx,t)&=\ee^{-i\mu t/\hbar} \sqrt{n_0(\vx)}\ee^{iS_0(\vx)}(1+\widetilde{\psi}(\vx,t)).
    \end{split}
\end{equation}
Then one has the correspondence
\begin{equation}
\tilde\psi=\frac{\eta(\vx,t)}{2n_0(\vx)}+i\varphi(\vx,t)
\end{equation}
which indicates that the contribution of density perturbations to the atomic field will be suppressed with respect to phase perturbations by the inverse of the background density.

Linear perturbations are described by expanding the GPE Lagrangian up to second order in the perturbations\footnote{Generally in field theory, the equations for linear perturbations are always described by the quadratic terms in the action expansion. The terms of the action which are linear in the perturbations are proportional to the background equations, so that they vanish when considering physical background solutions.}. The second order Lagrangian is
\begin{equation}
    \mathcal{L}_2=-\hbar n_0 \eta\lr{\partial_t+\vtv\cdot\vna}\varphi-\frac{\hbar^2 n_0}{2m}\lr{\frac{(\vna\eta)^2}{4}+(\vna\varphi)^2}-\frac{g n_0^2}{2}\eta^2,
\end{equation}
and its Euler-Lagrange equations for $\varphi$ and $\eta$ are respectively
\begin{equation}
\begin{split}
    &\lr{\partial_t +\vtv\cdot\vna}\eta+\frac{\hbar}{m}D^2_{n_0}\varphi=0\,,\\
    &\lr{\partial_t +\vtv\cdot\vna}\varphi+\frac{g n_0}{\hbar}\lr{1-\frac{\xi^2 }{4}D^2_{n_0}}\eta=0\,,
    \label{eq:DensPhasePerturbations}
\end{split}    
\end{equation}
where we have defined the background-dependent second-order differential operator 
\begin{equation}
    D_{n_0}^2\coloneqq \nabla^2+\frac{\vna n_0\cdot\vna}{n_0}.
\end{equation}
The above equations coincide with what would be obtained by taking the GPE in density-phase variables \eqref{eq:DensityPhaseGPE} and expanding to linear order in the perturbations, but we can get extra information from the Lagrangian. For instance, we see that the conjugate momentum to $\eta$ vanishes, and that of $\varphi$ is 
\begin{equation}
    \pi_{\varphi}=-\hbar n_0\eta.
    \label{eq:PhasPertMomentum}
\end{equation}
Thus, neither momenta depends on the velocities, and they impose two primary constraints which reduce the number of degrees of freedom, ending with only one degree of freedom described by the above system. 

One way of implementing the constraints at the level of the field equations is by formally solving for $\eta$ in the second equation by inverting the operator $(1-\xi^2D_{n_0}^2/4)$, which yields
\begin{equation}
    \label{eq:EtaPhi}
    \eta=\frac{\hbar}{gn_0}\lrsq{\frac{\xi^2}{4}D_{n_0}^2-1}^{-1}\lr{\partial_t+\vtv\cdot\vna} \varphi.
\end{equation}
Plugging this result into the first equation for the density-phase perturbations and performing some algebraic manipulations which make use of the background equations we find
\begin{equation}
\left\{\partial_t\lrsq{\frac{\hbar}{g}\lr{1-\frac{\xi^2}{4}D_{n_0}^2}^{-1}\lr{\partial_t +\vtv\cdot\vna}}+\vna\cdot\frac{\hbar}{g}\lr{\vtv \lrsq{1-\frac{\xi^2}{4}D_{n_0}^2}^{-1}\lr{\partial_t+\vtv\cdot\vna}-\cs^2\vna }\right\}\varphi=0.
\label{eq:PhasePertEqGen}
\end{equation}
We can rewrite this equation as $\partial_\mu f^{\mu\nu}\partial_\nu S_1=0$ if we define the operators
\begin{equation}
    \begin{split}
        &f^{tt}=\frac{\hbar}{g}\lrsq{1-\frac{\xi^2}{4}D_{n_0}^2}^{-1} 
        \hspace{3cm} 
        f^{ij}=\frac{\hbar}{g}\lr{\tv^i\lrsq{1-\frac{\xi^2}{4}D_{n_0}^2}^{-1}\tv^j-\cs^2\delta^{ij}}\\
        &f^{ti}=\frac{\hbar}{g}\lrsq{1-\frac{\xi^2}{4}D_{n_0}^2}^{-1} \tv^i 
        \hspace{2.71cm}
        f^{it}=\frac{\hbar}{g}\tv^i\lrsq{1-\frac{\xi^2}{4}D_{n_0}^2}^{-1},
        \label{eq:Cartesianfmetric}
    \end{split}
\end{equation}
which contain only spatial derivatives. From the definition of $D_{n_0}^2$, we see that this operator is sensitive to the wavelength of the density perturbations as well as the spatial gradient of the background density. Furthermore, we see that it only enters the perturbation equations multiplied by the healing length squared. This suggests looking into the regime where the scales over which the background density changes appreciably, as well as the wavelength of the perturbations, are both much larger than the healing length, since $D_{n_0}^2$ can be neglected in that case and $f^{\mu\nu}$ become background-dependent coefficients. If the characteristic length scale over which the background density varies significantly is $L$, and the wavelength of the perturbations is $\lambda$, we can neglect $D_{n_0}^2$ in the field equations if $\xi/\lambda\ll1$ and $\xi/L\ll1$. This is called the hydrodynamical approximation, and is suitable to describe BECs whose spatial inhomogeneities ---including background inhomogeneities and perturbations--- are large enough to become insensitive to its microscopic structure. 

In the hydrodynamic regime, we can rewrite the perturbation equations by defining the object $\sqrt{|g|}g^{\mu\nu}=-f^{\mu\nu}$, where $|g|$ is the determinant of $g_{\mu\nu}$ (the inverse of $g^{\mu\nu}$), which then read
\begin{equation}
\label{eq:KGeq}
    \Box_g S_1= g^{\mu\nu}\nabla_\mu\nabla_\nu S_1=\frac{1}{\sqrt{|g|}}\partial_\mu \sqrt{|g|}g^{\mu\nu}\partial_\nu S_1=0
\end{equation}
which is the Klein-Gordon equation of a massless scalar field in a curved spacetime described by the metric $g_{\mu\nu}$, also known as acoustic metric. If $|f|$ is the determinant of $f^{\mu\nu}$, we find $|f|=|g|^{\frac{D-2}{2}}$ (for $D>2$), where $D=d+1$ and $d$ is the number of spatial dimensions of the condensate. Hence $g_{\mu\nu}=|f|^{\frac{1}{D-2}}(f^{-1})_{\mu\nu}$, which yields
\begin{equation}
    g_{\mu\nu}=\lr{\frac{\hbar n_0}{m c_s}}^{\frac{2}{D-2}}
    \begin{pmatrix}
        -({\rm c}_{\rm s}^2-\tv^2)& -\vtv\\[10pt]
        -\vtv & \mathbb{I}_{d}
    \end{pmatrix}.
    \label{eq:AcousticMetric}
\end{equation}
For one dimensional condensates we have $g_{\mu\nu}=(f^{-1})_{\mu\nu}$, and the acoustic metric reads
\begin{equation}
    g_{\mu\nu}=\frac{m}{\hbar n_0}
    \begin{pmatrix}
        -({\rm c}_{\rm s}^2-\tv^2)& -\tv\\[10pt]
        -\tv & 1
    \end{pmatrix}.
    \label{eq:OneDimensionalAcousticMetric}
\end{equation}
We can see how, for all dimensions, the acoustic metric has Lorentzian signature, which implies the existence of a causal structure, with the lightcones of gravitational theories replaced by soundcones with inclination given by $c_s$. The derivation of the acoustic metric can also be obtained from the perturbations in the lab frame as in section \ref{sec:linearization-gpe}, though the process is much more involved. However, note that the form of the components of the metric will depend on the choice of coordinates, since the translation of the gradient operator $\vna$ in \eqref{eq:PhasePertEqGen} into $\partial_i$ in \eqref{eq:Cartesianfmetric} only holds in cartesian coordinates. One can derive the specific form of the components in other coordinates by going to \eqref{eq:PhasePertEqGen}, which is written in coordinate invariant form, writing the gradient operator in the desired coordinates, and reading out the corresponding components of $f^{\mu\nu}$ defined as the object such that $\partial_\mu f^{\mu\nu}\partial_\nu S_1=0$ is equivalent to equation \eqref{eq:PhasePertEqGen}.

As a final remark, we note that if the limit of short wavelength is chosen for the perturbations, the $D_{n_0}^2$ operators can be approximated by $-\tk^2$, where $\tk=2\pi/\lambda$, and one recovers the Bogoliubov dispersion relation from the equation \eqref{eq:KGeq}. This equation can be interpreted in terms of a $\tk$-dependent metric, also known as rainbow metric \cite{Lafrance:1994in,Magueijo:2002xx}. These are modifications to General Relativity which explore the possibility that the spacetime seen by observers depend on their momentum. The consequence is the absence of a universal causal structure for all perturbations, which renders the geometric analogy less powerful.

\section{Quantization of perturbations}
\label{sec:QuantumSound}

Ironically, we now want to undo our classicalization path and take the quantum nature of the system into account but only for the perturbations. This procedure of \textit{re-quantization} originates in the broken symmetry picture we adopted for BECs, and is the simplest way to deal with a quantized excitation field in generic (non-homogeneous) condensates.
Bear in mind that more rigorous approaches that do not break number conservation and do not require to quantize again have been developed \cite{castin1998low}. The interest in studying quantum perturbations stems from the fact that BECs can reach very low temperatures (order of nK), where quantum aspects of fluctuations can be observable. From our Analogue Gravity point of view instead, the quantization of the fluctuation field allows us to study a \textit{quantum} field theory in a curved spacetime.

To proceed with the quantization of the field, we focus here on the case of a stationary background and on the situations in which there are no zero-norm modes. We will comment on the quantization of dynamically unstable theories in Section \ref{sec:dynamical-instabilities}.

The quantization of Bogoliubov fluctuations in BECs is usually performed in the atomic field variable $\psi$, which is a complex scalar field. Usually, in QFT courses, one learns that the canonical quantization of a complex scalar field is equivalent to the canonical quantization of two real scalar fields (its real and imaginary part), leading to the description of two degrees of freedom which, heuristically,describe particles and antiparticles associated to such field.

As explained above, the dynamical content of a standard BEC and their collective excitations corresponds to a single scalar degree of freedom (there are no antiparticles). Thus, the choice of a complex field variable, although motivated from a physical perspective, leads to a quantization procedure which can look confusing for someone whose background comes from standard QFT courses. However, it is analogous to the quantization of a real Klein-Gordon field (see e.g. the Appendix in \cite{fulling1989aspects} for a case with negative- and zero-norm modes). The way to make this obvious is by canonically quantizing collective excitations in density-phase variables, since these are naturally a canonical  field-momentum pair for a real scalar field. We will first do so, and later on present the usual quantization of the atomic-field perturbations, as well as its relation to the density-phase quantization.

Notice that passing from the density-phase real scalar field representation to the atomic complex field representation is exactly as passing from position-momentum to creation and annihilation operators for a quantum harmonic oscillator.

\subsection{Quantization of density-phase perturbations}
\label{sec:QuantumDensPhase}

Here we will quantize linear perturbations of density-phase variables of the form $n=n_0(1+\eta)$ and $S=S_0+\varphi$. To that end, we recall that phase and density are canonically conjugated variables, with the phase perturbation $\varphi$ playing the role of a scalar field and the density perturbation playing the role of its conjugate momentum as given by \eqref{eq:PhasPertMomentum}. Therefore, their quantization proceeds as with a standard real KG field. To that end, we promote our classical field-momentum pair to operators acting on an abstract Hilbert space (to be defined) and impose equal time commutation relations
\begin{equation}
    [\hat\varphi(t,\vx),\hat\varphi(t,\vx')]=0,
    \qquad
    [\hat\eta(t,\vx),\hat\eta(t,\vx')]=0,
    \qquad
    [\hat\varphi(t,\vx),\hat\eta(t,\vx')]=-\frac{\ci}{n_0(t,\vx)} \delta(\vx-\vx'),
    \label{eq:CCRDensPhase}
\end{equation}
where we note that the conjugate momentum is given by \eqref{eq:PhasPertMomentum} (i.e. these are standard canonical commutation relations (CCR) for a real scalar field when written in terms of field and conjugate momentum).

We now want to find a representation of these operators in terms of annihilation and creation operators, as usual. To that end, we note that given any two complex solutions $\varphi_1$ and $\varphi_2$ (and their conjugate momenta $\pi_{\varphi}=-\hbar n_0 \eta$), there is a bilinear form, known as symplectic product, which is conserved during evolution with respect to laboratory time. This product takes the form of the standard Klein-Gordon symplectic product if written in terms of field and momentum, namely\footnote{Though it may seem odd to have an $\hbar$ in the definition of the symplectic product in a classical field theory, it is a convenient normalization choice so that the standard definition of creation and annihilation operators in terms of the Klein-Gordon product leads to the standard ladder operator algebra. See Problem \ref{prob:LadderOpsFromProd} for details.}
\begin{equation}
    (\varphi_1,\varphi_2)\coloneqq\frac{\ci}{\hbar}\int_\Sigma \dif \vx (\varphi_1^* \pi_2-\pi_1^* \varphi_2)=\ci\int_\Sigma \dif \vx\, n_0(\eta_1^* \varphi_2-\varphi_1^* \eta_2).
    \label{eq:SympProdDensPhase}
\end{equation}
We see that, although this product is not positive definite, it is hermitian in the sense that $$(\varphi_1,\varphi_2)=(\varphi_2,\varphi_1)^*\,,$$ and it satisfies linearity and conjugate linearity in its second and first arguments respectively. It is also straightforward to show that $(\varphi_1^*,\varphi_2^*)=-(\varphi_1,\varphi_2)$. Hence, this product defines a (pseudo-)norm in the space of complex solutions to the density-phase perturbation equations by $||\varphi||\coloneqq(\varphi,\varphi)$. From this definition, we see that conjugated solutions have norms with opposite signs $||\varphi^*||=-||\varphi||$. These properties all remind to the Bogoliubov inner product defined in section \ref{sec:BdGEigenvalue}. As we will see later, both products are equivalent ---indeed, they are the same mathematical object in different representations of the theory (see also \cite{Barcelo:2010bq} for a detailed discussion).

The procedure of canonical quantization in terms of creation and annihilation operators involves the space of complex solutions to the classical equations of motion \cite{Wald:1995yp}. For a (massless) Klein-Gordon field in a stationary and homogeneous spacetime, modes with different frequency and wavevector evolve independently. Hence, a convenient basis of such space is provided by plane waves $\ee^{\mp\ci(\omega t-\vk\cdot\vx)}$, where $\omega\geq0$. Their norm is specified by their temporal part so that $\ee^{-\ci\omega t}$ modes have positive norm and $\ee^{\ci\omega t}$ modes have negative norm. These are also known as \textit{positive frequency} and \textit{negative frequency} modes. Though in these lectures we will only consider stationary BEC configurations, where the frequency of perturbations is also conserved, we will write the quantization procedure in terms of a general basis of modes, so that the quantization procedure can be translated to non-stationary BECs (provided one finds a suitable basis of complex solutions).

Consider now a basis $\{(\varphi_i,\varphi_i^*)\}$ of the space of complex solutions to the density-phase perturbation equations \eqref{eq:DensPhasePerturbations} that is orthonormal with respect to the inner product defined above. This basis in the form of ordered pairs of orthonormal conjugated solutions always exists, since the reality of the coefficients of the density-phase equations ensures that if $\varphi$ is a solution so is $\varphi^*$, and they are orthogonal from \eqref{eq:SympProdDensPhase}. Without loss of generality, we label the basis elements so that all $\varphi_i$ have positive-norm  and, therefore, all $\varphi_i^*$ have negative norm (think of $\varphi_i$ as $\ee^{-\ci\omega t}$ modes and $\varphi_i^*$ as $\ee^{\ci\omega t}$ modes). We can thus write any real solution to the density-phase perturbation equations as follows\footnote{In cases where the label of the chosen basis is continuous, the sum should be replaced by the appropriate integrals, with Kronecker deltas replaced by Dirac deltas and so on. We remark, however, that the procedure presented here is general since the spaces of complex solutions are usually separable, so that a basis labeled by a discrete index can always be found.}
\begin{equation}
\begin{split}
        \varphi(t,\vx)=\sum_i a_i \varphi_i(t,\vx) + a_i^* \varphi_i^*(t,\vx),
        \\
        \eta(t,\vx)=\sum_i a_i \eta_i(t,\vx) + a_i^* \eta_i^*(t,\vx),
\end{split}
\end{equation}
where $a_i= (\varphi_i,\varphi)$ and $a_i^*= -(\varphi_i^*,\varphi)$, and where $\varphi_i$ and $\eta_i$ are not independent of each other, but related by \eqref{eq:EtaPhi}. Here $\varphi_i$ and $\varphi_i^*$ play the role of positive and negative frequency solutions respectively in standard KG quantization. Promoting the above expansion into operator form is achieved as usual by promoting the coefficients of the linear expansion to operators as
\begin{equation}
\begin{split}
        \hat\varphi(t,\vx)=\sum_i \hat a_i \varphi_i(t,\vx) +\hat a_i^\dagger \varphi_i^*(t,\vx),
        \\
        \hat\eta(t,\vx)=\sum_i \hat a_i \eta_i(t,\vx) + \hat a_i^\dagger\eta_i^*(t,\vx),
\end{split}
\label{eq:DPmodeDecompQuant}
\end{equation}
or equivalently by defining the operators $\hat{a}_i$ and $\hat{a}^\dagger_i$ in terms of the conserved product
\begin{equation}
\begin{split}
    &\hat{a}_i\coloneqq (\varphi_i,\hat{\varphi})=\ci\int_\Sigma \dif V n_0(\eta_i^* \hat{\varphi}-\varphi_i^* \hat{\eta})\,,
   \\
    &\hat{a}^\dagger_i\coloneqq -(\varphi_i^*,\hat{\varphi})=\ci\int_\Sigma \dif V n_0(\varphi_i \hat{\eta}-\eta_i \hat{\varphi})\,.
\end{split}
\label{eq:DefLadderOp}
\end{equation}
Taken together, the above definitions, the orthonormality of the basis of solutions $\{\varphi_i,\varphi_i^*\}$, and the CCR \eqref{eq:CCRDensPhase}, imply that these operators satisfy standard commutation relations for creation/annihilation operators (see Problem \ref{prob:LadderOpsFromProd})
\begin{equation}
    [\hat a_i,\hat a_j]=0,
    \qquad
    [\hat a_i^\dagger,\hat a_j^\dagger]=0,
    \qquad\text{and}\qquad
    [\hat a_i,\hat a_j^\dagger]=\delta_{ij}.
    \label{eq:CCRLadder}
\end{equation}
Thus $\hat a_i$ and $\hat a_i^\dagger$ are bosonic creation and annihilation operators associated to complex solutions of the classical field equations, as usual. The Hilbert space of our QFT for collective excitations is thus the usual Fock space of a real scalar field, with ${\hat a}_i$ and ${\hat a}_i^\dagger$ annihilating/creating one-particle states. The latter represent small phase fluctuations around the condensate, known as phonons. 

When the BEC is stationary and homogeneous, these phonons have a well-defined frequency $\omega$ and wavevector $\vk$ that satisfy the Bogoliubov dispersion relation. {Indeed, phonons are precisely the collective degrees of freedom described in section \ref{sec:CollectiveBogoApp}. From there, we recall that a phonon oscillating at frequency $\omega$ involves the presence of atoms that oscillate at frequencies $\mu/\hbar+\omega$ and $\mu/\hbar-\omega$. This is behind the widespread use of the term \textit{negative frequency} in descriptions of collective excitations in condensed matter systems. The reason why this term may sound misleading to readers who studied QFT in more fundamental settings is the following. In fundamental settings, one usually quantizes fields whose \textit{zero mode}, i.e. the mode that is constant in time, corresponds to $\omega=0$. However, when we consider low-energy excitations of a stationary condensate, the nonvanishing chemical potential implies that the zero phononic mode involves atoms oscillating in time at frequency $\mu/\hbar$. Thus, although the formalism is completely parallel to fundamental QFT descriptions from the phononic perspective (in the sense of an effective field theory \cite{Burgess:2007pt}), when connecting to the microscopic nature of these effective (composite) degrees of freedom, the term negative frequency becomes useful to refer to the presence of atomic excitations at frequencies $\mu/\hbar-\omega$, which are negative frequencies with respect to the zero phononic mode. In the reminder of the lectures, we will conveniently refer to $\ee^{-\ci\omega t}$ modes as positive frequency modes and $\ee^{\ci\omega t}$ modes as negative frequency modes if necessary (here $\omega>0$).}

As a final remark, let us comment on stationary cases so that our basis of complex solutions is made of pairs of positive and negative frequency modes. Then, as we will see later, in some cases where the BEC configurations considered have a sub-to-supersonic transition, the sign of the norms of modes of the forms $\ee^{-\ci\omega t}\varphi_\omega(\vx)$ and $\ee^{\ci\omega t}\varphi^*_\omega(\vx)$ modes can be reversed for some frequencies. Thus, if we follow our procedure strictly, our mode basis should be the set union of $\big\{\big(\ee^{-\ci\omega t}\varphi_\omega(\vx),\ee^{\ci\omega t}\varphi^*_\omega(\vx)\big)\big\}_{\omega\in\Omega^+}$ and $\big\{\big(\ee^{\ci\omega t}\varphi^*_\omega(\vx),\ee^{-\ci\omega t}\varphi_\omega(\vx)\big)\big\}_{\omega\in\Omega^-}$, where $\Omega^+$ and $\Omega^-$ are the sets of frequency values for which $\ee^{-\ci\omega t}$ have positive norm\footnote{{To be more precise, often, eigenspaces of the Bogoliubov operator of definite frequency have dimension larger than one, i.e. there is degeneracy in frequency eigenmodes. In such case, one has to include a complete basis for every frequency. The presence of extra symmetries like translation or rotation invariance suggests a particular choice of basis within each frequency eigenspace.}}.

Another way to circumvent this that is commonly found in the analogue gravity literature is as follows. One keeps the chosen basis as $\big\{\big(\ee^{-\ci\omega t}\varphi_\omega(\vx),\ee^{\ci\omega t}\varphi^*_\omega(\vx)\big)\big\}$ for all frequencies. If one blindly writes the field expansion with the usual definition of creation and annihilation operators \eqref{eq:DefLadderOp}, then one finds that the daggered operators associated to $\Omega^-$ modes destroy particles while undaggered create them, which ends up being unconfortable for most of us. A simple solution comes from noting that it is the sign of the norm of the mode, and not whether it is positive or negative frequency, what determines what type of operator it defines. In general, operators defined from positive norm modes as in \eqref{eq:DefLadderOp} satisfy the algebra of annihilation operators, while those defined from negative norm modes satisfy the algebra of creation operators. Hence, if one does not want to be careful in oredring the mode basis so that all first members of the pairs are positive norm and all second members are negative norm, but wants to keep daggered/undaggered operators as creation/annihilation operators, the resulting field expansion reads
\begin{equation}
\begin{split}
        &\varphi(t,\vx)=\sum_{\omega\in\Omega^+}  \ee^{-\ci\omega t}\varphi_\omega(\vx) \hat{a}_\omega+  \ee^{\ci\omega t}		\varphi^*_\omega(\vx) \hat{a}^\dagger_\omega + \sum_{\omega\in\Omega^-}  \ee^{\ci\omega t}\varphi^*_\omega(\vx) \hat{a}	_\omega+  \ee^{-\ci\omega t}\varphi_\omega(\vx) \hat{a}^\dagger_\omega        
        \\
        &\eta(t,\vx)=\sum_{\omega\in\Omega^+}  \ee^{-\ci\omega t}\eta_\omega(\vx) \hat{a}_\omega+  \ee^{\ci\omega t}			\eta^*_\omega(\vx) \hat{a}^\dagger_\omega + \sum_{\omega\in\Omega^-}  \ee^{\ci\omega t}\eta^*_\omega(\vx) \hat{a}		_\omega+  \ee^{-\ci\omega t}\eta_\omega(\vx) \hat{a}^\dagger_\omega        
\end{split}
\label{eq:ModeExpNormSplit}
\end{equation}
\subsection{Quantization in atomic-field variables}
\label{sec:RealComplexRelationClass}

In this section we quantize collective excitations of a condensate in terms of atomic-field variables. These variables are the most common ones in the BEC literature. However, standard treatments follow a quantization procedure which looks unfamiliar to the reader trained in standard QFT courses. The reason for this is that the quantization of a single degree of freedom in terms of a complex variable is uncommon in relativistic QFT.

Below, we derive the quantization in terms of atomic-field variables from that of density-phase perturbations, which we successfully quantized by following the canonical quantization prescription for a real scalar field, a familiar procedure for those trained in relativistic QFT. By relating both procedures, we hope to clarify that the usual way of quantizing collective excitations in terms of the complex field $\psi$ is fully consistent with standard techniques in relativistic QFT. Though our approach differs from the usual route found in the BEC literature, it shows consistency between different quantization prescriptions which are well established in their respective fields. For interested readers, we refer to \cite{pethick2008bose,pitaevskii2016bose,castin1998low} for the standard procedure to quantize BEC perturbations directly in atomic-field variables.

Consider linear perturbations to the atomic field of the form $\Psi=\Psi_0+\ee^{\ci S_0}\hat{\psi}$, which satisfy a linear system of PDEs which can be written as a BdG problem \eqref{eq:BdGProblem}. The transformation in \eqref{eq:DPtoAtomic} between the atomic field variable $\Psi$ and density-phase variables yield to the exact relation
\begin{equation}
\Psi_0+\ee^{\ci S_0}\hat\psi=\sqrt{n_0(1+\hat{\eta})}\ee^{\ci (S_0+\hat{\varphi})}
\end{equation}
which, neglecting terms beyond linearity, leads to the following relation between atomic-field perturbations and density-phase perturbations 
\begin{equation}
\hat{\psi}(t,\vx)=\sqrt{n_0} \lr{\frac{\hat{\eta}(t,\vx)}{2}+\ci\hat{\varphi}(t,\vx)}.
\end{equation}
Therefore, using the mode decomposition of density-phase perturbations \eqref{eq:DPmodeDecompQuant} we find
\begin{equation}
\begin{split}
    &\hat\psi(t,\vx)=\sqrt{n_0}\sum_i \hat{a}_i \lr{\frac{\eta_i(t,\vx)}{2}+\ci\varphi_i(t,\vx)} + \hat{a}_i^\dagger \lr{\frac{\eta_i^*(t,\vx)}{2}+\ci\varphi_i^*(t,\vx)}
    \\
    &\hat\psi^\dagger(t,\vx)=\sqrt{n_0}\sum_i \hat{a}_i \lr{\frac{\eta_i(t,\vx)}{2 }-\ci\varphi_i(t,\vx)} + \hat{a}_i^\dagger \lr{\frac{\eta_i^*(t,\vx)}{2 }-\ci\varphi_i^*(t,\vx)}.
\end{split}
\end{equation}
In the BEC literature, the above equations are usually written as
\begin{equation}
\begin{split}
    &\hat\psi(t,\vx)=\sum_i \hat{a}_i u_i(t,\vx) + \hat{a}_i^\dagger v_i^*(t,\vx),
    \\
    &\hat\psi^\dagger(t,\vx)=\sum_i \hat{a}_i v_i(t,\vx) + \hat{a}_i^\dagger u_i^*(t,\vx),
\end{split}
\label{eq:AtomicFieldPertExp}
\end{equation}
where we have defined the atomic field modes $u_i$ and $v_i$ in terms of density-phase modes through the invertible smooth map
\begin{equation}
 \label{eq:MapAtomicDP}
\begin{split}
        &u_i(t,\vx)=\sqrt{n_0(t,\vx)}\lr{\frac{\eta_i(t,\vx))}{2}+\ci\varphi_i(t,\vx)}\\
        &v_i(t,\vx)=\sqrt{n_0(t,\vx)}\lr{\frac{\eta_i(t,\vx))}{2}-\ci\varphi_i(t,\vx)}
\end{split}
\quad\Longleftrightarrow\quad
\begin{split}
        &\eta_i(t,\vx)=\frac{u_i(t,\vx)+v_i(t,\vx)}{\sqrt{n_0(t,\vx)}}\\
        &\varphi_i(t,\vx)=\frac{u_i(t,\vx)-v_i(t,\vx)}{2\ci\sqrt{n_0(t,\vx)}}
\end{split}
\end{equation}
Now, the orthonormality properties of the density-phase perturbations with respect to the symplectic product \eqref{eq:SympProdDensPhase} have a counterpart in $u_i$ and $v_i$ field variables. Indeed, from $(\varphi_i,\varphi_j)=\delta_{ij}$ we find
\begin{equation}
    \int_\Sigma dV \lr{u^*_i u_j-v^*_i v_j}=\delta_{ij}.
\end{equation}
As well, the bi-orthogonality relations \eqref{eq:BdGBiOrtho} can be found from $(\varphi_i^*,\varphi_j)=0$. Hence, we see that the symplectic product \eqref{eq:SympProdDensPhase} and the BdG inner product \eqref{eq:BdGInnerProd} are indeed equivalent.

For completeness, let us show the mode expansion in the atomic field basis in the case where we have a stationary condensate for which some positive frequency modes have negative norm, parallel to \eqref{eq:ModeExpNormSplit}. In this case, the expansion reads
\begin{equation}
\begin{split}
    &\hat\psi(t,\vx)=\sum_{\omega\in\Omega^+}\Big( \ee^{-\ci\omega t} u_\omega(\vx) \hat{a}_\omega+ \ee^{\ci\omega t} v^*_\omega(\vx) \hat{a}_\omega^\dagger\Big) +\sum_{\omega\in\Omega^-} \Big(\ee^{\ci\omega t} v^*_\omega(\vx) \hat{a}_\omega+ \ee^{-\ci\omega t} u_\omega(\vx) \hat{a}_\omega^\dagger\Big)
    \\
     &\hat\psi^\dagger(t,\vx)=\sum_{\omega\in\Omega^+}\Big( \ee^{-\ci\omega t} v_\omega(\vx) \hat{a}_\omega+ \ee^{\ci\omega t} u^*_\omega(\vx) \hat{a}_\omega^\dagger\Big) +\sum_{\omega\in\Omega^-} \Big(\ee^{\ci\omega t} u^*_\omega(\vx) \hat{a}_\omega+ \ee^{-\ci\omega t} v_\omega(\vx) \hat{a}_\omega^\dagger\Big).
\end{split}
\end{equation}
It is also worth to mention that some authors also perform the field expansion by summing for $\omega$ taking also negative values, while only taking into account positive frequency modes in the expansion. In the end, if one is careful, all of these procedures describe the same physics.

Once the field is quantized, the second order correction to the GP energy functional \eqref{eq:bogo-energy-functional} becomes a quadratic quantum Hamiltonian
\begin{equation}
    \hat{H}^{(2)}=\sum_{\omega\in\Omega^+} \hbar\omega\hat{a}^\dagger_{\omega}\hat{a}_{\omega}-\sum_{\omega\in\Omega^-}\hbar\omega\hat{a}^\dagger_\omega\hat{a}_{\omega}-\sum_{\omega}\hbar\omega\int_\Sigma dV|v_{\omega}(\vx)|^2\,.
\end{equation}
This Hamiltonian provides a quantum description for fluctuations on top of the classical stationary background. Its zero-point energy is the leading-order correction to the GP stationary state energy. The last term is the energy contribution of the quantum depletion of the condensate \eqref{eq:quantum-depletion}, that is the fraction of atoms that is not in the condensate due to quantum fluctuations. Notice that the Hamiltonian term for the negative-norm modes has a minus in front, so that if the frequencies of these modes are positive this Hamiltonian is unbounded from below, i.e. the system is energetically unstable. Of course this linear prescription breaks down if the population of these modes becomes large enough.

\subsection{Quantization in the presence of dynamical instabilities}
\label{sec:dynamical-instabilities}

What we said up to now is valid in the absence of dynamically unstable modes. When they are present instead, quantization is less straightforward since their norm vanishes. This is a solved problem in relativistic quantum field theories, see for example \cite{fulling1989aspects}, and was adapted to the BEC case in \cite{garay2001sonic,leonhardt2003theory}. The idea is that, while the orthogonality relation \eqref{eq:BiOrtthogonality} implies the vanishing of the norm of complex-frequency modes, this is not the case for the Bogoliubov inner product between a mode of complex frequency $\w_k$ and its pseudo-degenerate partner (of frequency $\w_k^*$). We can hence associate to each mode $k$ the dual mode $\bar{k}$ with $\omega_{\bar{k}}=\omega_k^*$, in terms of which the (non-constant) part of the quantized Bogoliubov Hamiltonian can be written
\begin{equation}\label{eq:hamiltonian-unstable-1}
	\hat{H}^{(2)}=\sum_k \omega_k \hat{a}_{\bar{k}}^\dagger \hat{a}_k
\end{equation}
with the commutation relations
\begin{equation}
	[\hat{a}_k,\hat{a}_l^\dag]=\delta_{\bar{k}l}\;, \hsp [\hat{a}_k,\hat{a}_l]=0.
\end{equation}
This means that for complex-frequency modes the canonical conjugate of the destruction operator is not its hermitian conjugate, i.e. the creation operator, so that they are not creation and annihilation operators in the usual sense. The quantum field theory in the presence of dynamically unstable modes does hence not have a \textit{particle interpretation}.

Further insight can be obtained by defining the operators
\begin{equation}
	\hat{c}_k=\frac{\hat{a}_k+\hat{a}_{\bar{k}}}{\sqrt{2}}\;,\hsp \hat{c}_{\bar{k}}=i\frac{\hat{a}_k^\dag-\hat{a}_{\bar{k}}^\dag}{\sqrt{2}}
\end{equation}
that instead behave as ordinary annihilation operators $[\hat{c}_k,\hat{c}_k^\dag]=[\hat{c}_{\bar{k}},\hat{c}_{\bar{k}}^\dag]=1$. Hamiltonian \eqref{eq:hamiltonian-unstable-1} can hence be rewritten as
\begin{equation}
	\hat{H}_{k,\bar{k}}=\Re(\w_k)\left[\hat{c}_k^\dag\hat{c}_k-\hat{c}_{\bar{k}}^\dag\hat{c}_{\bar{k}}\right] - \Im(\w_k)\left[\hat{c}_k^\dag\hat{c}^\dag_{\bar{k}}+\hat{c}_k\hat{c}_{\bar{k}}\right].
\end{equation}
This is a two-mode squeezing Hamiltonian, with the imaginary part of the frequency introducing the anomalous terms associated to the production of pairs of excitations. This kind of two-mode squeezing is very similar to what one obtains for spontaneous quantum pair production, as we are going to see in the following when discussing superradiance and the Hawking effect.

\bigskip

The above description of the quantum fluctuations of a BEC as linear perturbations is useful provided that their backreaction on the condensate can be neglected. These describe quantum acoustic excitations, or phonons, within the BEC. In the remainder of the notes, we will be interested in understanding their dynamics for a variety of flow profiles. The reason is that these profiles have interest as simulators of the dynamics of QFT in curved spacetimes. 

\section{Superradiance}\label{sec:SR}

Now that we introduced all machinery that we need to describe sound in BECs, we can start to look at some interesting physical phenomena, with a connection to curved spacetimes physics. While the most famous topic in Analogue Gravity is, for historical reasons, probably Hawking radiation, we will here first deal with another phenomenon occurring in field theories in curved spacetimes: superradiance. The reason is that, although involving more complex spacetimes, it is a simpler phenomenon.

Superradiance is an amplified scattering of radiation that is ubiquitous in field theories. It occurs not only in charged and/or rotating black holes, but also strong electric fields, rotating fluids, and others (see \cite{brito2020superradiance} for an extensive review). In general, we can define superradiance as an amplified scattering (reflection or transmission) of radiation occurring in time-independent physical situations.
Because of stationarity, superradiance occurs in non-homogeneous systems, in which distinct spatial regions can be identified, and requires the availability of negative-energy excitations in one of the regions. 
Being based on these simple ingredients, one can give a general characterization of superradiance, independent of the specific system, in terms of the scattering coefficients describing the solutions to a general wave equation \cite{richartz2009generalized,Delhom:2023gni}. Physical examples where dynamics is described by equations of this form are, e.g. Zeldovich's scattering on a rotating absorber, Klein Paradox for a scalar charged field, amplified scattering in rotating Kerr spacetime, etc. For a complete discussion see the review \cite{brito2020superradiance}.

Negative-energy excitations may sound like an exotic thing, but what they mean is simply that some part of the system we are considering is energetically unstable, i.e. it is in an excited state such that energy can be extracted from it. In the case of a superfluid, negative energy excitations are the idea at the basis of Landau's criterion, that identifies the critical velocity above which superfluidity breaks down because negative energy excitations start to be available. As we saw, in BECs the critical velocity is the speed of sound. We hence can expect superradiance to occur when a subsonic-to-supersonic transition is present. Remember that, in the analogue gravity language, supersonic regions are analogue ergoregions. This directly connects to black-hole superradiance, in which superradiant scattering can occur in the presence of an ergoregion in a Kerr (rotating) black hole.

Here we will first provide a general characterization of superradiance, independent on the origin of the negative energy excitations that are needed. We will then display its occurrence in two fluid setups, a shear layer and a vortex, commenting on their occurrence in BECs. We will then discuss what happens at the quantum level in systems that display superradiance, in particular showing quantum pair creation in the modes involved in the superradiant scattering. Finally, we will discuss the instabilities that can occur when superradiant scattering is present and discuss their characterization in terms of the spectrum of the system. Superradiance has been experimentally observed only with gravity waves in water \cite{torres_rotational_2017} and in nonlinear optics \cite{braidotti_measurement_2022}, and has not been widely considered in BECs. A superradiant origin of the instabilities of quantized vortices in BECs was put forward in \cite{Giacomelli:2019tvr}, and classical classical superradiance with Bogoliubov dispersion was considered in \cite{patrick_rotational_2020} and, in a shear layer configuration in \cite{giacomelli2021superradiant}, where also superradiant instabilities were considered. Also, spontaneous quantum superradiance in a BEC was considered in \cite{giacomelli2021spontaneous}.


\subsection{General treatment of superradiant scattering}
Let us first start by explaining the basic mechanisms behind superradiance in a general way. Consider a wave described by a 1+1 dimensional massless Klein-Gordon equation (the argument generalized to any number of dimensions in a straightforward manner)
\begin{equation}\label{eq:general-KG}
g^{\mu\nu}\nabla_\mu\nabla_\nu\Phi=0,
\end{equation} 
where $\nabla_\mu$ is a covariant derivative encoding the spacetime connection and a possible coupling to gauge fields. This equation has a conserved quantity defined by
\begin{equation}
(\Phi_1,\Phi_2)_{\rm KG}\coloneqq\frac{\ci}{\hbar}\int_{\Sigma_t} \lr{\Phi_1^*\Pi_2-\Pi_1^*\Phi_2},
\label{eq:KGNorm}
\end{equation} 
where $\Pi=\sqrt{-g}n^\mu\nabla_\mu\Phi$ is the conjugate momentum density and $n^\mu$ the timelike vector associated to the chosen foliation\footnote{A foliation is a splitting of space-time into space and time. One can think the leafs of the foliation $\Sigma_t$ as stacked one after the other continuously as the time parameter $t$ runs, so that spacetime has topology $\Sigma_t\times\mathbb{R}$. One can think of $\Sigma_t$ as the space at different instants of time $t$. $n^\mu$ is then a vector with a component that points toward the future of $\Sigma_t$. If $\Phi_1$ and $\Phi_2$ are solutions to the Klein-Gordon equation, the value of this quantity is independent of $t$ for any foliation.}. This quantity is in general complex, but it satisfies $(\Phi_1,\Phi_2)_{\rm KG}=(\Phi_2,\Phi_1)_{\rm KG}^*$, defining a (pseudo-)inner product in the space of complex solutions to the Klein-Gordon equation, also known as the symplectic product. {The inner product of a solution with itself is instead real, but} we see that it is not positive definite. This product thus defines a conserved (pseudo-)norm in the space of complex solutions to the Klein-Gordon equation (see, e.g., \cite{Wald:1995yp}). Given a solution $\Phi$, we define its norm as $N_\Phi\coloneqq{(\Phi,\Phi)_{\rm KG}}$, and we say that it is is normalized if $|N_\Phi|=1$. Notice the similarity of \eqref{eq:KGNorm} and the Bogoliubov inner product expressed in density-phase variables \eqref{eq:SympProdDensPhase}. What we said about positive and negative norms and energies for collective excitations in a BEC is hence analogous to what happens for solutions of a Klein-Gordon equation.

Let us now consider wave-propagation as described by the above equation, assuming that the problem is stationary. This is formalized by requiring that there is a Killing vector field (KVF) that is timelike in asymptotic regions. Thus, the frequency $\omega$ measured by asymptotic observers associated to this KVF is conserved during evolution. Consider a basis of solutions labeled by this frequency $\omega>0$. Since the equation is linear, the solutions form a vector space, and because it is second order in time and space, there are two linearly independent solutions for each value of $\omega$. A possible basis of solutions which are orthonormal with respect to the symplectic product is given by normalized right- and left-moving normalized wavepackets of a given width peaked at $\omega$, which we denote respectively $W_r$ and $W_l$. 

Let us now discuss wave propagation in a generic situation in which there are inhomogeneities in a compact region of space (due to the spacetime metric or to the presence of a gauge field), so that the waves propagate freely in the asymptotic regions, far from the inhomogeneities. To that end, define the IN mode basis as the basis of wavepackets which, in the asymptotic region, move towards the inhomogeneous region, and the OUT basis as those which, in the asymptotic region move away from it. Since we are in one dimension, these bases are fully characterized by right and left moving wavepackets. In general, an IN left-moving wavepacket at early times $W_{l}^{\rm in}$ will evolve into a complicated mathematical form due to scattering with the inhomogeneities. However, clarity can be gained by writing it as a linear combination of OUT modes. In this case, this linear combination will consist of a reflected right-moving wavepacket  $W_{r}^{\rm out}$ and a transmitted left-moving wavepacket  $W_{l}^{\rm out}$ at late times. This situation can be described as
\begin{equation}
W_{l}^{\rm in}\xrightarrow{\text{time}} r\, W_{r}^{\rm out} + t\, W_{l}^{\rm out}\,.
\end{equation}
The reason why we choose to represent the equality with an arrow\footnote{Note that this arrow is actually an equality. IN and OUT wavepackets are two different basis of solutions (which are therefore global in time). Therefore, one can write an element of the IN basis as a linear combination of the OUT basis elements. This is precisely what is done here, with $r$ and $t$ being the coefficients of the linear combination.} is to emphasize that the the decomposition in terms of OUT modes is useful at late times, after the scattering has occurred, when the modes have the clear physical interpretation of wavepackets traveling away from the scattering region (and the same for IN modes at early times). Since the wavepackets are normalized with respect to the symplectic product \eqref{eq:KGNorm}, conservation of their norm, linearity properties of the symplectic product, as well as orthogonality between left and right moving wave packets lead to the requirement
\begin{equation}
1=\frac{N_{r}^{\rm out}}{N_{l}^{\rm in}}|r|^2 + \frac{N_{l}^{\rm out}}{N_{l}^{\rm in}}|t|^2 \,,
\end{equation}
where $N^{\rm in}_{r,l}$ and $N^{\rm out}_{r,l}$ are the norms of the wavepackets $W^{\rm in}_{r,l}$ and $W^{\rm out}_{r,l}$ respectively. Analyzing the above eqaution, we notice that either or both of the quantities ${N_{l}^{\rm out}}/{N_{l}^{\rm in}}$ or ${N_{r}^{\rm out}}/{N_{l}^{\rm in}}$ have to be positive for it to be satisfied. This implies that at least one of the OUT modes has to have a norm of the same sign as the IN mode. We can now distinguish the two cases. If both out modes have the same sign, we arrive at
\begin{equation}
1=|t|^2 + |r|^2,
\end{equation}
indicating that both the amplitudes of the reflected and transmitted wave are smaller than that of the incident wave. This is a scattering process where there is no superradiance. On the contrary, if one of the OUT modes has norm of opposite sign than that of the incident mode, we arrive at
\begin{equation}
1=|t|^2 - |r|^2\hspace{1.5cm}\text{or}\hspace{1.5cm} 1= |r|^2- |t|^2,
\end{equation}
Indicating that either the amplitude of the transmitted wave or that of the reflected one must be bigger than that of the incident wave. This amplification in stationary scattering scenarios is what is usually known as superradiance \cite{brito2020superradiance}.

The above description of the scattering of wavepackets can be generalized to include all possible scattering events. The standard KG equation \eqref{eq:general-KG} describes (up to) two traveling modes for a given frequency, so that all scattering processes can be described with a $2\times 2$ scattering matrix containing the scattering coefficients\footnote{In  Appendix \ref{app:ComputationScattCoef}, we provide an algorithm that serves to compute the scattering amplitudes, namely the elements of the matrix $\bs{B}$, for any stationary scattering problem, provided one knows the form of the IN and OUT modes in the asymptotic regions (or at least an approximated form).} $T,R,t,$ and $r$, defined by
\begin{equation*}
\begin{matrix}
W_{r}^{\rm in}\xrightarrow{\text{time}}
T\, W_{r}^{\rm out} + R\, W_{l}^{\rm out}
\\[2pt]
W_{l}^{\rm in}\xrightarrow{\text{time}}
\, r\, W_{r}^{\rm out} + t\, W_{l}^{\rm out}
\end{matrix}
\end{equation*}
or in matrix notation
\begin{equation}
(W_{r}^{\rm in},W_{l}^{\rm in})
\xrightarrow{\text{time}}
(W_{r}^{\rm out},W_{l}^{\rm out})\cdot \bs{B}
\qquad\text{where}\qquad
\bs{B}=
\begin{pmatrix}
T & r\\
R & t
\end{pmatrix}\,.
\label{eq:GeneralModeEvolution}
\end{equation}
The most general scattering will be described by an IN wave of the form $a_l\,W^{\rm in}_{l} +a_r  \, {W^{\rm in}_{r}}$, and a scattered OUT wave of the form $b_r\,W^{\rm out}_{r} +b_l  \, {W^{\rm out}_{l}}$, with $a_i, b_i$ being complex wave amplitudes. The transformation of the wavepackets of the basis can be recast to a transformation of the amplitudes:
\begin{equation}
    (W_{r}^{\rm in},W_{l}^{\rm in})\cdot
    \begin{pmatrix}
        a_r\\
        a_l
    \end{pmatrix}
\xrightarrow{\text{time}}
(W_{r}^{\rm out},W_{l}^{out})\cdot\bs{B}\cdot
    \begin{pmatrix}
        a_r\\
        a_l
    \end{pmatrix}
    =
(W_{l}^{\rm out},W_{r}^{\rm out})\cdot \begin{pmatrix}
        b_r\\
        b_l
    \end{pmatrix}\,,
\end{equation}
which leads to the relation
\begin{equation}
    \begin{pmatrix}
        b_l\\
        b_r
    \end{pmatrix}
    =
     \bs{B}\cdot
     \begin{pmatrix}
        a_l\\
        a_r
    \end{pmatrix}
     \label{eq:GenScattProcess}
\end{equation}
Knowing the scattering matrix $\bs{B}$ provides a useful criterion, found in \cite{Delhom:2023gni}, to assess if a scattering process involves superradiance.

\begin{theorem}{\textbf{--- Characterization of superradiance}}

    A stationary scattering process is superradiant if and only if the matrix $\bs{B}$ describing the transformation between orthonormalized IN and OUT modes is not a unitary matrix.
    \label{thm:CharacterizationSR}
\end{theorem}

It is important to note that superradiant scattering can only occur if there are IN and OUT modes with different sign of their norms, for otherwise conservation of the norm would be broken (see Problem \ref{prob:SRTheorem} for details). This norm-mixing is at the heart of all amplification phenomena. In their quantum version, norm-mixing translates into mixing of creation and annihilation operators, which leads to quantum pair creation. Furthermore, for sufficiently pure IN states, the created pairs will be entangled. Generally, this norm-mixing processes between two modes can be described as two-mode squeezing operators \cite{IvanAnthony}.

\subsubsection{Characterization of superradiance through dispersion relations}

{Besides the above quantitative characterization in terms of the scattering matrix, a more intuitive and graphical method that does not require the knowledge of the scattering coefficients can be used to understand if a scattering event will be superradiant or not. If the scattering problem has asymptotic regions which are homogeneous enough, one can reason in terms of a dispersion relation for plane waves.

As an example, for a moving fluid, if the flow velocity $\vtv(\vx)$ varies over a typical length scale $L$ and we are interested in fluctuations with wavevectors $\abs{\vk}\gg 1/L$, we can consider a \textit{local} (WKB) dispersion relation. If we are interested in stationary scattering, the different frequency $\omega$ sectors will not mix and we can look for solutions of the form $\ee^{\ci\omega t}\varphi_\omega(x)$. This conserved frequency will select some wavenumbers for the modes in the different asymptotic regions, according to the local dispersion relation
\begin{equation}\label{eq:general-dispersion}
    \omega=\vtv(\vx)\cdot\vk_\omega \pm \cs \tk_\omega\lr{1+F(\tk_\omega)}.
\end{equation}
}
with $F(\tk_\omega)$ a function that vanishes in the hydrodynamic Klein-Gordon description, but is in general a non-linear function encoding the dispersion. This function is positive/negative definite for super-/sub-luminal media and, generally, it determines the number of modes per frequency (i.e. the dimension of the frequency eigenspaces). For the Bogoliubov dispersion relation in a BEC, $1+F(\tk)=\sqrt{1+\xi^2 \tk^2/4}$ and there are four solutions (i.e four wavevectors) per frequency.

One can check that the sign of the norm of the modes that solve the dispersion relation is positive for the $+$ branch, and negative for the $-$ branch, since it is given by the sign of $\omega-\vtv\cdot\vk_\omega$ (see Problem \ref{prob:SymProd}) . When the negative-norm branch raises to positive frequencies (or the other way around) the corresponding modes will have a negative energy. Since stationary scattering occurs only within modes of the same frequency, to see if a scattering event would be superradiant or not, it is enough to check whether in/outgoing propagating\footnote{We refer to modes with real wavevector as propagating modes, as opposed to evanescent modes. Generally, evanescent modes either vanish exponentially in spatial asymptotic regions, or are unphysical due to exponential growth. Even thought accounting for their presence can be relevant when solving the wave equations in the bulk and, therefore, to obtain the scattering matrix, they are not part of the IN and OUT basis.} modes of a given frequency have norms of opposite signs, and hence energy of opposite signs, in each asymptotic region.  This can be easily visualized by plotting the curves \eqref{eq:general-dispersion} in each asymptotic region side by side and drawing horizontal lines corresponding to different values of $\omega>0$. If there is any positive value of $\omega$ for which the horizontal line crosses the positive norm branch in one region and the negative norm branch in the other region, the scattering at that frequency will be superradiant, should it occur.

\begin{figure}
    \center
    \includegraphics[width=0.6\textwidth]{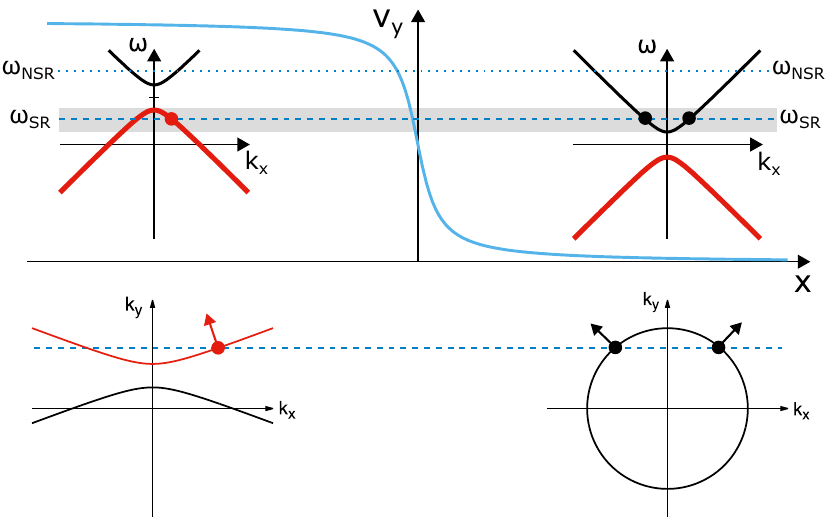}
    \caption{Upper part: velocity profile of the shear layer (light blue) and dispersion relations in the two asymptotic regions. The plotted lines are the positive (black upper curve) and negative (red lower curve) norm branches of the function $\omega=\tv_y \tk_y\pm\cs\sqrt{\tk_x^2+\tk_y^2}$ plotted as a function of $\tk_x$ for a fixed non-vanishing value of $\tk_y$, so that the effective mass gap (difference between the positive and negative branches) is $2\cs\tk_y$. Here $v_l>2\cs$, so that superradiant scattering is possible. The horizontal dashed line at $\omega_{\text{SR}}$ satisfies the superradiant condition \eqref{eq:SRcondShearLayer}, while the dotted line at $\omega_{\text{NSR}}$ does not. Lower part: same dispersion relation plotted at a fixed frequency in the superradiant interval. One can see the dependence on the transverse momentum. The arrows indicate the direction of the group velocities of the modes involved in the superradiant scattering process.}
    \label{fig:DispersionShear}
\end{figure}

A simple example of this graphical method is provided in Figure \ref{fig:DispersionShear}, and it will be applied in the following also for rotational superradiance and the analogue Hawking effect. This graphical method is more intuitive and simpler than computing the scattering coefficients, that can be cumbersome to obtain in general settings. However, this does not guarantee the occurrence of the scattering, nor will it provide the value of the scattering coefficients, which will strongly depend on the inhomogeneities of the problem. However, it is a quick and practical method to know a priori whether superradiant amplification can occur.

It is important to comment that our general treatment of superradiance so far supposes the existence of two \textit{open} asymptotic regions, i.e. we are imposing \textit{radiative boundary conditions} that do not backscatter radiation. These can be provided by infinite space or by a horizon in a black hole spacetime. If reflection at the boundaries is allowed, this usually results in \textit{superradiant instabilities}, that we will describe in Section \ref{sec:sr-instabilities}. In terms of the wavepackets we introduced above however, one can always think of scattering as we described up to now as long as the time window considered is smaller than the travel time for the waves to reach the boundaries after the scattering.

\subsection{Waves scattering on a shear layer}
\label{sec:shear-layer}

Let us start with what is probably the simplest example of superradiant scattering of sound waves, that will guide our intuition in the following. This occurs in a fluid in (at least) two spatial dimensions, and a \textit{shear layer}, that is a region of space in which the velocity is changing in module perpendicularly to its direction\footnote{Notice that such a configuration can display instabilities of the surface (such as Kelvin-Helmholtz instability), but we will not consider this physics here. There is, however (at least theoretically) a way to obtain such a tangential discontinuity in stable way in a BEC, as we are going to briefly discuss in the following.}.

Let us consider a two-dimensional compressible fluid of uniform density. We take its velocity to be directed along $y$ and varying along $x$, that is $\mathbf{v}=\tv_y(x) \hat{\mathbf{y}}$. Furthermore, we will assume that the velocity profile is asymptotically constant, in particular $\tv_y(x\to -\infty)={\tv}_l$ and $\tv_y(x\to +\infty)=0$, as depicted in the upper part of Figure \ref{fig:DispersionShear}. We ignore for the moment dispersive effects and assume that sound propagation is described by a standard Klein-Gordon equation (i.e. we are in the hydrodynamic regime), with the acoustic metric given by
\begin{equation}\label{eq:shear-layer-metric}
	g_{\mu\nu} = \frac{n_0}{c_s} 	
	\begin{bmatrix}
		-({\rm c}_{\rm s}^2-\tv_y^2) & 0 & -\tv_y\\
		0 & 1 & 0\\
		-\tv_y & 0 & 1
	\end{bmatrix}.
\end{equation}
The interesting regime is obtained for $\tv_l>c_s$, for which the $x\to-\infty$ asymptotic region is an ergoregion. If the increase in $\tv_y$ is monotonous, then at the position where $\tv_y(x)=c_s$ we have a \textit{planar ergosurface}.

We consider stationary solutions to the Klein-Gordon equation in this metric. A useful basis of solutions will be of the form $\Phi(t,x,y)=\ee^{i\tk_yy-i\omega t}\varphi(x)$, since $\tk_y$ is conserved given the translational invariance of the setup along the $y$ direction. We can hence write the Klein-Gordon equation for a fixed frequency and $\tk_y$ as
\begin{equation}\label{eq:KG-shearlayer}
    \lrsq{\frac{\partial^2}{\partial x^2}+\lr{\frac{\omega-\tv_y \tk_y}{\cs}}^2-\tk_y^2}\varphi(x)=0.
\end{equation}
We are interested in scattering solutions for this equations, that is solutions in which a plane wave is incoming from one the two uniform regions. In particular, take a plane wave coming from $+\infty$, in the two asymptotic regions the solution will have the form
\begin{equation} 
	\begin{aligned}
		\varphi(x\to+\infty)&=e^{i\tk_{\rm in} x} + R e^{-i\tk_{\rm in} x}\\
		\varphi(x\to-\infty)&=T e^{i\tk_{\rm tr} x}.
	\end{aligned}
\end{equation}
The wavevectors $\tk_{\rm in}$ and $\tk_{\rm out}$ are the solutions to the dispersion relations in the two asymptotic regions for fixed $\tk_y$ and $\omega$, that need to be selected so that the $x$ component of the group velocity $\tv_{g,x}=\partial_{\tk_x}\omega(\vk)$ of the plane wave is pointing in the right direction.

An intuitive understanding can be obtained simply from plots of the dispersion relation, that is well defined in the asymptotic regions because of the asymptotic homogeneity of the fluid. This is given by the Doppler-shifted dispersion relation
\begin{equation}
    \omega-\tv_y \tk_y=\pm\cs\sqrt{\tk_x^2+\tk_y^2}.
\end{equation}
This dispersion relation can be understood as giving $\omega$ as a function of $\tk_x$ so that the considered wave is a solution of \eqref{eq:KG-shearlayer}, where $\tk_y$ is a conserved quantity that enters as a parameter in the dispersion. In the upper part of Figure \ref{fig:DispersionShear}, we plot such a dispersion for a fixed $\tk_y$ in the two asymptotic regions. When seen this way, the conserved wavenumber $\tk_y$ generates an effective mass gap, so that at $x\to\infty$ (where only positive-energy modes exist) propagating modes must satisfy $\omega>\cs\tk_y$. The Doppler shift due to the transverse velocity is also responsible for a vertical shift of the dispersion relation, so that at $x\to-\infty$, where the fluid is supersonic, negative-energy waves are available, visible as propagating negative-norm modes at positive frequencies $0<\omega<\tk_y(\tv_l-\cs)$. One can already see that there exists a frequency interval (grey area) in which only positive-energy modes are available at $x\to\infty$ and only negative-energy ones are available at $x\to\-\infty$. Conservation of norm then implies that, if a positive-energy mode in this frequency range coming from $+\infty$ is transmitted to $-\infty$, it will also be reflected with a larger amplitude than that of the incoming mode. This is the essence of superradiant scattering.

From this argument one can already derive the frequency range over which superradiant scattering is possible
\begin{equation}
    \tk_y \cs<\omega<\tk_y(\tv_l-\cs),
    \label{eq:SRcondShearLayer}
\end{equation}
which can only be satisfied if $\tv_l>2\cs$. This is due to the effective mass gap, as can be seen in figure $\ref{fig:DispersionShear}$. For $\tk_y=0$ modes, the gap vanishes, but there are no negative-energy modes. Physically, one can understand why the supersonic region must be moving at least at two times the speed of sound from the translational invariance of the configuration along $y$; in fact, for smaller velocities one can always perform a Galilean transformation to a frame in which the full fluid is subsonic. As we are going to see, this condition is absent in rotating configurations, that are geometrically more cumbersome but only require the existence of a supersonic flow to display superradiance (if open boundary conditions are assumed).

A more quantitative analysis, that is useful in more complicated cases, can be obtained by taking advantage of the fact that, as a consequence of norm conservation for the Klein-Gordon equation\footnote{Notice that the expression of the Wronskian for the complex conjugate solutions has the shape of a probability current. The conservation of this current is related to the conservation of the Klein-Gordon (symplectic) product.}, for \eqref{eq:KG-shearlayer} the \textit{Wronskian} $W(\varphi_1,\varphi_2):=\varphi_1(\partial_x\varphi_2)-(\partial_x\varphi_1)\varphi_2$ of two solutions $\varphi_1$ and $\varphi_2$ is constant in space. By computing it with $\varphi_1=\varphi$ and $\varphi_2=\varphi^*$ in the two asymptotic regions one obtains the relation between the reflection and transmission coefficients
\begin{equation}\label{eq:shear-layeramplification-condition}
	1-\abs{R}^2 = \frac{\tk_{\rm tr}}{\tk_{\rm in}}\abs{T}^2.
\end{equation}
As we already said, the wavenumbers in the two regions must be chosen so that the group velocity is in the right direction, that is towards the shear layer for $\tk_{\rm in}$ and away from the shear layer for $\tk_{\rm tr}$. In Figure \ref{fig:DispersionShear} we indicate with dots the modes involved in the superradiant scattering process. The $x$ component of the group velocity is given by the slope of the curves in the upper plots. The crucial point is that, for negative energy modes, the group velocity has the opposite sign of $\tk_x$. This implies that the ratio appearing in \eqref{eq:shear-layeramplification-condition} will be negative if the transmitted wave is a negative-energy one, and hence that $\abs{R}>1$.

\subsubsection{Electrostatic equivalence}
\label{sec:kleinparadox}

It is interesting to compare the Klein-Gordon equation in the shear-layer metric for a fixed transverse momentum $\tk_\ty$ \eqref{eq:KG-shearlayer} with the one for a one-dimensional massive charged scalar field in an electrostatic potential
\begin{equation}\label{eq:KG-electrostatic}
    \lrsq{\frac{\partial^2}{\partial x^2}+\lr{\frac{\omega-eA_0/\hbar}{c}}^2-\frac{m^2c^2}{\hbar^2}}\varphi(x)=0,
\end{equation}
where $A_0(x)$ is the electrostatic potential, $m$ is the mass of the field and $c$ is the speed of light. The two equations are mapped into each other with
\begin{equation}\label{eq:parallelism}
	\frac{m^2c^2}{\hbar^2}\longleftrightarrow \tk_y^2 \hspace{0.8cm} \frac{eA_0}{\hbar}\longleftrightarrow \tk_y \tv_y \hspace{0.8cm} c\longleftrightarrow \cs\,.
\end{equation}
That is, the transverse momentum $\tk_y$ in the shear layer can be thought as both a mass and a charge for the reduced one-dimensional field, while the flow velocity is giving the electrostatic field.

This provides a mapping to what is known as the bosonic Klein paradox \cite{brito2020superradiance}. A treatment of the scattering analogous to the one we performed in this section reveals amplified scattering for the charged field. If the electrostatic potential jump between the two asymptotic regions is $\bar A_0$, the condition $\tv_l>2\cs$ that we found above becomes here $e\bar A_0>2mc^2$, i.e. the electrostatic potential energy must be larger than the mass gap. This suggests that in the Klein paradox the amplification is due to the production of particles and antiparticles, i.e. negatively charged antiparticles are transmitted to allow amplification with conservation of the total charge.

This parallelism between the two-dimensional flowing background and an electrostatic potential in one dimension is also relevant in black hole physics. In fact, superradiant scattering can occur not only in rotating black holes (as we are going to see in the next section), but also in non-rotating charged static Reisser-Nordstr\"om black holes. In this case the negative energy states are not provided by gravity (or by the background flow), but by a strong enough electric field, while the black hole spacetime is important only to provide the correct boundary conditions to the scattering problem.

The superradiant instabilities that we are going to discuss in \ref{sec:sr-instabilities} have parallels for charged fields. In fact, a precursor to the study of superradiant instabilities was the discovery of instabilities for a relativistic bosonic field in an electrostatic potential box, known as the Schiff-Snyder-Weinberg effect \cite{schiff1940existence}. This was connected to superradiant instabilities in black holes in \cite{fulling1989aspects}.

\subsubsection{Shear layer model in a BEC}
\label{sec:shear-layer-BEC}
The shear layer profile we are considering is often considered in hydrodynamics. However, from the point of view of superfluids it is not natural since a velocity $\vtv=\tv_y(\tx)\hat{\vy}$ is not irrotational, i.e. it does not have vanishing curl. In the particular case of a BEC it is not a stationary solution of the GPE. However, it is possible to circumvent this limitation and realize a stationary condensate with this velocity profile by working with what are known as \textit{synthetic gauge fields} \cite{dalibard2011colloquium,goldman2014light}. We will not go into the details, but a possible idea is to take advantage of the internal degrees of freedom of the atoms by coupling them to external lasers to create a dressed state whose dispersion relation has a minimum at a non-zero value of $\vk$. The resulting condensate is described by a modified GPE of the form
\begin{equation}\label{eq:gpe-gaugefield}
    \ci\hbar\partial_t\Psi=\lrsq{\frac{(-i\hbar\vna-\bs{A})^2}{2m}+V_{\rm ext}+g|\Psi|^2}\Psi.
    \end{equation}
which can be understood as a standard condensate minimally coupled to a synthetic gauge field. By resorting to the hydrodynamical representation we can see how the synthetic gauge field modifies the relation between the phase of the condensate and its velocity in the lab frame, which is now given by $\tv=(\hbar \vna\theta-\bs{A})/m$. A finite curl of the velocity of the condensate can hence be given by the curl of the synthetic gauge field, which is externally imposed. Hence, one can engineer non-irrotational flow profiles.

A shear layer with a synthetic vector potential jumping discontinuously from zero to a nonzero value was investigated in \cite{giacomelli2021understanding}. This shear layer turns out to be stable, differently from analogous classical fluids, displaying for example Kelvin-Helmholtz instability. Moreover, the effect of the superluminal Bogoliubov dispersion was considered. Importantly, while there are qualitative and quantitative deviations from the dispersionless case, the excitation modes involved in superradiance are analogous. We will see that this is not the case for the Hawking effect, where the microscopic description is fundamentally different in the dispersive case.

\subsection{Rotational superradiance in a 2D vortex}\label{sec:SR-vortex}

After treating the geometrically simpler shear layer case, let us now consider a rotating fluid. Superradiance was originally considered in rotating media, and is a feature of rotating black holes \cite{brito2020superradiance}. In Section \ref{sec:vortex-geometry} we introduced the vortex geometry, which is an analogue rotating black hole. The physics of this kind of analogue spacetime has been extensively studied with gravity waves in water: superradiant scattering was first observed in \cite{torres_rotational_2017} and also quasi-normal modes \cite{torres2020quasinormal} and backreaction \cite{patrick2021backreaction} have been experimentally investigated. Superradiant scattering was also observed for a vortex in a nonlinear optics experiment \cite{braidotti_measurement_2022}.

Consider the Klein--Gordon equation in the $(2+1)$-dimensional vortex metric \eqref{eq:AcousticMetricPolar}. Because of the stationarity of the flow and its symmetry under rotations, we can take a solution of the form
\begin{equation}
\phi(t,r,\theta)=e^{-i\w t}e^{im\theta}R(r),
\end{equation}
where $m$ is the integer azimuthal number. The Klein--Gordon equation for $R(r)$ turns out to be
\begin{equation}\label{eq:kg-vortex}
\begin{split}
\frac{1}{r}\left(1-\frac{A^2}{{\rm c}_{\rm s}^2r^2} \right)\frac{d}{dr}& \left[ r \left(1-\frac{A^2}{{\rm c}_{\rm s}^2r^2} \right) \frac{d}{dr}\right] R(r)+\\
&+\left[\omega^2 - \frac{2Bm\omega}{c_s r^2} - \frac{m^2}{r^2} 
\left(1-\frac{A^2+B^2}{{\rm c}_{\rm s}^2r^2}\right) \right] R(r)  =  0.
\end{split}
\end{equation}

Since we are interested in scattering solutions, it is convenient to have well defined asymptotic regions. It is hence useful to introduce the \textit{tortoise coordinate} $r^*$
\begin{equation}
dr_*=\left(1-\frac{A^2}{{\rm c}_{\rm s}^2r^2}\right)^{-1}dr\ \Longrightarrow\ r_*=r+\frac{|A|}{2c_s}\log\left|\frac{r-|A|/c_s}{r+|A|/c_s}\right|,
\end{equation}
whose main effect is to push the horizon position to $r_*\to -\infty$. It is also convenient to rewrite the radial problem in terms of $G(r_*)= r^{1/2}R(r)$, so that equation (\ref{eq:kg-vortex}) becomes
\begin{equation}\label{eq:kg-vortex-rewritten}
\frac{d^2G(r^*)}{dr^{*2}}+\underbrace{\left[Q(r)+\frac{1}{4r^2}\left(1-\frac{A^2}{{\rm c}_{\rm s}^2 r^2}\right)^2 - \frac{A^2}{{\rm c}_{\rm s}^2r^4}\left(1-\frac{A^2}{{\rm c}_{\rm s}^2 r^2}\right)\right]}_{V_{\mathrm{eff}}(r^*)}G(r^*)=0,
\end{equation}
where we defined
\begin{equation}
Q(r)= \frac{1}{r^4}\left[(A^2+B^2-{\rm c}_{\rm s}^2r^2)m^2-2Bm\w r^2+\w^2 r^4\right].
\end{equation}

This equation has a Schr\"odinger-like shape, with an \textit{effective potential} that depends on the parameters of the vortex flow, and on the frequency and azimuthal number of the field. This potential takes constant values in the asymptotic regions 
\begin{equation}
\begin{cases}
    V_{\mathrm{eff}}(r_*\to\infty)=\w^2 \\
    V_\mathrm{eff}(r_*\to-\infty)=\left(\w-m\Omega_H\right)^2
\end{cases},
\end{equation}
where we defined the angular velocity of the horizon $\Omega_H:=c_sB/A^2$. Since the potential is constant in these regions, we can find asymptotic solutions as a superposition of plane waves
\begin{equation}\label{eq:vortex-asymp-waves}
	\begin{aligned}
		G(r^*\to+\infty)&=e^{i\tk_{+\infty} r^*} + R e^{-i\tk_{+\infty} r^*}\\
		G(r^*\to-\infty)&=T e^{i\tk_{-\infty} r^*},
	\end{aligned}
\end{equation}
where, as discussed in Section \ref{sec:shear-layer}, the wavenumbers need to be chosen so that the corresponding plane wave as a group velocity pointing in the right direction. In particular, $\tk_{-\infty}$ has a negative group velocity.

Just as in Section \ref{sec:shear-layer}, a relation between the reflection and transmission coefficients can be obtained from the conservation of the Wronskian of solution \eqref{eq:vortex-asymp-waves} and its complex conjugate:
\begin{equation}\label{eq:refl-coeff-vortex}
    \abs{R}^2=1-\frac{\tk_{-\infty}}{\tk_{+\infty}}\abs{T}^2=1-\frac{\omega-m\Omega_H}{\omega}\abs{T}^2.
\end{equation}
This implies that transmission to negative-energy waves, i.e. superradiant scattering can occur for
\begin{equation}\label{eq:sr-condition-vortex}
    \omega<m\Omega_H
\end{equation}
$\omega<m\Omega_H$. This superradiant condition involving frequency and angular momentum is very general and comes out in different physical contexts \cite{brito2020superradiance}.

\begin{figure}
    \center
    \includegraphics[width=0.7\textwidth]{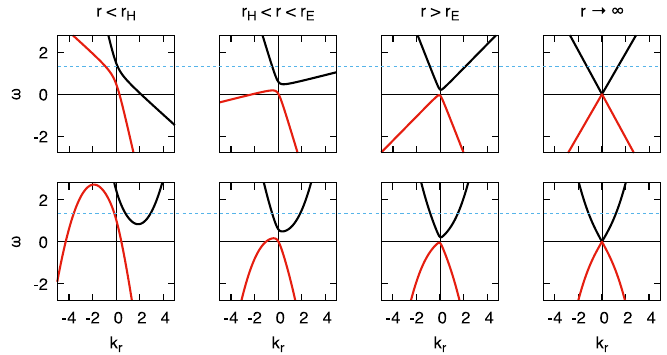}
    \caption{Local Klein-Gordon (upper panels) and Bogoliubov (lower panels) dispersion relations in the draining vortex geometry. Here an azimuthal number $m=1$ is fixed and the dispersion is plotted as a function of the radial wavenumber with the local value of the velocity at different positions. The parameters of the flow velocity \eqref{eq:vortex-geometry-flow} are $A=3\cs$ and $B=4\cs$.}
    \label{fig:dispersion-relations-vortex}
\end{figure}

It is interesting to comment that the superradiant condition \eqref{eq:sr-condition-vortex} on the frequencies that can be amplified comes from imposing a one-way \textit{radiative} boundary condition at the horizon, that is a one-way barrier from which waves cannot go back. This can be seen in the first panel of the upper row of Figure \ref{fig:dispersion-relations-vortex}, in which one sees that only left-moving waves are available inside the horizon.

One can also consider a situation in which there is no horizon, for example if the flow is purely azymuthal ($A=0$). In this case the frequencies at which negative energy modes are available inside the ergoregion is not bounded by \eqref{eq:sr-condition-vortex}. However, the choice \eqref{eq:vortex-asymp-waves} of a single ingoing wave in the interior region is not justified. Since the system is now \textit{closed} on the inside, allowed interior solutions are standing waves, leading to a discrete spectrum. If the frequency of the incident wave does not coincide with that of an allowed standing wave, total reflection of the incoming wave $\abs{R}^2=1$ will occur, and no amplification will take place. On the other hand, if the incident wave has the precise frequency of one of the standing waves, some transmission will occur, leading to an amplified reflected wave. Notice however that this is not the full story. This simple picture is given by the fact that we are looking for stationary scattering solutions. In fact, as we discuss in Section \ref{sec:sr-instabilities}, in this case dynamical instabilities emerge, related to solutions with a complex frequency.

As a last comment, here we considered a constant-density vortex and non-dispersive sound waves. In a Bose-Einstein condensate a couple of differences can be expected. First, vortices with a drain do not occur naturally as solutions to the GPE. Moreover, in the drain-less case, vortices are stationary solutions, but will have a quantized circulation and the supersonic region (i.e. the ergosurface) will also correspond to a dip in the density. However, different tricks can in principle be played to go beyond these constraint, like evaporating atoms at a point to create a radial flow or working with multi-component condensates and focusing on the \textit{spin} modes that see a flat density profile. Another way of bypassing this shortcoming is to use dissipative fluids with in-built dissipative mechanisms, such as polaritons \cite{Delhom:2023gni,Guerrero:2025kdn}.

Second, the superluminal Bogoliubov dispersion relation will change the available propagating modes in the different parts of the analogue spacetime. A comparison can be seen in Figure \ref{fig:dispersion-relations-vortex}. One can see that, apart from the fact that a momentum-dependent group velocity makes the ergosurface position ill-defined (in the sense that high momentum waves can still travel against the flow within supersonic regions), the mode structure at a fixed frequency is qualitatively the same for the Klein-Gordon and Bogoliubov fields as long as $r>r_H$. In particular the propagating modes are always up to two, and inside the ergoregion one can see negative energy modes available in both cases. Since superradiant scattering basically depends on the mode structure outside of the horizon, one can see that dispersion does not fundamentally change the picture.

This is not the case for the region inside the horizon, where one can see that important qualitative differences emerge in the available modes at a fixed frequency emerge. In particular, left-moving modes exist and there can be up to four propagating modes. This changes crucially the behavior of the horizon as a boundary condition, and we will see plays a major role in the microscopic explanation of the analogue Hawking effect. In other words, while superradiant scattering is not essentially altered by dispersion, the stationary analogue of the Hawking effect fundamentally depends on it. Superradiance, even though it involves more complex spacetimes, is \textit{simpler} from the analogy point of view, and this is why we chose to treat it first.

\subsection{Quantum superradiance}\label{sec:QuantumSR}

We will now extend the classical formalism of superradiant scattering described above to the quantum realm. In order to do this, we follow the quantization procedure described in section \ref{sec:QuantumDensPhase}. We will use two different basis to quantize, the IN or OUT bases described in the previous section. We will assume that the solutions labeled as  $W^{\rm in,out}_{l,r}$ are linear combinations of positive-frequency plane waves, and those labeled by $W^{\rm in,out}_{l,r}{}^*$ are linear combinations of negative-frequency plane waves. Then, we associate a creation and annihilation operator $\hat{a}_{l,r}{}^{\rm in,out}$ to each normalized wavepacket through the symplectic product, as described in \eqref{eq:DefLadderOp}. If we are careful in associating annihilation operators to positive-norm wavepackets and creation operators to negative-norm ones, the defined operators will satisfy canonical commutation relations. 

For each of the four wave packets involved in the scattering process, we define a pair of such operators, where creation and annihilation operators for the left- and right-moving IN modes commute, as well as for the OUT modes. As for the wavepackets, IN and OUT creation/annihilation operators associated to wavepackets of fixed central frequency $\omega$ are related by a linear combination which can generally be written as
\begin{equation}
    \hat{\mathbf{A}}_{\rm out}=\mathbf{S}\cdot \hat{\mathbf{A}}_{\rm in}\, ,
    \label{eq:TransfCreationOpsScatt}
\end{equation} 
where  
\begin{equation}
    \label{vecA}
    \hat{\mathbf{A}}_{\rm out}=\left(\begin{array}{cc} \hat{a}^{\rm out}_{r} \cr \hat{a}^{\rm out}_{l} \cr \hat{a}^{\rm out\, \dagger}_{r}\cr \hat{a}^{\rm out\, \dagger}_{l} \end{array}\right)
    \qquad\text{and}\qquad
    \hat{\mathbf{A}}_{\rm in}=\left(\begin{array}{cc} \hat{a}^{\rm in}_{r} \cr \hat{a}^{\rm in}_{l} \cr \hat{a}^{\rm in\, \dagger}_{r}\cr \hat{a}^{\rm in\, \dagger}_{l} \end{array}\right)\, ,
\end{equation} 
and $\mathbf{S}$ is a complex $4\times 4$ matrix which is fully determined by the classical scattering coefficients ---i.e. the elements of $\bs{B}$. 

To obtain $\mathbf{S}$, we use the definition of creation and annihilation operators, as well as the transformation of IN modes into OUT modes provided by $\bs{B}$ in \eqref{eq:GenScattProcess}. In the process, we must distinguish superradiant from non-superradiant modes, since the sign of the norm of positive- and negative-frequency modes (used to build the corresponding wavepackets) differs in each case. Therefore, one must be careful in appropriately assessing the norms of the IN and OUT basis elements during this computation. Let us detail this computation. Take for instance the first row of \eqref{eq:TransfCreationOpsScatt}, which reads
\begin{equation}
    \hat{a}_r^{\rm out}=S_{11}\hat{a}_r^{\rm in}+S_{12}\hat{a}_l^{\rm in}+S_{13}\hat{a}_r^{\rm in}{}^\dagger+S_{14}\hat{a}_l^{\rm in}{}^\dagger\,.
    \label{eq:OperatorModeRelationExample}
\end{equation}
The relation between IN and OUT wavepackets \eqref{eq:GenScattProcess} allows to write IN creation/anihilation operators in terms of the OUT by using their definition in terms of the symplectic product \eqref{eq:DefLadderOp}. However, in order to find $S_{ij}$ from \eqref{eq:OperatorModeRelationExample}, we need to write OUT operators in terms of IN operators. The simplest way to do so is by inverting the relation between IN and OUT modes, which yields
\begin{equation}
(W_{r}^{\rm in},W_{l}^{\rm in})\cdot \bs{B}^{-1}
\xrightarrow{\text{time}}
(W_{r}^{\rm out},W_{l}^{\rm out})\,.
\end{equation}
Taking the symplectic product with the Klein-Gordon field at early and late times, and exploiting its conservation, one finds the sought expression for OUT operators in terms of IN operators, which allows to read out the $S_{ij}$ components. Here it becomes necessary to distinguish between superradiant and non-superradiant cases.

\noindent\underline{Non-superradiant scattering:}\\
$B_{\rm NSR}$ is unitary, so we can write $\bs{B}_{\rm NSR}^{-1}=\bs{B}_{\rm NSR}^\dagger$. Hence we find the following expressions for OUT modes in terms of IN modes
\begin{equation}
W_{r}^{\rm out}\xrightarrow{\text{time}}T^*W_{r}^{\rm in}+r^*W_{l}^{\rm in}
\qquad\text{and}\qquad
W_{l}^{\rm out}\xrightarrow{\text{time}}R^*W_{r}^{\rm in}+t^*W_{l}^{\rm in}
\end{equation}
Taking the symplectic product with $\hat\Phi$ and using its properties we find
\begin{equation}
\begin{split}
    &\hat{a}_r^{\rm out}=(W_r^{\rm out},\hat\Phi)=T\, \hat{a}_r^{\rm in} + r\, \hat{a}_l^{\rm in}\\
    &\hat{a}_l^{\rm out}=(W_l^{\rm out},\hat\Phi)=R\, \hat{a}_r^{\rm in} + t\, \hat{a}_l^{\rm in}\\
    &\hat{a}_r^{\rm out}{}{}^\dagger=-(W_r^{\rm out}{}^*,\hat\Phi)=T^* \hat{a}_r^{\rm in}{}^\dagger + r^* \hat{a}_l^{\rm in}{}^\dagger\\
    &\hat{a}_l^{\rm out}{}^\dagger=-(W_l^{\rm out}{}^*,\hat\Phi)=R^*\hat{a}_r^{\rm in}{}^\dagger + t^* \hat{a}_l^{\rm in}{}^\dagger
\end{split}
\end{equation}
which leads to
\begin{equation}
    \mathbf{S}_{\rm NSR}= \begin{pmatrix}
T & r & 0 & 0 \\
R & t & 0 &0 \\
0 & 0 &T^* & r^* \\
0 & 0 &R^*& t^*   
\end{pmatrix}\,,
 \label{eq:NSSmatrix}
\end{equation}
\noindent\underline{Superradiant scattering:}\\
In this case we have $\bs{B}_{\rm SR}^{-1}\neq\bs{B}_{\rm SR}^\dagger$. However, using the constraints satisfied by the coefficients $T,R,t,$ and $r$ for superradiant scattering\footnote{They are found by taking all possible products of IN modes, which are $\pm1$ or $0$ due to orthonormality, and expressing them in terms of OUT modes and the scattering coefficients.}, we find
\begin{equation}
    \bs{B}_{\rm SR}^{-1}=\pm
    \begin{pmatrix}
        -T^* & R^*\\
        r^* & -t^*
    \end{pmatrix}\,,
 \label{eq:InverseBSR}
\end{equation}
where we have used $\text{Arg}[rR]=\text{Arg}[tT]$ and $\text{det}(\bs{B})=tT-rR=\ee^{\ci\text{Arg}[\pm r^*R^*]}$. The $\pm$ signs distinguish the two different cases regarding the signs of the norms of the modes involved: case 1) the modes in each spatial asymptotic region have the same sign of the norm, so the coefficients satisfy $|R|^2-|T|^2=1$, and $\bs{B}_{\rm SR}^{-1}$ goes with the $+$ sign. case 2) there is a mode with each sign of the norm in each of the asymptotic spatial regions, so the coefficients satisfy $|T|^2-|R|^2=1$ and $\bs{B}_{\rm SR}^{-1}$ goes with the $-$ sign. We will assume case 1), since it is typical to encounter this setup in systems displaying superradiance, while case 2) looks more unnatural from a physical perspective. However, the process that we will outline leads to a symplectic matrix that describes a norm-mixing scattering process in both cases. In the case of both modes in each asymptotic region having the same sign of the norm we find
\begin{equation}
\begin{split}
&W_{r}^{\rm out}= - T^* W_{r}^{\rm in} + r^*\, W_{l}^{\rm in}
\\
&W_{l}^{\rm out}= R^*\,W_{r}^{\rm in} - t^*\,W_{l}^{\rm in}\,.
\end{split}
\end{equation} 
We still can choose which region involves positive/negative norm modes for a given positve frequency. This choice is a matter of convention, so we will choose the asymptotic region $x\to-\infty$ as the one carrying the negative norm modes. This corresponds to $N_{W_r^{\rm in}}=N_{W_l^{\rm out}}=-1$ and $N_{W_l^{\rm in}}=N_{W_r^{\rm out}}=1$. Proceeding as above, in this case, we find the relations
\begin{equation}
\begin{split}
    &\hat{a}_r^{\rm out}=(W_r^{\rm out},\hat\Phi)=T\, \hat{a}_r^{\rm in}{}^\dagger + r \, \hat{a}_l^{\rm in}\\
    &\hat{a}_l^{\rm out}=(W_l^{\rm out}{}^*,\hat\Phi)=R^* \hat{a}_r^{\rm in}{} + t^* \hat{a}_l^{\rm in}{}^\dagger\\
    &\hat{a}_r^{\rm out}{}{}^\dagger=-(W_r^{\rm out}{}^*,\hat\Phi)=T^*\, \hat{a}_r^{\rm in}{} + r^*\, \hat{a}_l^{\rm in}{}^\dagger\\
    &\hat{a}_l^{\rm out}{}^\dagger=-(W_l^{\rm out}{},\hat\Phi)=R\,\hat{a}_r^{\rm in}{}^\dagger + t\, \hat{a}_l^{\rm in}
\end{split}
\label{eq:SRINOUTOpRel}
\end{equation}
which leads to
\begin{equation}
    \mathbf{S}_{\rm SR}= \begin{pmatrix}
0 & r & T & 0 \\
R^* & 0 & 0 & t^* \\
T^* & 0 & 0 & r^* \\
0 & t & R & 0   
\end{pmatrix}\,,
 \label{eq:GeneralSuperradiantMatrix}
\end{equation}
where $T,R,t,$ and $r$ are the elements of $\mathbf{B}$ as defined in \eqref{eq:GenScattProcess}. Again, we emphasize that this particular form of the entries of the matrix $\bs{S}$ depends on the sign of the norms in each asymptotic region (in particular, it is only valid for the case when the modes at $x\to-\infty$ have negative norm).

It is worth noting that $\mathbf{S}_{\rm NSR}$ is a unitary matrix, while $\mathbf{S}_{\rm SR}$ is not. Indeed, one can reformulate the theorem characterizing superradiance in terms of the matrix $\bs{S}$, having that a stationary quantum scattering process is superradiant if and only if the matrix $\bs{S}$ relating IN to OUT creation and anihilation operators is not a unitary matrix. 

From a physical perspective, $\mathbf{S}_{\rm NSR}$ and $\mathbf{S}_{\rm SR}$ describe qualitatively different processes. The unitary nature of $\mathbf{S}_{\rm NSR}$ has physical implications: it preserves the vacuum state and the total number of quanta, conserving energy. $\bs{S}_{\rm NSR}$ is the matrix representation of a beam-splitter operator. In contrast, $\mathbf{S}_{\rm SR}$ does not preserve either the vacuum state or the total number of particles, but it leaves the number difference unchanged, namely $\hat{N}_r^{\rm out}-\hat{N}_l^{\rm out}=\hat{N}_r^{\rm in}-\hat{N}_l^{\rm in}$ implying the creation of excitations in pairs. $\mathbf{S}_{\rm SR}$ is the matrix representation of a two-mode squeezing operator. This implies that superradiance is able to spontaneously create particles out of the vacuum, and to generate entnaglement between the OUT quanta provided the IN state is sufficiently pure.

\begin{figure}
    \center
    \includegraphics[width=0.5\textwidth]{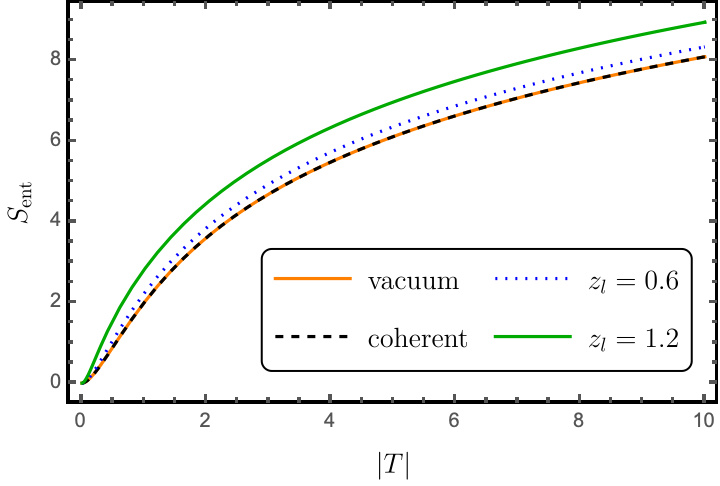}
    \caption{Entanglement entropy of the OUT modes produced in superradiant scattering for different input states. $z_l$ lines correspond to a single-mode squeezed state for $W^{\rm in}_l$ of the in modes with squeezing intensity $z_{l}$ and vacuum for $W^{\rm in}_l$. Plot extracted from \cite{Delhom:2023gni}.}
    \label{fig:EntropyGeneralSR}
\end{figure}

With the symplectic matrix for superradiant scattering at hand, we can now show how, besides amplification, superradiance leads to particle creation as well as entanglement generation. To that end, we assume that the initial state is vacuum for the IN modes. Using the relations between IN and OUT creation and annihilation operators \eqref{eq:SRINOUTOpRel}, we find that the mean number of OUT quanta is given by
\begin{equation}
    \langle \hat{N}_r^{\rm out}\rangle=\langle \hat{N}_l^{\rm out}\rangle=|T|^2.
\end{equation}
This implies a non-trivial population of OUT quanta in the absence of IN quanta, i.e. spontaneous pair creation. We can also study quantum correlations among the created particles. To do so, since the state of the system is pure (vacuum states of quadratic Hamiltonians are pure and gaussian), we compute the entanglement entropy for the bipartition of the system corresponding to left and right moving wavepackets as subsystems. The result is (see \cite{IvanAnthony} to learn how to do this computation)
\begin{equation}
    S_{\rm ent}=(1+|T|^2) \log_2 \left(1+|T|^2\right)-|T|^2\log_2 |T|^2,
\end{equation}
which is positive and vanishes only if no quanta are created ($|T|=0$). This shows that besides amplification, superradiance also generates entanglement \cite{Delhom:2023gni}. Though we only discussed the case of vacuum input, these properties are robust, and not a specific result of vacuum initial states. Indeed, both the creation of quanta as well as entanglement can be amplified by appropriately choosing the initial state, as can be seen in ~\ref{fig:EntropyGeneralSR} (see \cite{Delhom:2023gni} for details). 

\subsubsection{A quick workaround to compute the S matrix}
Before finishing this section, we will outline a trick that allows to compute the matrix $\bs{S}$ directly from the transformation of modes which is valid for any of the above cases. The trick is noticing that the quantum field $\hat\Phi_\omega(t,\vx)$ is decomposed into a sum of $\omega$ sectors that do not mix with each other, and which roughly have the form $\hat{\Phi}_\omega(t,\vx)=\bs{W}\cdot\hat{\bs{A}}$ where $\hat{\bs{A}}$ is a column vector of creation/annihilation operators and $\bs{W}$ is a row vector of the space-time dependent mode functions\footnote{Because of stationarity, the time dependence in $\bs{W}$ is incorporated with the corresponding $\ee^{\pm \ci\bs{\omega}}$ factor for each mode} and their conjugates. For a fixed ordering of creation and annihilation/operators within the vector $\hat{\bs{A}}$, each component of the mode vector has to be a positive/negative norm mode if that component of the operator vector is an annihilation/creation operator respectively.

This decomposition of the quantum field into modes and associated creation/annihilation operators can either be done for IN or OUT modes. Recall that both IN and OUT modes are solutions of the field equations at all times. They are not, however, the same solutions: an IN mode will be localized in a given spatial region at early times, and propagating towards the observer, but it can have a complicated spatial support at late times. On the contrary, an OUT mode can have a complicated spatial support at early times, but will be localized in a given spatial region at late times, and propagating away from the observer. The scattering picture comes from decomposing IN modes at late times in terms of OUT modes, so that one can have a clear physical interpretation of the outcome of the scattering problem in terms of propagation of energy and other physical quantities. This decomposition is exactly what is described by Eq. \eqref{eq:GeneralModeEvolution}. Thus, nothing forbids to decompose the quantum field in terms of one or the other basis of solutions. Hence, we will have $\hat\Phi(t,\vx)=\bs{W}^{\rm in}\cdot\hat{\bs{A}}^{\rm in}=\bs{W}^{\rm out}\cdot\hat{\bs{A}}^{\rm out}$. Since we have $\bs{W}^{\rm in}=\bs{W}^{\rm out}\cdot\bs{S}$ we must have as well $\hat{\bs{A}}^{\rm out}=\bs{S}\cdot\hat{\bs{A}}^{\rm in}$. Hence, we can compute the matrix $\bs{S}$ relating IN and OUT operators in equation \eqref{eq:TransfCreationOpsScatt} from the transformation of the IN and OUT modes and their conjugates. This is a shortcut that avoids the computation of the inverse of the $\bs{B}$ matrix, which can be tedious as it has to be done in a case by case basis.

Let us see how the trick works in the two examples considered. Consider the operator vectors $\bs{A}^{i}=(\hat{a}^{i}_r,\hat{a}^{i}_l,\hat{a}^{i}_r{}^\dagger,\hat{a}_l^{i}{}^\dagger)^\top$, where $i=\text{in,out}$. For the non-superradiant scattering,  since all modes $W^{i}_r$ and $W^{i}_l$ are of positive norm, we build the mode vectors as $\bs{W}^i=(W^i_r,W^i_l,W^i_r{}^*,W^i_l{}^*)$. From the decomposition of IN modes in terms of OUT modes in \eqref{eq:GeneralModeEvolution}, and $\bs{W}^{\rm in}=\bs{W}^{\rm out}\cdot\bs{S_{\rm NSR}}$, we find again matrix \eqref{eq:NSSmatrix} as expected. Hence, the trick reproduces well the non-superradiant case. For the superradiant scattering, in the chosen case, we have that the positive norm IN modes are $W^{\rm in}_r{}^*$ and $W^{\rm in}_l$, with their conjugates $W^{\rm in}_r$, and $W^{\rm in}_l{}^*$ being of negative norm. Therefore, we build $\bs{W}^{\rm in}=(W^{\rm in}_r{}^*,W^{\rm in}_l,W^{\rm in}_r,W^{\rm in}_l{}^*)$. For the OUT modes, we have that $W^{\rm out}_r$ and $W^{\rm out}_l{}^*$ are the positive norm modes and their conjugates $W^{\rm out}_r{}^*$, and $W^{\rm out}_l$ are the negative norm ones. Hence we build $\bs{W}^{\rm out}=(W^{\rm out}_r,W^{\rm out}_l{}^*,W^{\rm out}_r{}^*,W^{\rm out}_l)$. Again, from the decomposition of IN modes in terms of OUT modes in \eqref{eq:GeneralModeEvolution}, and $\bs{W}^{\rm in}=\bs{W}^{\rm out}\cdot\bs{S_{\rm NSR}}$, we find matrix \eqref{eq:GeneralSuperradiantMatrix} reproducing the expected result for the superradiant case with the chosen norm structure. One can check that if another norm structure is chosen in the superradiant case, both procedures yield consistent results for the components of $\bs{S}$. This also generalizes to cases with more modes, provided that we order the operator vector and the mode vector in a consistent way, with creation operators corresponding to negative norm modes, and annihilation operators to positive norm ones.


\subsection{Superradiant instabilities}
\label{sec:sr-instabilities}

\begin{figure}[t]
    \center
\includegraphics[width=0.5\textwidth]{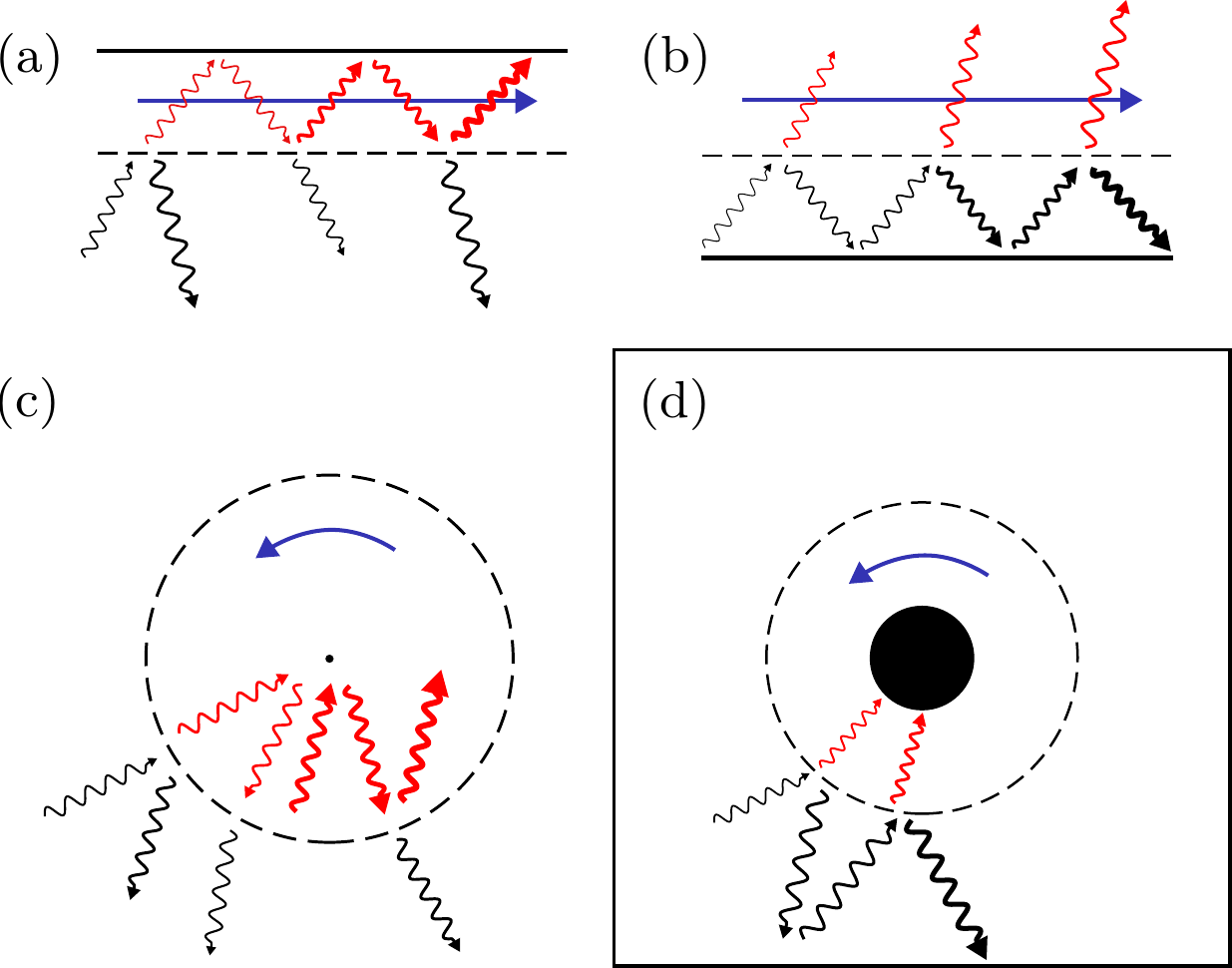}
    \caption{Pictorial representation of superradiant instabilities. (a) and (b) show two possible instability processes for a shear layer, when a reflecting boundary condition is introduced in the supersonic or in the subsonic region. (c) and (d) show the correspondent processes in a rotating black hole configuration. The trapping of the negative-energy waves give rise to an ergoregion instability, the trapping of the positive-energy ones gives rise to the o black hole bomb instability.}
    \label{fig:superradiant-instabilities}
\end{figure}

As we discussed, superradiant scattering relies on the existence of negative-energy modes, i.e. in BEC language on the system being energetically unstable. The presence of these modes can also give rise to dynamical instabilities, which have a much more dramatic effect on the system. Here we give a quick discussion and a numerical example applied to a shear layer in a BEC. For a complete account of superradiant instabilities see the review \cite{brito2020superradiance}, and for a more complete pedagogical account in analogues see \cite{giacomelli2021understanding,giacomelli2021superradiant}.

The amplified scattering we treated up to now supposes that ingoing waves (wavepackets) come from far away toward the ergosurface and outgoing waves are free to move away from it. In superradiant scattering from a rotating black hole the amplified positive-energy outgoing wave propagates away from the black hole, while the partner outgoing negative-energy wave falls into the black hole. That is, one has radiative/open boundary conditions on both sides of the ergosurface. If the boundary conditions of the problem change the outgoing waves can be fed back to the ergosurface, where they undergo superradiant scattering again. The result is that there exist trapped in a region undergoing repeated amplification. This means that the amplified scattering cannot occur as a stationary phenomenon. Instead, the amplitude of the mode exponentially increases, as can be understood intuitively with the following reasoning.

Consider the case in which a reflecting boundary condition is present in the supersonic side of the shear layer configuration we considered in Section \ref{sec:shear-layer}, as depicted in Fig.\ref{fig:superradiant-instabilities}(a). If the amplitude of the first transmitted negative-energy wavepacket is $\mathcal{A}_0$, when this reaches the interface again, it will undergo superradiant scattering, i.e. it will be reflected with an amplitude $\mathcal{A}_0\cdot R$, with $|R|>1$. After $N$ bounces, the trapped wave will have an amplitude $\mathcal{A}_0\cdot(R)^N$, i.e. it exponentially increases with the number of bounces. As we are going to explicitly see in a moment, this reflects with the presence of complex-frequency modes in the spectrum of excitations. The imaginary part of the frequency, determining the rate of exponential growth in time of the amplitude of the dynamically unstable mode, is roughly given by $\Im{\omega}\sim R*\Delta t$, where $\Delta t$ is the time the wave takes to travel from the interface to the boundary condition and back. An analogous mechanism occurs if the reflecting boundary condition is in the subsonic region, resulting in a trapping of the positive-energy waves, while the negative-energy ones are free to propagate away, as represented in Fig.\ref{fig:superradiant-instabilities}(b).

Going back to a rotating flow, which is analogous to rotating spacetimes in general relativity, these two configurations correspond to two different instabilities. The situation in panel (a) corresponds to a spacetime in which there is an ergoregion but not a horizon (e.g. in a ergostar), and the corresponding instability is called an \textit{ergoregion instability}, as depicted in panel (c). The situation in panel (b) is instead parallel to the case in which negative-energy waves fall in the horizon and are not backscattered, but the positive-energy waves are sent back to the black hole to be amplified again, for example by a mirror outside a black hole\footnote{A physically relevant \textit{mirror} can occur for massive fields, that have a mass gap that forbids the propagation of excitations away from the black hole.}. In this case the instability is known as \textit{black hole bomb}.

In a BEC this exponential growth can continue while the amplitude of the excitations is small enough to justify the linearization at the basis of the Bogoliubov approach. When the the amplitude grows out of this regime, the excitations start to behave nonlinearly and to \textit{back react} on the stationary state, driving in away from the starting configuration. In more field theoretic words, this tells us that the assumed stationary background is unstable due to a mechanism similar to a tachyonic instability \cite{Delhom:2022vae}. To give a specific example of this process, multiply quantized vortices in a BEC display ergoregion instabilities, whose eventual effect is to split the vortex in smaller ones \cite{Giacomelli:2019tvr}. This is analogous to superradiant instabilities determining modifications of the spacetime they develop on, e.g. by reducing the angular momentum of a rotating body \cite{brito2020superradiance}.

Given the parallelism between superradiant scattering and the bosonic Klein paradox we discussed in Section \ref{sec:kleinparadox}, superradiant instabilities can occur also for a charged relativistic field, specifically when there is an electrostatic potential \textit{box}. In this case the resulting dynamical instability is known as Schiff-Snyder-Weinberg effect \cite{schiff1940existence,fulling1989aspects}.

These instabilities are also very similar to a kind of instability that was widely studied in analogue gravity: the black hole lasing instability. As we are going to see in section \ref{sec:black-hole-lasing}, differently from superradiant instabilities, black hole lasing needs a supersonic dispersion relation to occur.

From the point of view of the physics of BECs, the mechanism of superradiant instabilities can play a role in several situations with supersonic flow, like the splitting of multiply quantized vortices \cite{giacomelli2021spontaneous}, vortex nucleation around a strong defect, and interplay with other hydrodynamical instabilities, like the Kelvin-Helmholtz one \cite{giacomelli2023interplay}.

\begin{figure}[t]
    \center
\includegraphics[width=0.5\textwidth]{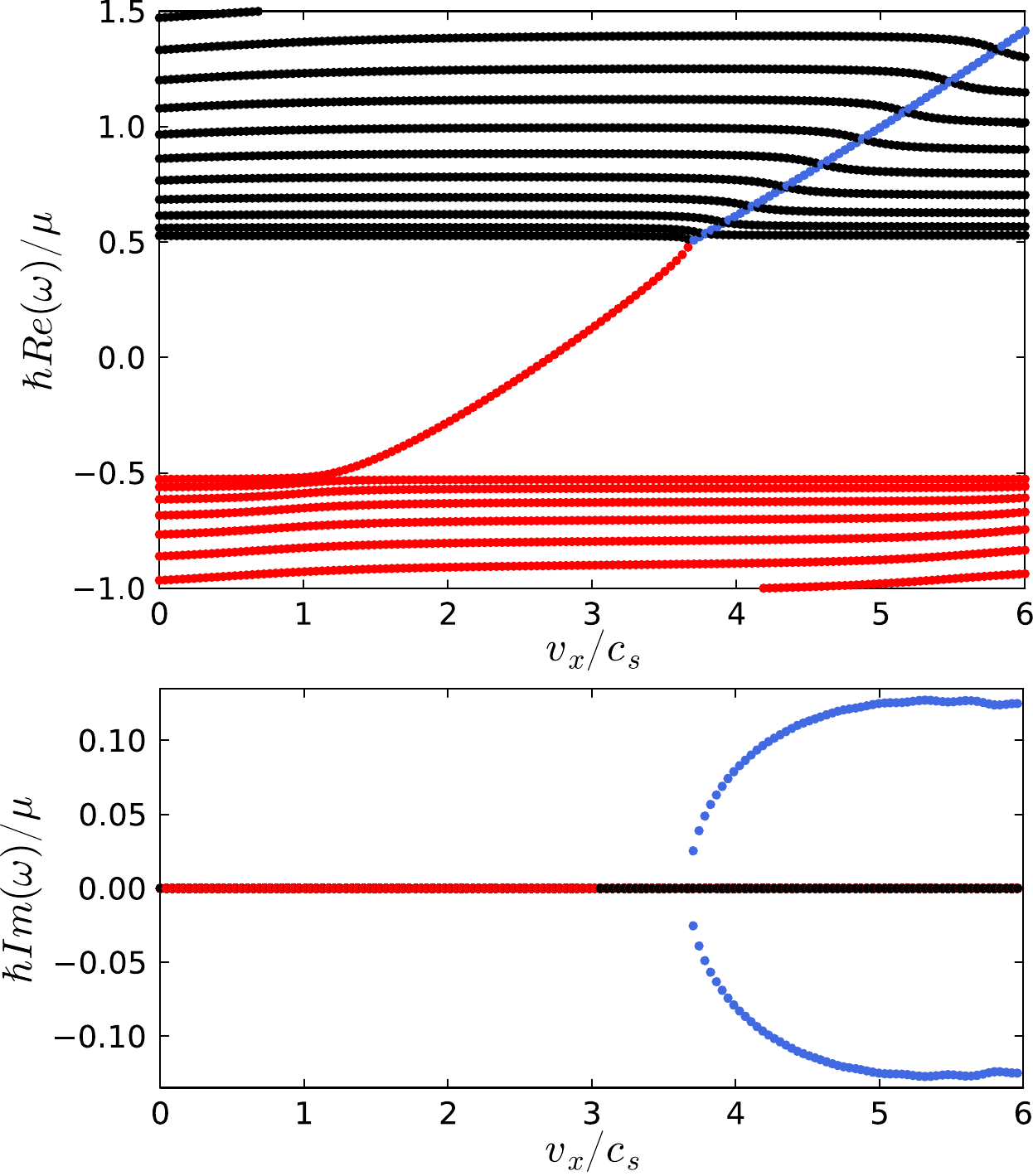}
    \caption{Real and imaginary parts of the eigenfrequencies of the Bogoliubov problem for a shear layer configuration, for a fixed transverse momentum of the excitations $k_x$. Different values of the velocity of the moving region are shown (horizontal axis). Black (red) points are positive (negative) norm modes, blue points are modes of zero norm. Here we took a system of length $L_y=30\xi$ in the direction transverse to the flow. The condensate is moving along $x$ with a velocity $\tv_x=-A_x/m$ in a region $y\in[L_y-2\xi,L_y]$. Dirichlet boundary conditions are imposed at the edges. on the Bogoliubov problem. The plot refers to the solution for a fixed transverse momentum $k_x=0.5\xi^{-1}$.}
    \label{fig:SSW-spectrum}
\end{figure}

\subsubsection{Detecting dynamical instabilities with the numerical diagonalization of the Bogoliubov problem}
\label{sec:bogo-diag-SR}

Let us present in detail a concrete example of the occurrence of a dynamical instability. To do that, we will analyze numerically the superradiant instabilities of the shear layer we just discussed. A more complete discussion of the physics of this instability can be found in \cite{giacomelli2021understanding,giacomelli2021superradiant}. This serves also as an example of how to numerically find the eigenmodes of the Bogoliubov problem and study their stability, that can be useful also in more complicated setups. Writing the code that diagonalizes the Bogoliubov problem is the content of Problem \ref{prob:BogoShearLayer}, whose solution is attached to these lecture notes as a Jupyter notebook written in Julia \cite{Julia-2017}.

We want to obtain the spectrum of linear excitations on top of the stationary state given by the shear layer. As we discussed in section \ref{sec:shear-layer-BEC}, this can be obtained by introducing a synthetic gauge field. We consider the GPE equation \eqref{eq:gpe-gaugefield} with $\bs{A}(x,y)=A_x\theta(y)\hat{\bs{x}}$. To have stationary states with a constant density, we also introduce an external potential $V_\mathrm{ext}(x,y)=-\theta(y)A_x^2/(2m)$. In particular, the constant-density real wavefunction $\Psi_0(x,y,t)=\ee^{-i\mu t}\sqrt{n_0}$ is a stationary solution, corresponding to the shear layer in which the lower part of the system is not moving, while the upper part has a velocity $\tv_x=-A_x/m$ fixed by the synthetic gauge field.

Considering fluctuations around this steady state of the form \eqref{eq:DefPertGPE}, we can take advantage of the translational invariance along $x$ to decompose the fluctuation field in plane waves along that direction., i.e. $\psi(x,y,t)=\ee^{i\tk_x x}\chi(y,t)$. The resulting Bogoliubov problem is
\begin{equation}\label{eq:bogo-problem-gauge}
	i\hbar\partial_t 
	\begin{pmatrix}
		\chi\\\chi^*
	\end{pmatrix}
	=
	\begin{bmatrix}
		D_+ & gn_0\\
		-gn_0 & -D_-
	\end{bmatrix}
	\begin{pmatrix}
		\chi\\\chi^*
	\end{pmatrix},
\end{equation}
where
\begin{equation}
	D_\pm=-\frac{\hbar^2}{2m}\partial_y^2+\frac{(\pm \tk_x-A_x)^2}{2m}+2gn_0+V_\mathrm{ext}-\mu\,.
\end{equation}
We can hence solve the original two-dimensional problem as a set of one-dimensional ones at fixed transverse momentum $\tk_x$. This is not dissimilar to considering separate problems with different angular momenta in a rotating configuration.

The numerical representation of this problem can be made on a discrete grid of positions, with spatial discretization $\Delta y$. If this grid has $N$ points, the Bogoliubov two-component wavefunction can then be represented as a $2N$ dimensional vector $(u_1,\dots,u_N,v_1,\dots,v_N)^T$, and the Bogoliubov matrix will hence be $2N\times 2N$. The off-diagonal $N\times N$ blocks will simply be diagonal (in position), while the diagonal blocks will have a diagonal part given and an off diagonal terms given by the Laplacian. The simplest finite difference approximation one can take for the second derivative is\footnote{For arbitrary precision on arbitrary grids refer to \cite{fornberg1988generation}.} 
\begin{equation}
    \partial_y u(y)= \frac{u(y+\Delta y)-2u(y)+u(y-\Delta y)}{\Delta y^2}+o(\Delta y^2).
\end{equation}
The diagonal blocks of the Bogoliubov matrix will hence be tri-diagonal matrices. The way one treats the points on the border of the grid determines the boundary conditions one imposes. For example, here we will take non left (right) neighboring point for the first (last) point, and this results in Dirichlet reflecting boundary conditions, but periodic boundary conditions can easily be implemented. This matrix can then be diagonalized numerically to obtain the eigenfrequencies and eigenmodes. One can compute the Bogoliubov norm of the eigenmodes $\sum_{j=1}^N |u_j|^2-|v_j|^2$.

In Figure \ref{fig:SSW-spectrum} we show the real and imaginary parts of the eigenfrequencies for different values of the velocity of the moving region and a fixed value of the transverse momentum (parameters in the caption). Each point on the horizontal axis is the result of a different diagonalization. The color of the points indicate the norm of the eigenmode. One can see that, for large enough values of the transverse velocity, the system starts to be energetically unstable (negative norm modes at positive frequencies), and for even higher velocities it becomes dynamically unstable, as can be seen by the vanishing of the norm (blue points) and by the nonzero imaginary part of the frequency. 

The dynamical instability emerges when the energetically unstable mode becomes resonant with a positive energy one, and the two merge in a zero norm branch. This is a general property of pseudo-Hermitian problems, that for eigenvectors of opposite norm do not display the familiar avoided crossing of quantum mechanics, but gives rise to this \textit{level sticking}. The corresponding dynamical instabilities correspond to the continuous production of the two original modes with equal and opposite energies. For more details on this general property see e.g. \cite{giacomelli2021superradiant}.

\section{The stationary analogue Hawking effect}
\label{sec:Hawking}

In the previous section we discussed superradiance and saw that it occurs around ergosurfaces, giving rise to pair creation and entanglement generation at the quantum level. The realization of this phenomenon is conceptually straightforward, relying only on the possibility of mixing of positive and negative energy modes at fixed frequency. Dispersive effects play here only a secondary role: they changing the region of parameter space in which superradiant scattering can occur, but they are not needed for the effect to happen. For instance, in rotating setups, superluminal dispersion limits the maximum angular momentum for which amplified scattering is possible \cite{giacomelli2021understanding,patrick_rotational_2020,Giacomelli:2019tvr}.

The situation is less trivial for the most sought-after phenomenon in analogue gravity (and also the first one to be investigated \cite{Unruh:1980cg}): the Hawking effect. Even though its stationary analogue is very similar to spontaneous superradiance, and can also be investigated in simple, effectively one-dimensional spacetimes, we are going to see that its occurrence heavily relies on dispersive effects. 

We are going to discuss the analogue Hawking effect in stationary BECs, that is, with a supersonic (superluminal) dispersion relation. For subsonic dispersion relations the qualitative description is different, but the effect also occurs, and the procedure to derive it outlined below will apply similarly \cite{robertson2012theory}.

Below, we will first describe the stationary analogue of the Hawking effect as a stationary scattering problem, and we will then compare it to the astrophysical case, which is conceptually different due to its occurrence in the dynamical scenario of a collapsing gravitational object.

\subsection{The analogue Hawking effect as a stationary scattering problem}

For simplicity, we will focus on one-dimensional systems, but this approach can be generalized to higher dimensional systems. An effectively one-dimensional BEC, that can be approximately described with a GPE with one spatial dimension, can be obtained with what is known as a \textit{cigar-shaped} configuration \cite{pitaevskii2016bose}, i.e. using an elongated trap that is tightly confining in the transverse direction, leading to an approximate cylindrical shape. In this configuration, an analogue black hole can be obtained by moving a step potential through the system, thus creating a \textit{waterfall} along which the condensate is accelerated from subsonic to supersonic speed. This is the strategy that was followed in a series of experiments \cite{steinhauer2016observation,munoz2019observation,kolobov2021observation}, although other configurations have been proposed (see \cite{Almeida:2022otk} for a historical account).

\begin{figure}[t]
    \center
\includegraphics[width=0.4\textwidth]{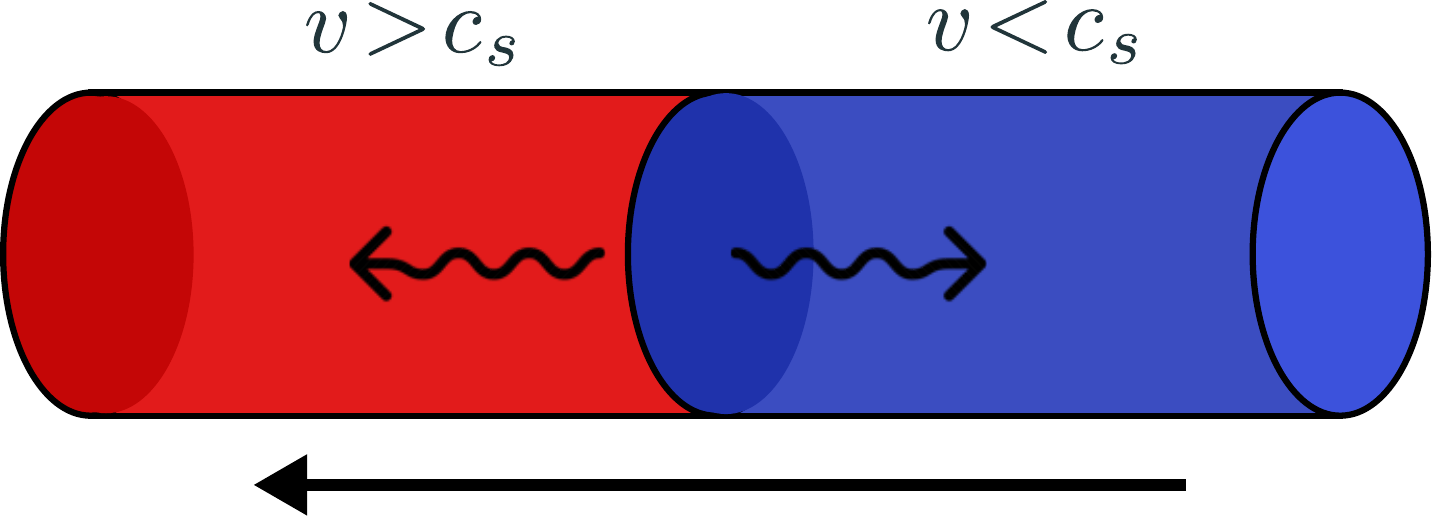}
    \caption{Pictorial representation of the one-dimensional BEC black hole, consisting of a subsonic and a supersonic region. The quantitative physics will depend on the specific implementation, but the general structure of the effect does not. The Hawking effect can be expected as the spontaneous production of pairs at the analogue horizon.}
    \label{fig:hawking-toy-model}
\end{figure}

Here, we will not focus on the details of the experimental realizations. We will instead work with a \textit{toy model} that is described by the one-dimensional GPE and consists of two asymptotic regions that are respectively subsonic and supersonic, as depicted in Figure \ref{fig:hawking-toy-model}. We will first give a conceptual explanation, independent on the details of the realization, of how the stationary analogue of the Hawking effect arises. After that, we will discuss the technical aspects of the derivation and provide a concrete example.

\subsubsection{Mode structure of the transsonic condensate}
Perturbations to the condensate are described by the BdG equations, as explained in section \ref{sec:linearization-gpe}.
If the spatial variation of the condensate takes place over sufficiently large lengths, collective excitations can be well understood from their local dispersion relation, as we already did in the study of superradiant scattering. For a 1D condensate flowing at constant velocity $\vtv=\tv \hat{\vx}$, the Doppler-shifted Bogoliubov dispersion relation \eqref{eq:doppler-shift} is of the form  
\begin{equation}
    \omega=\tv\tk\pm\cs|\tk|\sqrt{1+\frac{\xi^2}{4}\tk^2},
    \label{eq:DispersionHawking1D}
\end{equation}
where now $\vk=\tk \hat{\vx}$, so that $\tk$ and $\tv$ take negative values when $\vk$ and $\vtv$ point respectively towards the left. At a fixed frequency, this equation has 4 (generally complex) solutions $\tk(\omega)$. Real solutions describe modes that propagate energy and momentum (propagating modes) in the direction of their group velocity $\vtv_{\rm g}=\partial_{\tk}\omega$, while complex solutions either describe evanescent waves or are unphysical due to exponential growth in spatial directions, depending on the boundary conditions. 

\begin{figure}[t]
    \center
    \includegraphics[width=0.7\textwidth]{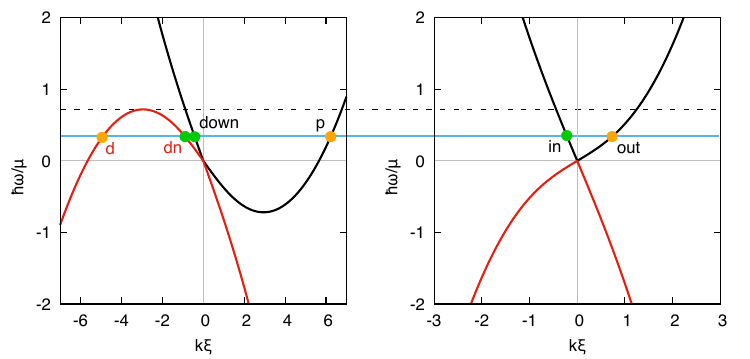}
    \caption{Dispersion relation in the asymptotic regions inside (left) and outside (right) of the black hole for transsonic 1D flow moving towards the left. The curved lines are the positive (blue curve) and negative (red curve) branches of the dispersion \eqref{eq:DispersionHawking1D}, plotted as a function of $\tk_x$. The horizontal lines represent constant values of the frequency. The dashed line is the threshold frequency $\omega_{\rm max}$, obove which there is no stationary analogue Hawking effect. We have that the flow velocity in the left asymptotic region satisfies $v_l>\cs$, so that there are 4 propagating modes for $\omega<\omega_{\rm max}$.}
    \label{fig:DispersionHawking1D}
\end{figure}

Let us provide a qualitative analysis of the solutions of the above dispersion relation (see Fig~\ref{fig:DispersionHawking1D}). The different regimes are essentially controlled by the flow velocity and the non-linearities in the dispersion. In subsonic regions $\tv<\cs$, two solutions are real and two are complex, with the two real solutions propagating in opposite directions. On the other hand, supersonic regions $\tv>\cs$ display a richer structure. This is seen in Fig.~\ref{fig:DispersionHawking1D}, where black/red curves correspond to the $+/-$ sign in $\eqref{eq:DispersionHawking1D}$ respectively. Propagating modes in each of these branches have respectively positive and negative symplectic norm, as given by the symplectic or BdG product \eqref{eq:BdGInnerProd}. The effect of increasing the velocity from sub- to supersonic is to deform both branches so that, for $|\tv|>\cs$ they develop a critical point at $\tk=\tk_{\rm max}$. This critical point is a maximum with value $\omax$ for the negative-norm branch and a minimum with value $-\omax$ for the negative-norm branch. Thus, in supersonic regions, the mode structure depends crucially on whether the considered frequency is above or below the threshold $\omax$ (see Problem \ref{prob:wmax}). 

For $\omega>\omax$ there are two real and two complex solutions, completely parallel to the subsonic case. On the other hand, for $0\leq\omega<\omax$ there are four real solutions. Two of these solutions propagate in the direction of the flow, while two other solutions propagate against the flow. These counter-propagating solutions only exist because of the non-linearities in the dispersion relation. In fact, one can check that they disappear in the limit $\xi\to0$, where the standard relativistic dispersion is recovered (see Problem \ref{prob:wmax}). All this can be easily visualized in Fig.~\ref{fig:DispersionHawking1D}, where the intersections with a horizontal line are real solutions (propagating modes) of a given laboratory frequency, and their group velocity is the slope of the curve at the intersection points. 

The mode structure shown in Fig.~\ref{fig:DispersionHawking1D} allows a qualitative analysis of the stationary analogue of the Hawking effect. The two panels are the dispersion relations in the two asymptotic regions of the one-dimensional analogue black hole of Fig.~\ref{fig:hawking-toy-model} (assuming that far enough from the interface the velocity profile is homogeneous enough). In particular, the condensate moves to the left and is subsonic for $x>0$ and supersonic for $x<0$. In the subsonic region, for all positive frequencies, two propagating modes exist, both with a positive norm (and hence positive energy). In the supersonic region, the same is true for $\omega>\omax$, but four propagating modes exist below this frequency: two of positive energy and two of negative energy. 

In this second case, for each frequency $\omega<\omax$, each branch of the dispersion hosts one mode propagating in each direction. While the low $|\tk|$ modes both propagate towards the left, away from the scattering region, the two solutions with higher values of $|\tk|$ propagate towards the right, that is against the supersonic flow. As we mentioned, these solutions only exist due to the higher-order terms in the dispersion relation disappear in the hydrodynamic limit $\xi\tk\to0$, for which the solutions of the dispersion relation are always two. 

Analogously to what we did for superradiant scattering, here we describe stationary scattering on a subsonic-to-supersonic transition. Analogously to spontaneous superradiance, pair creation will come from norm mixing. As explained in section \ref{sec:SR}, stationary scattering processes conserve frequency, and hence norm mixing can only occur if there are both negative and positive norm modes at the same frequency. To describe a scattering problem, one usually considers two basis of solutions to the wave equations: one of modes propagating towards the scattering region, known as IN basis, and another one of modes moving away from the scattering region, known as OUT basis. Since both are bases of solutions, any scattering solution (or global mode) is a linear combination of modes of either basis.

In the scattering picture, initial conditions are naturally expressed in terms of IN modes, and one is usually interested in the properties of the scattered wave when decomposed in terms OUT modes. In this transsonic BEC configuration, at $\omega<\omax$, we have three IN modes, labeled\footnote{The labels come from the analogy with the gravitational Hawking effect (as explained later). A nomenclature sometimes used in the literature is the one introduced in \cite{recati2009bogoliubov},  in which ingoing modes are called respectively $u-in,d1-in,d2-in$ and outgoing ones $u-out,d1-out,d2-out$.} as $(W_{\rm in},W_{\rm p},W_{\rm d})$, and three OUT modes, labeled by $(W_{\rm out},W_{\rm down},W_{\rm dn})$. One of the IN and one of the OUT modes have negative norm at positive frequencies, while the rest have positive norm. This implies that positive and negative norm IN modes can mix leading to creation of entangled excitations in the OUT modes. In other words, the vacuum of the IN modes is not a vacuum for OUT excitations, just as in the case of superradiant scattering we already considered.

In fact, this mode structure implies the occurrence of amplified scattering also in this one-dimensional analogue black hole, as can be inferred solely from the dispersion curves. Consider a (negative-energy) wave in the $d$ ingoing mode. After the scattering, the three outgoing modes will be polulated: the negative-energy $dn$, and the positive-energy $down$ and $out$. Since energy must be conserved, if any amplitude ends up in the last two modes, the reflected negative-energy wave will have a greater amplitude than the ingoing one.

Whether this amplification occurs, and correspondingly any quantum pair production, depends on the quantitative details of the transsonic flow. These are encoded in the coefficients of the scattering matrix, which allows to express any of the OUT modes as a linear combination of IN modes. In particular, if any of the coefficients relating positive- and negative-energy modes is nonzero, the scattering matrix will be non-unitary, leading to superradiant scattering (in the sense of stationary scattering displaying amplification) and quantum pair production. 

Notice that the relevant physics only occurs for $\omega<\omax$. In fact, for larger frequencies the negative-energy modes $W_{\rm d}$ and $W_{\rm dn}$ are no longer available, and there are only two positive-norm propagating solutions in the supersonic region. The scattering problem is hence analogous to the one between two subsonic regions and no norm mixing can occur.

In the following sections we first explicitly characterize the scattering matrix, and we then provide an explicit solution of the scattering coefficients for a simple step-like transsonic flow.

\subsubsection{Scattering matrix of the analogue black hole}

As anticipated, here we explicitly build the scattering matrix for a stationary transsonic configuration such as the one in Fig.~\ref{fig:hawking-toy-model}, and we show that it is non-unitary for $\omega<\omax$, and unitary for $\omega>\omax$. Theorem \ref{thm:CharacterizationSR} then guarantees that there is superradiant scattering for $\omega<\omax$, which leads to pair production and entanglement.

We will proceed here as in Section \ref{sec:QuantumSR}, where we treated quantum superradiance. Also here, since the flow is stationary, different $\omega$ sectors will not mix, and the full quantum $S$-matrix will be block diagonal in frequency eigenspaces, with blocks $\bs{S}_\omega$. For simplicity, we will refer to these single blocks just as the quantum $\bs{S}$-matrix. To obtain the matrix we again expand every $\omega$ sector of our quantum field in the IN and OUT mode basis as $\hat\phi(t,\vx)=\bs{W}^{\rm in}\cdot\hat{\bs{A}}^{\rm in}=\bs{W}^{\rm out}\cdot\hat{\bs{A}}^{\rm out}$. Following the procedure detailed in section \ref{sec:QuantumDensPhase}, we Kkeping into account the norm of the modes in ordering the modes vectors in $\bs{W}^{\rm in,out}$, so that positive norm modes occupy the first half of the entries of $\bs{W}$ and the corresponding negative-norm modes occupy the last half. In this way, the operator vectors $\hat{\bs{A}}$ contain annihilation operators in the fist half of the entries and the corresponding creation operators in the last half. Thus, for $\omega<\omax$ we write
\begin{equation}
\begin{split}
\bs{W}^{\rm in}_<=(W_{\rm{in}},W_{\rm{p}},W_{\rm{d}}^*,W_{\rm{in}}^*,W_{\rm{p}}^*,W_{\rm{d}})
&\qquad\text{and}\qquad
\bs{A}^{\rm in}_<=(\hat{a}_{\rm{in}},\hat{a}_{\rm{p}},\hat{a}_{\rm{d}},\hat{a}_{\rm{in}}^\dagger,\hat{a}_{\rm{p}}^\dagger,\hat{a}_{\rm{d}}^\dagger)^\top\,,
\\
\bs{W}^{\rm out}_<=(W_{\rm{out}},W_{\rm{down}},W_{\rm{dn}}^*,W_{\rm{out}}^*,W_{\rm{down}}^*,W_{\rm{dn}})
&\qquad\text{and}\qquad
\bs{A}^{\rm out}_<=(\hat{a}_{\rm{out}},\hat{a}_{\rm{down}},\hat{a}_{\rm{dn}},\hat{a}_{\rm{out}}^\dagger,\hat{a}_{\rm{down}}^\dagger,\hat{a}_{\rm{dn}}^\dagger)^\top\, ,
\end{split}
\label{eq:ModeExpInsideBH}
\end{equation}
while for $\omega>\omax$
\begin{equation}
\begin{split}
\bs{W}^{\rm in}_>=(W_{\rm{in}},W_{\rm{p}},W_{\rm{in}}^*,W_{\rm{p}}^*)
&\qquad\text{and}\qquad
\bs{A}^{\rm in}_>=(\hat{a}_{\rm{in}},\hat{a}_{\rm{p}},\hat{a}_{\rm{in}}^\dagger,\hat{a}_{\rm{p}}^\dagger,)^\top
\\
\bs{W}^{\rm out}_>=(W_{\rm{out}},W_{\rm{down}},W_{\rm{out}}^*,W_{\rm{down}}^*)
&\qquad\text{and}\qquad
\bs{A}^{\rm out}_>=(\hat{a}_{\rm{out}},\hat{a}_{\rm{down}},\hat{a}_{\rm{out}}^\dagger,\hat{a}_{\rm{down}}^\dagger)^\top.
\end{split}
\label{eq:ModeExpOutsideBH}
\end{equation}
With this notation, the quantum scattering matrix relates the two bases as $\bs{W}^{\rm in}=\bs{W}^{\rm out}\cdot\bs{S}$ and $\hat{\bs{A}}^{\rm out}=\bs{S}\cdot\hat{\bs{A}}^{\rm in}$. Hence, we can compute the matrix $\bs{S}$ as the change of basis from IN to OUT modes (and their conjugates). As in Section \ref{sec:QuantumSR}, all the results can be expressed in terms of the smaller \textit{classical} scattering matrix $\bs{B}$, whose coefficients are the scattering amplitudes relating the IN and OUT mode bases. Thus, as in equation \eqref{eq:GeneralModeEvolution}, i.e. $(W_{\rm{in}},W_{\rm{p}},W_{\rm{d}})=(W_{\rm{out}},W_{\rm{down}},W_{\rm{dn}})\cdot\bs{B}$ for $\omega<\omax$; and $(W_{\rm{in}},W_{\rm{p}})=(W_{\rm{out}},W_{\rm{down}})\cdot\bs{B}$ for $\omega>\omax$. For simplicity, we also introduce indices for the modes \textit{branches}, i.e. $1$ indicates the modes $in$ and $out$, $2$ the modes $p$ and $down$, and $3$ the modes $d$ and $dn$. The resulting scattering matrices for $\omega<\omax$ and $\omega>\omax$ are
\begin{equation}
    \mathbf{S}_{<}= \begin{pmatrix}
B_{11} & B_{12} & 0 & 0 & 0 & B_{13} \\
B_{21} & B_{22} & 0 & 0 & 0 & B_{23} \\
0 & 0 & B_{33}^* & B_{31}^* & B_{32}^* & 0 \\
0 & 0 & B_{13}^* & B_{11}^* & B_{12}^* & 0 \\
0 & 0 & B_{23}^* & B_{21}^* & B_{22}^* & 0 \\
B_{31} & B_{32} & 0 & 0 & 0 & B_{33} \\
\end{pmatrix}
\qquad\text{and}\qquad
    \mathbf{S}_{>}= \begin{pmatrix}
B_{11} & B_{12} & 0 & 0 \\
B_{21} & B_{22} & 0 & 0 \\
0 & 0 & B_{11}^* & B_{12}^*  \\
0 & 0 & B_{21}^* & B_{22}^*  \\
\end{pmatrix}.
\end{equation}
As in the superradiant case, these matrices encode the Bogoliubov transformations between IN and OUT modes, that is the linear combinations of creation an annihilation operators, for example, for $\omega<\omax$,
\begin{equation}\label{eq:bogo-transf-hawking}
    \hat a_{out} = B_{11} \hat a_{in} + B_{12}\hat a_{p} + B_{13}\hat a_{d}^\dag.
\end{equation}
The $\bs{S}$-matrix must be symplectic in order to to preserve the commutation relations. Classically, this corresponds to the elements of $\bs{B}$ assuring the conservation of the symplectic norm during the scattering. This condition can be compactly expressed in terms of the matrix $\bs{B}$ (provided that the modes are orthonormalized with respect to the symplectic product). For $\omega>\omax$ the matrix is unitary, i.e. $\bs{B}^\dag=\bs{B}^{-1}$, and in turn $\bs{S}_>$ is both symplectic and unitary. This means that it preserves the vacuum (does not mix creation and annihilation operators) and no pair creation occurs. For $\omega<\omax$ instead, the matrix $\bs{B}$ is no longer unitary, but is pseudo-unitary with respect to the diagonal matrix encoding the norms of the basis modes $\eta=\mathrm{diag}(1,1,-1)$, namely we have $\bs{B}^{-1}=\eta^{-1}\cdot\bs{B}^\dagger\cdot\eta$. This implies that $\bs{S}_<$ is symplectic but non-unitary, so that it gives rise to pair production and entanglement generation.

With these relations, late time observables can be computed (see Appendix appendix \ref{app:ComputationScattCoef} for a method to explicitly compute the scattering matrix coefficients) . Knowing the initial state of the system, one can directly compute any observable related to the OUT modes from the Bogoliubov transformation $\hat{\bs{A}}^{\rm out}=\bs{S}_<\cdot\hat{\bs{A}}^{\rm out}$. Of course the interesting case is $\omega<\omax$, for which the stationary analogue Hawking effect occurs. Here, we simply compute the number of emitted quanta for a vacuum input. From \eqref{eq:bogo-transf-hawking}, one easily obtains for the number of quanta in the $out$ mode (at a fixed frequency) is given by
\begin{equation}
    \expval{\hat a_{out}^\dag\hat a_{out}} = |B_{11}|^2\expval{\hat a_{in}^\dag\hat a_{in}} + |B_{12}|^2 \expval{\hat a_{p}^\dag\hat a_{p}} + |B_{13}|^2 \expval{\hat a_{d}\hat a_{d}^\dag}.
\end{equation}
If the expectation value is taken on the vacuum of IN modes\footnote{This gives the zero temperature result, a more faithful description of experiments would include a thermal input, or even a coherent input in experiments looking for the stimulated Hawking effect.}, i.e. $\hat A^{\rm in}_i\ket{0}_{IN}=0$ for $i\in\{1,2,3\}$, the only nonzero term is the last one. The same computation can be done for the other OUT modes, so that the spectra of the emitted quanta for vacuum input are given by

\begin{equation}
\expval{\hat{N}_{\rm out}}= |B_{13}|^2\quad,\quad \expval{\hat{N}_{\rm down}}= |B_{23}|^2\quad\text{and}\qquad\expval{\hat{N}_{\rm dn}}= |B_{33}|^2,
\end{equation}

In the analogy, what can be called Hawking radiation are the emitted quanta in the $out$ mode. Notice that this is given by the scattering amplitude for an ingoing negative-energy $d$ mode that is transmitted to the $out$ mode. This ingoing mode is absent in the hydrodynamic limit, so that this procedure heavily relies on the presence of dispersive effects\footnote{While qualitatively different, the same is true also for subluminal dispersion \cite{robertson2012theory}.}, and cannot hence be used to derive pair production in the non-dispersive case, that is the one usually considered in a gravitational setting. We will comment more about this in the following.

\subsection{Explicit solution for the step interface horizon}\label{sec:step-horizon}

\begin{figure}
    \center
    \includegraphics[width=\textwidth]{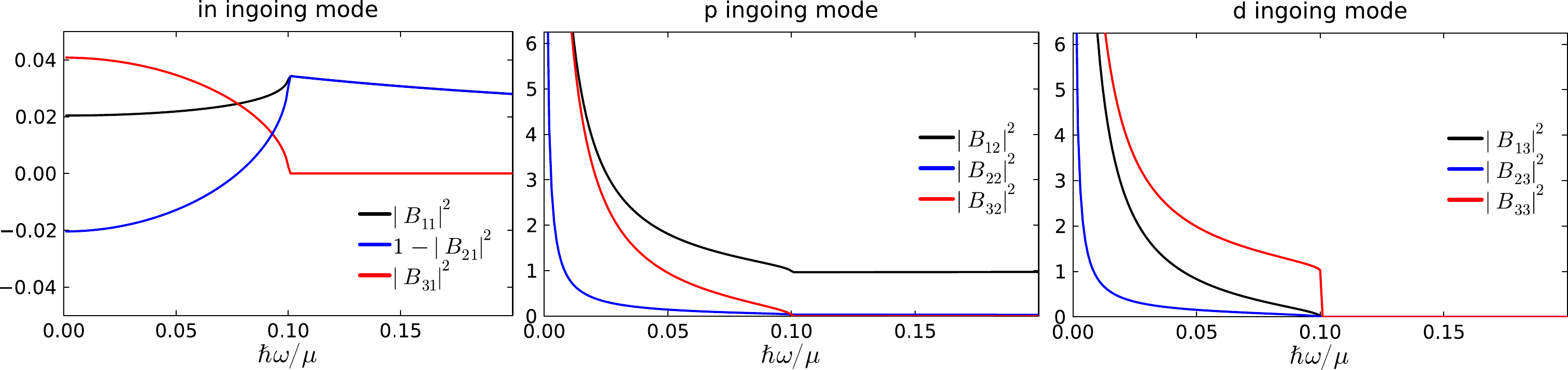}
        \caption{Scattering matrix elements obtained from the numerical solution of the step-interface horizon. Each panel is obtained selecting a different ingoing wave. The three colors indicate the three different outgoing modes. Above $\omax\sim 0.1\mu/\hbar$ the $d$ ingoing mode is no longer available. The parameters are $\tv=0.75 {\rm c}_{\rm s}^u$ and ${\rm c}_{\rm s}^d=0.3{\rm c}_{\rm s}^u$.}
    \label{fig:hawking-scattering}
\end{figure}

In the previous section we saw that, in the presence of a subsonic asymptotic region and a supersonic one, emission of quanta occurs at low enough frequencies. The number of emitted quanta is determined by the scattering amplitudes, that depend on the quantitative details of the full flow profile, not only on the asymptotic regions.

Here we give an explicit example of the scattering amplitudes for a step-like interface between two uniform subsonic and supersonic regions. This kind of configuration was considered in  \cite{balbinot2008nonlocal,carusotto2008numerical,recati2009bogoliubov,larre2012quantum}, and can be obtained for example by considering an effectively 1D condensate with a uniform velocity $\vtv=-|\tv| \hat{\vx}$ and a space dependent speed of sound. The flow can be made locally supersonic by changing the speed of sound $\cs=\sqrt{gn/m}$ by tuning the interaction constant $g$. Moreover, one can think of keeping the density constant by introducing an external potential to make the chemical potential $\mu=gn+V_\mathrm{ext}$ constant in space. This can be done by changing abruptly the speed of sound from the upstream one ${\rm c}_{\rm s}^u>|\tv|$ for $x>0$ to the downstream one ${\rm c}_{\rm s}^d<|\tv|$ for $x<0$.

While technically challenging, this configuration serves as a minimal realistic theoretical toy model, easy enough to analyze while displaying the important physics. More realistic discontinuous but non-homogeneous configurations that are also treatable analytically were studied in \cite{larre2012quantum}, where an analytical solution of the scattering problem for the configuration we are considering was given. Here we will resort to a numerical solution to display an alternative approach. Writing the code to perform this solution is the content of Problem \ref{prob:HawkAmplitudes}, whose solution is provided as an attached Jupyter notebook written in Julia \cite{Julia-2017}.

We are hence dealing with two uniform regions, in which fluctuations follow the Doppler-shifted dispersion relation \eqref{eq:DispersionHawking1D}, with the two behaviors displayed in Fig.~\ref{fig:DispersionHawking1D}. In each region, the fluctuation field at a fixed frequency can be expanded in terms of the four plane waves whose wavenumbers solve these dispersion relation. We now want to match these two expansions at the sharp interface by imposing the continuity of the fluctuation field and its spatial derivative. When only two real solutions exist, the other two will be complex wavenumbers, that is waves that grow or decay exponentially in space. While these do not enter the scattering matrix, they are important for the matching procedure. Of course, in each region only the decaying complex root should be allowed, i.e. the evanescent wave. Boundary conditions must be placed so that the exponentially growing mode is not excited \cite{Macher:2009nz}.

Explicitly, in the subsonic upstream region the atomic fluctuation field ($\hat\psi$ in section \ref{sec:RealComplexRelationClass}) can be expanded as
\begin{equation}\label{eq:inout-modes}
	\begin{pmatrix}
		u(x)\\ v(x)
	\end{pmatrix}_{\w,u}
	=
	A^{in}
	\begin{pmatrix}
		u_{\tk^{in}}\\ v_{\tk^{in}}
	\end{pmatrix}
	\frac{e^{i\tk^{in}x}}{\sqrt{|v_g^{in}|}}
	+
	A^{out}
	\begin{pmatrix}
		u_{\tk^{out}}\\ v_{\tk^{out}}
	\end{pmatrix}
	\frac{e^{i\tk^{out}x}}{\sqrt{|v_g^{out}|}}
	+
	A^{ev}
	\begin{pmatrix}
		u_{\tk^{ev}}\\ v_{\tk^{ev}}
	\end{pmatrix}
	e^{i\tk^{ev}x}.
\end{equation}
In the downstream supersonic region, for $\omega>\omax$, there will be an analogous expansion, while for $\omega<\omax$ all the four real roots must be considered
\begin{equation}
	\begin{pmatrix}
		u(x)\\ v(x)
	\end{pmatrix}_{\w,d}
	=
    \sum_{I}
	A^{I}
	\begin{pmatrix}
		u_{\tk^{I}}\\ v_{\tk^{I}}
	\end{pmatrix}
	\frac{e^{i\tk^{I}x}}{\sqrt{|v_g^{I}|}}.
\end{equation}
The two component vectors of the Bogoliubov modes are normalized as $|u_{\tk^I}|^2-|v_{\tk^I}|^2=\pm 1$, and the propagating modes are further normalized with their group velocity $v_g^{I}=\partial_\tk\omega(\tk)|_{\tk=\tk^I}$ as seen in \eqref{eq:PlaneWaveNormBdG}. This is because, since the plane waves are not strictly normalizable, we want their current to be normalized. As shown in \cite{dalfovo1996quantum} (Appendix B), a plane wave solution of the Bogoliubov problem has an associated \textit{norm} current $j^I=v_g^I\left(|u_{\tk^I}|^2-|v_{\tk^I}|^2\right)$, so one needs to divide by the group velocity to have a normalized current.

The matching conditions that need to be satisfied at the interface (say $x=0$) are the continuity of the solution and its first derivative, i.e.
\begin{equation}
    \begin{pmatrix}
		u(x=0)\\ v(x=0)
	\end{pmatrix}_{\w,u}
    =
    \begin{pmatrix}
		u(x=0)\\ v(x=0)
	\end{pmatrix}_{\w,d}
    \hspace{0.5cm}
    \mbox{and}
    \hspace{0.5cm}
    \frac{d}{dx}
    \begin{pmatrix}
		u(x)\\ v(x)
	\end{pmatrix}_{\w,u}\Bigg|_{x=0}
    =
    \frac{d}{dx}
    \begin{pmatrix}
		u(x)\\ v(x)
	\end{pmatrix}_{\w,d}\Bigg|_{x=0}.
\end{equation}
The full scattering solution can be solved by choosing a non-vanishing amplitude for only one of the normalized IN modes while the rest vanish. The amplitudes of the normalized OUT and evanescent modes is then determined by the four conditions above. Then, the scattering amplitudes $B_{ij}$, where $j$ is the index of the chosen IN mode $I$ and $i$ runs over OUT modes, are found as ratios between the selected IN amplitude and the corresponding amplitudes for the OUT modes. One then repeats the process going over all the IN modes to fully reconstruct $\bs{B}$, thus finding all the scattering amplitudes characterizing the process (see Appendix \ref{app:ComputationScattCoef}). 

\begin{equation*}
    \tilde{\tk}=\xi_u\tk\ ,\quad \tilde{\omega}=\hbar\omega/g_u n\ ,\quad \tilde{\tv}\coloneqq \tv/c_u\ \text{, and}\quad\tilde{c}_i=c_i/c_u.
\end{equation*}

In Fig.~\ref{fig:hawking-scattering}, we plot some numerically obtained scattering amplitudes as a function of the frequency. As explained above, the amplitudes corresponding to the ingoing $d$ mode (index 3) give the number of emitted quanta in the different outgoing waves, so that $|B_{3j}|^2$ is the emission spectrum in mode $j$.

Notice that all scattering amplitudes with an ingoing negative-energy wave $d$ diverge at small frequencies. In particular, the reflection coefficient for the $d$ mode is always larger than one, i.e. its scattering is superradiant, as we concluded on general grounds in the previous section. Notice that also the reflection of the $p$ mode can be superradiant at small enough frequencies, and that amplification can occur in transmission for an ingoing $in$ wave. Finally, in the first panel of Fig.~\ref{fig:hawking-scattering} we can see that there is some reflection for an ingoing $in$ mode from \textit{outside} the black hole. This means that our black hole is not really black. In fact, some reflection is to be expected with a sharp change in space of the background flow and with our step profile we are very far from the hydrodynamic limit in which the analogy strictly holds.

Above $\omax$ the $d$ mode does not exist anymore. The analogue Hawking spectrum $|B_{31}(\omega)|^2$ (black line in the third panel of Fig.~\ref{fig:hawking-scattering}) must hence vanish above that frequency. As we will be clearer after we review the gravitational Hawking effect in the next section, this is something that is not expected for a non-dispersive relativistic theory, for which the spectrum is predicted to be thermal, and hence be nonzero at all frequencies. We will comment more on the comparison in the following.

\subsection{Analogy with the gravitational Hawking effect}

In this section, we discuss the analogy between the stationary analogue Hawking effect described above and the original effect predicted by Hawking in a scenario where an astrophysical object collapses to form a black hole. To that end, we will first provide a quick overview of the essential aspects of the original prediction by Hawking, and then discuss similarities and differences with the stationary analogue described above.

\subsubsection{Essential aspects of the astrophysical Hawking effect: a dynamical collapse scenario}

\begin{figure}[t]
    \center
\includegraphics[width=0.3\textwidth]{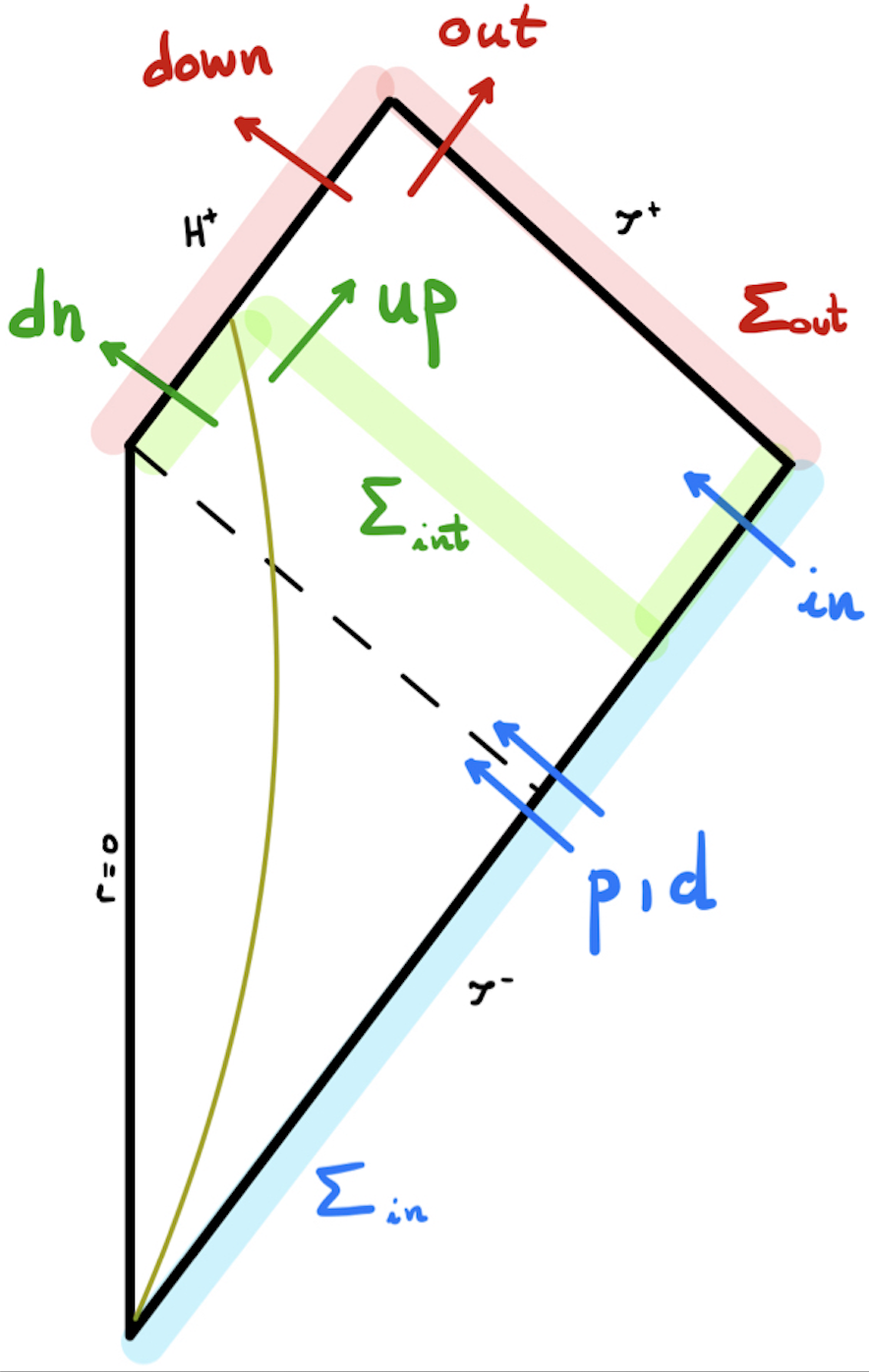}
    \caption{Penrose diagram of the formation of a black hole, with the relevant modes for the Hawking effect indicated with arrows.}
    \label{fig:penrose}
\end{figure}

The Hawking effect was originally derived in a dynamical scenario, where an astrophysical object undergoes a full gravitational collapse, eventually forming a black hole. The collapse can be represented in a conformal (Penrose) diagram as in Fig.~\ref{fig:penrose}, in which time flows vertically, the radius grows horizontally, and light moves in 45-degree lines. Here, the thin curved line represents the surface of a collapsing body\footnote{We consider the simplest case of a spherically symmetric non-rotating body. The generalization of the rotating case has some subtleties that are not relevant to understanding the basic mechanisms behind the Hawking effect.}.

In his original work \cite{Hawking:1975vcx}, Hawking considered the evolution of a test quantum field (with a standard relativistic dispersion relation) in a dynamical spacetime where a black hole forms from a collapsing star. Starting in the vacuum state at early times, before the collapse, he found that a static observer placed far away from the BH at late times would detect a steady flux of radiation, independently of the details of the collapse process. Concretely, this steady flux is detected at late times in future null infinity $\mathscr{I}^+$ (top right in Fig.~\ref{fig:penrose}), and can be thought of as originating near the BH horizon as black body (thermal) radiation at its Hawking temperature. This temperature is uniquely determined by properties of the spacetime near the horizon. Concretely, for a massless bosonic field, an observer at spatial infinity would measure a spectrum of quanta
\begin{equation}
\label{eq:HawkingSpecAstro}
    n(\omega)=\frac{\Gamma(\omega)}{e^{\omega/T_H}-1},
\end{equation}
 in natural units, where $T_H=\kappa/2\pi$ is the Hawking temperature, and $\kappa$ the  surface gravity of the BH (see \eqref{eq:SurfGravDV}). Lastly, $\Gamma(\omega)$ is the \textit{greybody factor}, i.e. the modification of the black-body spectrum observed at infinity due to the scattering of the emitted radiation with the gravitational potential of the black hole, which acts as a filter.

Let us now provide some detail on the fundamental degrees of freedom of the quantum field that interact to produce the radiation, to allow a comparison with the stationary analogue case we studied in the previous sections. The spacetime represented in Fig.~\ref{fig:penrose} can be understood as composed of two regions: 1) a dynamical (time-dependent) region where the collapse is taking place and no timelike Killing vector exists. This region is the triangle below the dashed black line. 2) a stationary (time-independent) region where the collapsing body has crossed its Schwarzschild radius $r_{\rm S}\coloneqq2GM/c^2$ and the black hole has already formed. In this region, which is the rectangular region above the dashed line, the vector $\partial_t$ is a timelike Killing vector field (which becomes null at the horizon, the $H^+$ line in the diagram). The mechanism underlying Hawking pair production takes place in the dynamical region, specifically near the formation of the horizon. In this region, the absence of a timelike Killing vector field implies that there is no associated conserved frequency. In particular, the frequency measured by static asymptotic observers, associated to $\partial_t$, is not conserved in evolution. As a consequence, $\ee^{-\ci\omega t}$ and $\ee^{\ci\omega t}$ modes mix during this period, producing the Hawking radiation. On the other hand, outward propagation of the emitted radiation takes place in the static geometry outside of the formed black hole, where $\partial_t$ is a timelike Killing vector field, and modes with different frequency evolve independently of each other. In this region, the produced blackbody radiation interacts with the stationary potential barrier of the black hole, filtering the radiation that arrives to future null infinity $\mathscr{I}^+$ as dictated by the greybody factors $\Gamma(\omega)$ (these are in fact the absolute value squared of the transmission coefficients associated to the potential barrier of the BH).

We want to emphasize that time dependence is hence the key ingredient in the spontaneous production of quanta. If $\partial_t$ was a global timelike Killing vector field\footnote{Here, having a global timelike Killing vector would mean having a Killing vector which is timelike everywhere outside the horizon. As a remark, note that in the presence of ergoregions, even though there can be Killing vector fields which are asymptotically timelike, they are not timelike everywhere. This is why, in the case of a stationary rotating black hole, there is norm-mixing related to superradiant scattering.}, the norm-mixing process would not occur, since modes with different $\omega$ values would evolve independently and norm-mixing would not occur. However, due to the dynamical region, $\partial_t$ is not a global timelike Killing vector. As a consequence, modes with different frequency mix during the whole process, and Hawking showed that a positive-norm mode $\ee^{-\ci\omega t}$ at $\mathscr{I}^+$ originates from a linear combination of infinitely many $\ee^{\pm\ci\omega t}$ at $\mathscr{I}^-$, including a contribution from negative-norm $\ee^{\ci\omega t}$ modes\footnote{In fact, there is a non-negligible contribution from modes which have trans-Planckian frequencies at $\mathscr{I}^-$, giving rise to the debated trans-Planckian problem \cite{Barbado:2011ai}.}. Wald showed that one can in fact identify two ``progenitor'' modes of single $\omega$ Hawking quanta, often called $p$ and $d$ \cite{Wald:1975kc}. The $up$ modes and their purifying (negative-norm) modes $dn$, also known as partner modes, are produced from a two-mode squeezing transformation of the progenitor modes near the horizon formation region. The emitted black-body $up$ quanta are then scattered by the potential barrier of the BH, which acts as a beam splitter. As a result, reflected $down$ quanta are sent back into the BH and transmitted $out$ quanta reach observers at $\mathscr{I}^+$ as grey-body Hawking radiation, with the spectrum given by \eqref{eq:HawkingSpecAstro}.
\subsubsection{Differences and similarities}

From the above discussion it appears that the gravitational Hawking effect (GHE) and the pair creation in a stationary BEC with a transsonic flow configuration (which we call stationary Hawking effect (SHE)) differ in the fundamental fact that the first relies on time-dependence, while the second is derived as a stationary effect. In other words, the GHE arises from non-conservation of the frequency (measured by far away observers), while the SHE occurs while preserving the frequency measured by laboratory observers through the whole process.

The analogue case can, in principle, be engineered to mimic the dynamical formation of an acoustic black hole, as is the case in most experiments, and as was done for example in \cite{carusotto2008numerical,Barcelo:2008qe}. In this time dependent case the frequency of excitations is not conserved. Importantly, the results for this case are similar to the ones obtained with the stationary configuration (e.g. the density-density correlations) at low frequencies.

Still, the GHE cannot be predicted in a stationary setting, and the main difference lays in the dispersive effects present in the analogue,  i.e. the higher order spatial derivatives in the wave equation. These provide the ingoing modes that propagate against the supersonic flow, which are essential for the calculation of the SHE, and whose presence has been experimentally verified in various fluid platforms  \cite{steinhauer2014observation,Euve:2018uyo,Falque:2023ctx}. Such modes are not present for a field with a standard relativistic dispersion, and the Bogoliubov transformation at fixed frequency cannot hence be constructed, since the IN and OUT scattering modes are in different numbers.

Indeed, one can see how the calculations of the SHE start to be less physical when the hydrodynamic limit $\xi^{-1} k\to 0$ is approached. The healing length of the condensate $\xi$ can be thought as the scale of ultraviolet physics, and the effect of sending it to infinity on the supersonic dispersion of Fig.~\ref{fig:DispersionHawking1D} is to push the needed ingoing modes $d$ and $p$ to infinite momenta, {not dissimilarly to the fact that the Hawking progenitors $p$ and $d$ in the GHE contain infinitely large frequencies. In other words, the dispersive scale $\xi$ works as a \textit{regularizator} of the ultraviolet physics, and can be thought as an analogue of the Planck scale.}

It is hence interesting to draw a qualitative parallelism between the modes of the quantum fields involved in the pair production in the two cases. To this end, we chose appropriate names for the modes in Fig.~\ref{fig:DispersionHawking1D} that match the ones usually employed for the GHE. One can think of the ingoing $p$ and $d$ modes as the progenitors that undergo a two mode squeezing to produce the Hawking radiation $out$ and the partners $dn$ and $down$.

Despite these subtle points, it is remarkable that the calculation for the SHE gives a prediction the the emitted spectrum of $out$ quanta which is close to the thermal Hawking spectrum at small enough frequencies. The expectation for an acoustic black hole in the hydrodynamic limit is a thermal spectrum with the Hawking temperature given by the surface gravity of the acoustic horizon \cite{barcelo2011analogue}
\begin{equation}\label{eq:hawking-temp-analogue}
    T_H=\frac{\kappa}{2\pi }\qquad\text{where}\qquad \kappa=\frac{1}{2}\frac{\dd (c^2-v^2)}{\dd x}\Bigg|_\mathrm{hor}.
\end{equation}

This above result is in exact correspondence to the Hawking temperature found from gravitational collapse into a BH with surface gravity $\kappa$. However, we know that the spectrum of the SHE cannot be exactly thermal, since it is exactly zero above $\omax$. Still, looking for example at the spectrum given by the black line in the third panel of Fig.~\ref{fig:hawking-scattering}, at low frequencies it goes as $\omega^{-1}$, as a thermal spectrum. This is interesting also because, strictly speaking, the analogy with standard QFT in curved spacetimes does not hold for the step interface horizon we considered, since we are outside the hydrodynamic limit (the surface gravity \eqref{eq:hawking-temp-analogue} is formally infinite). Besides the cutoff of the emission spectrum, dispersive effects also cause the position and the properties of the horizon to become mode-dependent. For realistic velocity profiles, this causes also the analogue Hawking temperature to depend on frequency \cite{isoard2020departing,isoard2021bipartite}.

From a wave-propagation viewpoint, the  thermality of the spectrum in the hydrodynamic limit can be traced back to the relation between the affine parameters describing propagation of sound rays (i.e. their ``peeling''). This relation is affected by dispersion, but becomes identical to that occurring in a gravitational collapse into a BH in the hydrodynamic limit. The reason is that, while dispersion makes the horizon to be a mode-dependent concept, the acoustic horizon becomes a true causal barrier in such limit. Hence, hydrodynamic modes traveling against the flow right outside of the horizon feel an exponential peeling similar to that felt by outgoing modes in the vicinity of a BH event horizon. It is precisely this exponential peeling what is at the heart of the thermality of the Hawking spectrum \cite{Barcelo:2010pj}, with the temperature related to the surface gravity, which effectively controls ``how strong the peeling is''.

\subsection{Density correlations as signatures of the analogue Hawking effect}
\begin{figure}
    \center
    \includegraphics[width=0.5\textwidth]{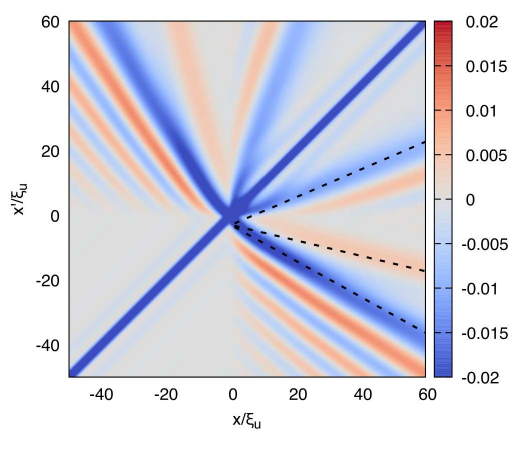}
    \caption{Density-density correlations of the quantum emission for the step interface horizon. Here the constant velocity is $\tv=0.75c_u$, while the downstream speed of sound is $c_d=0.3c_u$. The dashed lines are the locations where strongest correlations are expected, as explained in the text.}
    \label{fig:hawking-correlations}
\end{figure}

One of the problems of observing Hawking radiation in an astrophysical setting is that the Hawking temperature for solar-mass black holes (or bigger) is much smaller than the temperature of the cosmic microwave background. In a BEC, the temperature of the analogue Hawking radiation is closer to the temperature of the condensate, but the direct detection of the radiation is still difficult.

However, a clear advantage of studying the Hawking effect in analogue systems is the possibility to access information in the interior of the analogue black hole. This possibility lead to the proposal of using correlations between the density of atoms at points in opposite sides of the horizon as an observable signature of Hawking radiation \cite{balbinot2008nonlocal,carusotto2008numerical}. This proposal has been used extensively in the literature, and is at the basis of the observations of the analogue Hawking effect in BEC systems \cite{steinhauer2016observation,munoz2019observation,kolobov2021observation}. Here we will just summarize the idea, we refer to \cite{recati2009bogoliubov,larre2012quantum} for a complete account of the computations. These correlations, besides being a smoking gun of the analogue Hawking emission, are a powerful tool that can be used, for example, to extract information about the spectrum of emission and its entanglement.

The idea is that excitations are produced in pairs near the horizon, one traveling inwards and the other outwards. If a pair is produced at time $t_0$, after a time interval $\Delta t$, there should be a correlation between the density perturbations at points which are separated by a distance given by the velocity of propagation of each of the excitations times $\Delta t$. This correlation should lead to a visible signature in the equal-time two-point correlation function. Intuitively, considering for example the emission of a wavepacket in the $out$ mode and one in the $dn$ mode, after a time $\Delta t$ the two wavepackets will be centered at positions given by the respective group velocity, i.e. $x_{out}=v_g^{out}\Delta t$ and $x_{dn}=v_g^{dn}\Delta t$. If one plots the position of one wavepacket as a function of the other, one obtains a straight line $x_{dn}=(v_g^{dn}/v_g^{out})x_{out}$, such as the dashed lines plotted in Fig.~\ref{fig:hawking-correlations}.

A relevant quantity to compute to pursue this idea is the equal time density-density correlation function
\begin{equation}
    G_2(t,\tx,\tx')=\expval{:\hat{n}(t,\tx)\hat{n}(t,\tx'):}-\expval{\hat{n}(t,\tx)}\expval{\hat{n}(t,\tx')}.
\end{equation}
Using $\hat{n}=\hat{\Psi}^\dagger\hat{\Psi}$, and decomposing the field into a background plus a perturbation $\hat{\Psi}=\ee^{-\ci\mu t}(\Psi_0\hat{\mathbb{I}}+\hat{\psi})$ as in \eqref{eq:DefPertGPE},  we can write the above correlation function to leading order in perturbations as
\begin{equation}
    G_2(t,\tx,\tx')=\Psi_0^*(\tx)\Psi_0(\tx')\expval{\hat{\psi}^\dagger(t,\tx)\hat{\psi}(t,\tx')}+\Psi_0^*(\tx)\Psi_0^*(\tx')\expval{\hat{\psi}(t,\tx)\hat{\psi}(t,\tx')}+ \text{h.c.}.
\end{equation}
The fluctuation field $\hat \psi$ can be expanded in terms of the propagating IN and OUT modes in each region, and their corresponding creation and annihilation operators in \eqref{eq:ModeExpInsideBH} and \eqref{eq:ModeExpOutsideBH}. The OUT operators can then be expressed in terms of IN operators by transforming with the $\boldsymbol{S}$-matrix. Averages over specific IN states can then be computed, with the vacuum of these modes giving the correlations of the spontaneous emission (i.e. emission at zero BEC temperature).

Here we do not derive the explicit expression and refer to \cite{recati2009bogoliubov,larre2012quantum} for the full computation in the atomic basis. Expressions can be obtained that depend only on the Bogoliubov amplitudes of the plane waves in the two regions and on the scattering matrices. In Fig.~\ref{fig:hawking-correlations} we show an example of the resulting $G_2(x,x')$ for the step interface horizon considered in Section \ref{sec:step-horizon}. The lower right (or upper left) quadrant of this graph gives information about the correlations between the inside and the outside of the black hole. The upper right and lower left ones give instead the $in-in$ and $out-out$ correlations. The dashed lines are computed with the ratios of the group velocities of the three outgoing waves, as explained above, and from top to bottom correspond to the pairs of OUT modes $down-dn$, $down-out$ and $dn-out$, and match well with the features in the plot. The last pair in particular, is the Hawking \textit{mustache} that was observed in experiment \cite{steinhauer2016observation,munoz2019observation,kolobov2021observation}.

\subsection{Black hole lasing}
\label{sec:black-hole-lasing}

As discussed above, the stationary analogue Hawking effect is essentially a manifestation of superradiant physics. In fact we already pointed out how amplified scattering can occur in the one dimensional black hole we considered, as also signaled by the scattering coefficients in Figure \ref{fig:hawking-scattering} that largely exceed one. As we saw in Section \ref{sec:sr-instabilities}, superradiant scattering can give rise to dynamical instabilities when the amplified waves are trapped and undergo repeated amplification. This is the case also here.

Instabilities here arise when one consider a flow profile that passes from subsonic to supersonic, and then to subsonic again. The second interface is an analogue \textit{white hole} horizon. This situation is what one obtains when trying to create a sonic black hole on a torus, as was considered for example in \cite{garay2001sonic}. In particular, in the supersonic region one has two counter-propagating negative-energy modes that are not present in the subsonic region, so that each scattering of this wave will be amplified. 

Notice that, differently from the superradiant case, the occurrence of black hole lasing is strongly dependent on the superluminal nature of the Bogoliubov dispersion relation, i.e. on the presence of modes that propagate against the flow in the supersonic region. This kind of instability was first found by studying fields with a superluminal high-frequency dispersion in spacetimes with both an outer and an inner horizon (like charged and rotating black holes) \cite{corley1999black}.

This instability was studied experimentally in \cite{steinhauer2014observation,kolobov2021observation}, but in those setups it is probably not the main source of instability (see \cite{kolobov2021observation} for a discussion and references).

\subsubsection{Numerical time evolution of the GPE}
\label{sec:numerical-evolution-gpe}

\begin{figure}
    \center
    \includegraphics[width=0.8\textwidth]{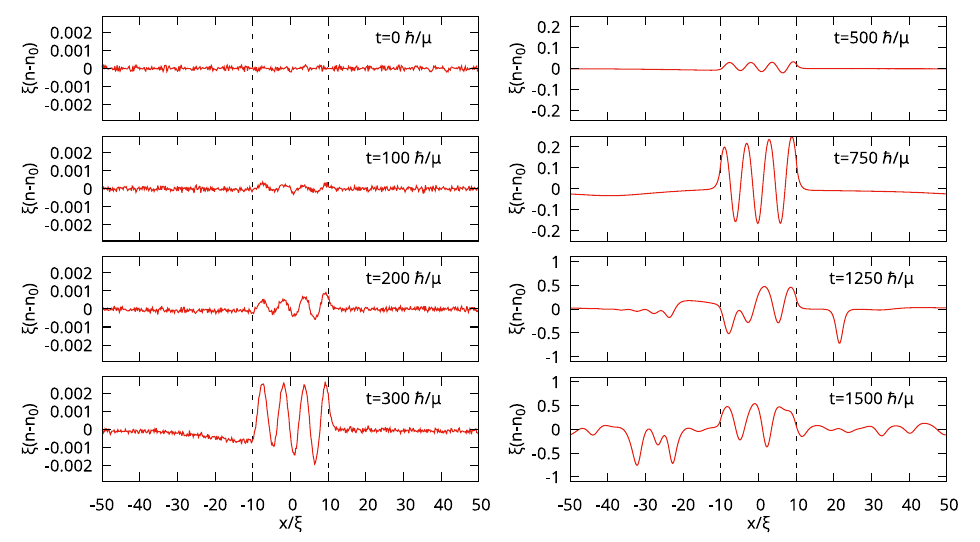}
    \caption{Snapshots of the numerical time evolution of the GPE displaying a black hole lasing instability. The initial configuration is a plane wave with flow velocity $\tv_x\simeq 0.75 c_u$, with $c_u$ the speed of sound outside of the central region. The speed of sound between the dashed lines is taken to be $c_d=0.25 c_u$.}
    \label{fig:blackhole-lasing}
\end{figure}

This instability can be detected numerically by diagonalizing the Bogoliubov problem, as we did in Section \ref{sec:bogo-diag-SR}, but this time with periodic boundary conditions (this can be done as a follow-up exercise to Problem \ref{prob:BogoShearLayer}). One can for example check that the instability rate decreases when the size of the supersonic region is increased. Here we want to display another way to observe this physics by numerically solving the GPE in time. Writing a code that does this, using the following discussion, is the content of Problem \ref{prob:BHLasing}. A Jupyter notebook written in Julia \cite{Julia-2017} is provided as a solution.

To do this, we will use what is known as a split-step pseudo-spectral method of integration. The basic idea is to perform the following approximation on the propagator
\begin{equation}
    \Psi(\vx,t+\Delta t)=\ee^{-\ci H_{GP}\Delta t}\Psi(\vx,t)=\ee^{-\ci\mathcal{W}\Delta t/2}\ee^{-\ci\mathcal{K}\Delta t}\ee^{-\ci\mathcal{W}\Delta t/2}\Psi(\vx,t)+o(\Delta t^2),
\end{equation}
where $H_{GP}=\mathcal{K}+\mathcal{W}$, with $\mathcal{K}=-\hbar^2\nabla^2/2m$ the kinetic part of the GPE Hamiltonian, and $\mathcal{W}$ all the rest. The useful property of this kind of splitting is that the exponential involving $\mathcal{W}$ is diagonal in position space (i.e. it acts by multiplication), while the exponential involving the kinetic term is diagonal in momentum space. One can hence represent $\Psi(\vx,t)$ on a spatial grid and then use the following steps to perform the time evolution of $\Delta t$:
\begin{itemize}
    \item Multiply the wavefunction at time $t$ by the first part of the split propagator: 
    
    \noindent$\Psi_1(\vx,t):=\ee^{-\ci\mathcal{W}(\vx)\Delta t/2}\Psi(\vx,t)$
    \item Perform a discrete Fourier transform of $\Psi_1(\vx,t)$ to obtain its representation in momentum space $\widetilde\Psi_1(\vk,t)$
    \item Multiply by the kinetic part of the split propagator $\widetilde\Psi_2(\vk,t):=\ee^{\ci(\hbar^2\vk^2/2m)\Delta t}\widetilde\Psi_1(\vk,t)$
    \item Invert the Fourier transform to obtain $\Psi_2(\vx,t)$
    \item Multiply by the last part of the split propagator $\Psi(\vx,t+\Delta t):=\ee^{-\ci\mathcal{W}(\vx)\Delta t/2}\Psi_2(\vx,t)$.
\end{itemize}
This kind of algorithm is convenient for the GPE because it allows to avoid finite difference formulas for the derivatives and automatically preserves the normalization of the wavefunction. Notice that the use of the Fourier transform also automatically implements periodic boundary conditions at the edges of the numerical box.

To apply this to the black hole lasing we are discussing, let us consider a one dimensional GPE on a domain $\tx\in[-L/2,L/2]$ with periodic boundary conditions. Let us take the interaction constant $g$ to depend on space, with a value $g_2$ in an interval $[-\ell/2,\ell/2]$ and value $g_1$ outside of it, with $g_2<g_1$, i.e. lowering locally the speed of sound. As explained above, we also introduce an external potential in the central region, so that a plane wave is a stationary solution of the GPE. In particular, we take as initial condition the stationary solution $\Psi(\tx,t=0)=\sqrt{n_0}\ee^{\ci \frac{2\pi}{L} q\, \tx}$, with $q$ an integer, so that the plane wave is compatible with the compact domain we are considering. We also add at each point in space a small random number $\delta(\tx)$, such that, for example, $\delta(\tx)/\sqrt{n_0}\sim 10^{-4}$. This is a trick to seed the instability by giving a nonzero population to all the small amplitude eigenmodes. To this initial condition we then apply the above algorithm to compute the time evolution of this initial condition.

In Figure \ref{fig:blackhole-lasing}, we show some snapshots of the time evolution of the density difference with respect to the steady state $n_0$, in the case in which the condensate is supersonic in the central region, and subsonic outside (parameters in the caption). In the first column one can see how the noisy initial condition evolves in a standing wave mode living inside the supersonic region. This is a dynamically unstable mode that grows exponentially in time, as long as its amplitude is small enough. In the second column we show longer times. In the first panel the same standing wave mode has a much larger amplitude (notice the scale that is two orders of magnitude larger than in the first column). At this point nonlinear effects start to play a role, and the system is outside of the validity of the Bogoliubov approximation. The condensate has at this point a very "turbulent" behavior, with density variations of the order of the initial uniform density. In the panel for $t=1250\hbar/\mu$ one can see that a pronounced dip is present in the right subsonic region: this is a soliton, a purely nonlinear traveling excitation \cite{pitaevskii2016bose}.

\section{Conclusion}
In these lecture notes we saw in detail how elementary excitations of BEC behave as a scalar quantum field. In particular, they are governed by a generalized Klein-Gordon equation with non-relativistic terms, that becomes relativistic in the long-wavelength limit, recovering the standard Klein-Gordon in a curved spacetime dictated by the flow profile. This fact enables the use of atomic BECs as quantum simulators of field theories in curved spacetimes, i.e. as analogue gravity systems. Here we focused on field theory effects around black holes: superradiance and the Hawking effect. Other effects have also been investigated, such as, for instance, particle production in expanding universes \cite{Parker:1968mv,Fischer:2004bf,Fedichev:2003bv,Barcelo:2003wu,Weinfurtner:2008if,Eckel:2017uqx,GomezLlorente:2019mvt,Eckel:2020qee,Eckel:2020qee,Bhardwaj:2020ndh,Steinhauer:2021fhb,Banik:2021xjn,Viermann:2022wgw,Bhardwaj:2023squ,Agullo:2024lry,Gondret:2025fdc,Chunn:2025zlq}, the Unruh effect \cite{Oliveira:2018ckz,Leonhardt:2017lwm,Gooding:2020scc,Gooding:2020scc,Barros:2020zkt,Biermann:2020bjh,Tian:2022gfa,Bunney:2023vyj}, the dynamical Casimir effect \cite{Lahteenmaki:2011cwo,Faccio:2011nwa,Wilson:2011rsw,Jaskula_2012,Rego:2014wta,Doukas:2014bja,GarciaMartin-Caro:2024qpk,Tettamanti:2024nzj}, vacuum decay \cite{Billam:2018pvp,Braden:2019vsw,Lagnese:2021grb,Billam:2021nbc,Jenkins:2023npg,Batini:2023zpi,Jenkins:2023eez,Zhu:2024dvz,Caneletti:2024kww,Cominotti:2025qia}, superposition of spacetimes \cite{Barcelo:2021nhs}, or formation of spectral cascades relevant to reheating scenarios \cite{Gregory:2024ogi} among others.

The use of analogues for these phenomena provides not only possible experimental verifications of predictions of quantum field theory, but also a conceptual playground to understand the crucial ingredients of the different phenomena and how they depend on modifications of the purely relativistic field theory description. The \textit{microscopic} physics of the (non-relativistic) analogue systems provides in fact some modification to the ultraviolet physics of the fields, such as the superluminal dispersion of Bogoliubov excitations. Moreover, analogue systems allow to gain insight on the \textit{back-reaction} of the quantum field on the spacetime, a very difficult problem to approach in a gravitational setting. Even though results in this directions will not be descriptive of the gravitational scenario (the GPE cannot reproduce Einstein's equations), one could hope to understand some universal mechanisms ocuring in backreaction scenarios which could shed light into the gravitational problem. Steps in this direction are being taken in the study of the superradiant instability and its quenching mechanisms \cite{Oliveira:2018ckz,Giacomelli:2019tvr,Delhom:2023gni,Oliveira:2024quw,Berti_2024,Patrick:2025tpk,Guerrero:2025kdn}. Understanding the existence of such mechanisms could be key in understanding the physics of rotating extreme compact objects \cite{brito2020superradiance,Cardoso:2017cqb}.  

\section*{Acknowledgements}
The authors thank Ivan Agullo, Iacopo Carusotto, and Maxime J. Jacquet for multiple insightful discussions on several topics contained in these lecture notes.

\paragraph{Funding information}
AD acknowledges support from the
project PID2022-139841NB-I00 and Grant PR28/23 ATR2023-145735 funded by MCIN/AEI/10.13039/501100011033 as well as NSF grants PHY-2409402 and PHY-2110273 and by the Hearne Institute for Theoretical Physics.

\begin{appendices}
\numberwithin{equation}{section}

\section{Computation of scattering coefficients in stationary problems}\label{app:ComputationScattCoef}

We will now outline a semi-analytical method to obtain the scattering coefficients in a general two-mode scattering scenario. The argument is quite general, and can be applied to any scattering problem provided that there is satationarity, also including scenarios with arbitrary finite number of modes. In general, the modes spanning the IN and OUT basis can be labeled by quantities that are conserved in time evolution due to symmetries of the problem. In the present section, we will only assume stationarity, but we will later see some specific examples with further symmetries.

According to \eqref{eq:GenScattProcess}, the time evolution of the elements of the IN basis can be written as a combination of elements of the OUT basis as
\begin{equation}
    W^{\rm in}_r \overset{\text{time}}{\longrightarrow} T_{\omega}\, W^{\rm out}_r + R_{\omega}\, W^{\rm out}_l, \qquad\text{and}\qquad
    W^{\rm in}_l \overset{\text{time}}{\longrightarrow} t_{\omega}\, W^{\rm out}_l + r_{\omega}\, W^{\rm out}_r.
\end{equation}

Now, note that IN and OUT wavepackets are peaked far from the interaction region at early and late times respectively, and are usually built out of linear combinations of exact solutions which behave as in- and outgoing waves of a given frequency far from the interaction region. These exact solutions typically have divergent norm, i.e. they are unphysical, though the wavepackets have finite norm. In order to obtain the scattering coefficients, we will work with these unphysical solutions, obtaining the sought frequency-dependent coefficients as if these were the in- and outgoing waves in a realistic sceanrio. This trick, which allows to simplify the problem greatly, yields meaningful results provided that the actual wavepackets have a spread in frequency which is much smaller than the range over with the scattering coefficients vary.

For simplicity, we will also assume that we are in one spatial dimension, so that  the assymptotic regions far from the scatterer are $x\to+\infty$ and $x\to-\infty$. Due to stationarity, the wave equation is separable in space and time, with the time dependence given by \(\ee^{\ci\, \omega t}\). Let us call $\varphi_{r}^{\rm in}$, $\varphi_{l}^{\rm in}$, $\varphi_{r}^{\rm out}$, and $\varphi_{l}^{\rm out}$ to the asymptotic form of the spatial solutions far from the interaction region, so that $W_{r}^{\rm in}$, $W_{l}^{\rm in}$, $W_{r}^{\rm out}$, and $W_{l}^{\rm out}$ at early and late times can be written as an integral over frequency of \(\ee^{\ci\, \omega t}\) times the corresponding $\varphi$. Also, let us assume that the $\varphi$'s are Dirac-delta-normalized.

For each frequency, we solve the spatial differential equation twice. First we solve it using boundary data corresponding to the function \(\varphi^{\rm in}_l\) and its first derivative at $x\to+\infty$. We propagate the solution until \(x\to-\infty\) where it becomes a linear combination of the two independent solutions \(\varphi^{\rm in}_r\) and \(\varphi^{\rm out}_l\). The asymptotic behavior of this solution is
\begin{equation}
    \varphi^{(1)}_{\omega}(x)=
    \begin{cases}
        a_{\omega} \, \varphi^{\rm out}_{l} + b_{\omega} \, \varphi^{\rm in}_{r} &x\rightarrow -\infty\\[10pt]
         \varphi^{\rm in}_{l} &x\rightarrow +\infty
    \end{cases}
\end{equation}
We then extract the coefficients \(a_{\omega}\) and \(b_{\omega}\) by fitting the numerical solution near $x\rightarrow -\infty$ to a linear combination of \(\varphi^{\rm out}_l\) and \(\varphi^{\rm in}_r\). Then we repeat the process with boundary data specified by \(\varphi^{\rm out}_r\) at \(x\rightarrow +\infty\), we find a spatial solution with the assymptotic behavior
\begin{equation}
\begin{aligned}
    \varphi^{(2)}_{\omega}(x) &=
    \begin{cases}
        c_{\omega}\, \varphi^{\rm in}_{r} + d_{\omega}\, \varphi^{\rm out}_{l} &x\rightarrow -\infty\\[10pt]
        \varphi^{\rm out}_{r} &x\rightarrow +\infty
    \end{cases},
\end{aligned}
\end{equation}
from which we can obtain \(c_{\omega}\) and \(d_{\omega}\). Adding the time dependence \(e^{-\ci \omega t}\) to these spatial functions, we obtain two ``stationary'' solutions to the wave equation, in the sense that they do not propagate energy, but rather oscillate eternally due to their harmonic time dependence \(e^{-\ci \omega t}\). As explained above, the trick to transform these stationary solutions into propagating waves that can scatter is to build wavepackets by integrating them in a frequency range as 

\begin{equation}
\label{eq:WavepacketConstruction}
\frac{1}{\sqrt{\epsilon}}\int_{j\epsilon}^{(j+1)\epsilon} \dd\omega \,
 \ee^{\ci\omega \frac{n}{\epsilon}}\, \ee^{-\ci\omega\, t}\varphi^{(i)}_{\omega}(x)\,,
\end{equation}

where \(i=1,2\), \(j>0\) and \(n\) are integers, and \(\epsilon\) is a real number with dimensions of frequency. For \(i=1\), this function describes the scattering process
\begin{equation}
\begin{aligned}
    b_{\omega}\, W^{\rm in}_r + W^{\rm in}_l \xrightarrow{\text{time}}  a_{\omega}\,  W^{\rm out}_l \, ,
\end{aligned}
\end{equation}

where \(W^{\rm in}_r, W^{\rm in}_l\) and \(W^{\rm out}_l\) are wave packets peaked at frequency \(\omega_{\rm peak}=(j+1/2)\epsilon\) and with spread \(\Delta \omega=\epsilon\). The integer \(n\) controls the region in space where these packets are supported at each instant of time. One can select \(\epsilon\) to be small enough so that the coefficients \(a_{\omega}\) and \(b_{\omega}\) do not change significantly with \(\omega\) within in interval \(\omega_{\rm peak}\pm\Delta \omega\). Similarly, for \(i=2\), the function \eqref{eq:WavepacketConstruction} describes the scattering process
\begin{equation}
\begin{aligned}
    c_{\omega}\, W^{\rm in}_r \xrightarrow{\text{time}}  d_{\omega}\, W^{\rm out}_l  +  W^{\rm out}_l \, .
\end{aligned}
\end{equation}
Since the dynamics are linear, we can take linear combinations of these two scattering processes to obtain information about other scattering process of interest. In particular, there are two linear combinations which produce 
\begin{equation}
\begin{aligned}
    W_r^{\rm in} \xrightarrow{\text{time}} \frac{d_{\omega}}{c_{\omega}}\, W_l^{\rm out}+ \frac{1}{c_{\omega}}\, \Phi_r^{\rm out}\qquad\text{and}\qquad  W_l^{\rm in}\xrightarrow{\text{time}} \lr{a_{\omega}-\frac{b_{\omega}d_{\omega}}{c_{\omega}}}W_l^{\rm out} + \frac{b_{\omega}}{c_{\omega}}W_r^{\rm out}\,
\end{aligned}
\end{equation}
which provide the sought scattering coefficients as
\begin{equation}
\begin{aligned}
    R_{\omega} = \frac{d_{\omega}}{c_{\omega}}\quad,\quad
    T_{\omega} = \frac{1}{c_{\omega}}\quad,\quad
    t_{\omega} = a_{\omega}-\frac{b_{\omega}d_{\omega}}{c_{\omega}}\quad\text{and}\quad
    r_{\omega} = \frac{b_{\omega}}{c_{\omega}}\,.
\end{aligned}
\end{equation}
This strategy allows us to determine the scattering coefficients provided that we know how to Dirac-delta normalize the unphysical definite-frequency solutions. This can in general be a difficult task to perform analytically. However, a way around the analytical computation of the Dirac-delta normalization is to use the set of constraints provided by norm conservation (see Problem \ref{prob:SymProd}). If the norms of some of these modes are known, or a relation between them, is known a priory, these constraints provide a set of equations that can be used to determine the unknown normalization factors ---see appendix D of \cite{Delhom:2023gni}. 

\section{Problems}
\label{app:Problems}

\begin{problem}
\label{prob:QuantumGPE}
\textbf{Quantum GPE}

\noindent\textit{Show that Heisenberg time evolution of the atomic field according to the Hamiltonian \eqref{eq:GPHamiltonian} yields the quantum GPE \eqref{eq:heisenberg-eom}.}

\medskip\noindent\textit{Solution:}

In the Heisenberg picture, evolution is encoded in operators. An operator $\hat{A}$ evolves under a hamiltonian $\hat{H}$ according to $\ci\hbar\partial_t{\hat{A}}=[\hat{A},\hat{H}]$. From the GPE Hamiltonian \eqref{eq:GPHamiltonian} and the commutation relations $[\hat{\Psi}(\vx,t),\hat{\Psi}^\dagger(\vy,t)]=\delta(\vx-\vy)$
\begin{equation}
\begin{split}
	\hat{H}=\int_\Sigma \dd^d\tx\, \Big[\frac{\hbar^2}{2m}\vna\hat{\Psi}^\dagger(t,\vx)\cdot\vna\hat{\Psi}(t,\vx)&+\hat{\Psi}^\dagger(t,\vx)V_\textrm{ext}(\vx)\hat{\Psi}(t,\vx)
     \\
     &+\frac{g}{2}\hat{\Psi}^\dagger(t,\vx)\hat{\Psi}^\dagger(t,\vx)\hat{\Psi}(t,\vx)\hat{\Psi}(t,\vx)\Big].
\end{split}
\end{equation}
we find
\begin{equation}
\begin{split}
    [\hat{\Psi}(\vy,t),\hat{H}]=&\int_\Sigma \dd^d\tx\, \Big[\frac{\hbar^2}{2m}\vna_\vx\delta(\vx-\vy)\cdot\vna_\vx\hat{\Psi}(t,\vx)+\delta(\vx-\vy)V_\textrm{ext}(x)\hat{\Psi}(t,\vx)
    \\
    &\hspace{1.3cm}+g\delta(\vx-\vy)\hat{\Psi}^\dagger(t,\vx)\hat{\Psi}(t,\vx)\hat{\Psi}(t,\vx)\Big]
    \\
    =&-\frac{\hbar^2}{2m}\vna_\vy^2\hat{\Psi}(t,\vy)+V_\textrm{ext}(\vy)\hat{\Psi}(t,\vy)+g\hat{\Psi}^\dagger(t,\vy)\hat{\Psi}(t,\vy)\hat{\Psi}(t,\vy),
\end{split}
\end{equation}
where we have used $\int_\Sigma \dd^d\vx f(\vx)\vna\delta(\vx)=-\int_\Sigma \dd^d\vx \delta(\vx)\vna f(\vx)$. This result implies that the Heisenberg equations of motion reproduce the quanutm GPE
\begin{equation}
   \ci\hbar\partial_t\hat\Psi = \lrsq{-\frac{\hbar^2}{2m}\vna^2+V_\textrm{ext}+g\hat{\Psi}^\dagger\hat{\Psi}}\hat{\Psi}.
\end{equation}
\end{problem}


\begin{problem}
\label{prob:FluctHam}
\noindent\textbf{Hamiltonian for low-energy excitations of a homogeneous BEC}

\noindent\textit{Show that, for a condensed and homogeneous weakly interacting dilute gas, its full Hamiltonian \eqref{eq:GPHamiltonian} can be approximated by \eqref{eq:QuantumLowEnergyHam} to leading order in the Bogoliubov approximation.}

\medskip\noindent\textit{Solution:}

For a homogeneous condensate we need a constant potential, which we take $V_{\rm ext}=0$---a constant external potential can be absorbed into the chemical potential. To start the derivation, we first put the system in a $d$-dimensional rectangular box characterized by $d$ edges of lengths $L_i$, so its volume is $V=\Pi_{i=1}^d L_i$, with periodic boundary conditions. We will take the continuum limit at the end.

Within the box, the atomic field is expanded as
\begin{equation}
    \hat\Psi(\vx)
    =
    \frac{1}{\sqrt V}
    \sum_{\vk}
    \ee^{\ci\vk\cdot\vx}
    \hat b_{\vk},
    \qquad
    \hat\Psi^\dagger(\vx)
    =
    \frac{1}{\sqrt V}
    \sum_{\vk}
    \ee^{-\ci\vk\cdot\vx}
    \hat b^\dagger_{\vk},
    \label{eq:BoxFieldExpansion}
\end{equation}
where the allowed momenta are discrete and given by
\begin{equation}
    \vk
    =
    \lr{
        \frac{2\pi n_1}{L_1},
        \ldots,
        \frac{2\pi n_d}{L_d}
    },
    \qquad
    n_i\in\mathbb{Z},
\end{equation}
and mode orthonormality is encoded in the identity
\begin{equation}
    \int_V \dd^d \vx\,\ee^{\ci(\vk-\vq)\cdot \vx}=V\delta_{\vk\vq}\,.
    \label{eq:ModeOrthonormaityBox}
\end{equation}
As a consequence of the canonical commutation relations (CCR) $[\hat{\Psi}(\vx),\hat{\Psi}^\dagger(\vy)]=\delta(\vx-\vy)$, $[\hat{\Psi}(\vx),\hat{\Psi}(\vy)]=[\hat{\Psi}^\dagger(\vx),\hat{\Psi}^\dagger(\vy)]=0$, and of mode orthonormality, the ladder operators satisfy the standard CCR
\begin{equation}
    \lrsq{
        \hat b_{\vk},
        \hat b^\dagger_{\vq}
    }
    =
    \delta_{\vk\vq}\,,
    \qquad
    \lrsq{
        \hat b_{\vk},
        \hat b_{\vq}
    }
    =\lrsq{
        \hat b^\dagger_{\vk},
        \hat b^\dagger_{\vq}
    }
    =0\,.
    \label{eq:BoxCommutator}
\end{equation}
We now start from the homogeneous Gross-Pitaevskii Hamiltonian
\begin{equation}
    \hat H
    =
    \int_V \dd^d\vx\,
    \lrsq{
        \frac{\hbar^2}{2m}
        \vna\hat\Psi^\dagger(\vx)
        \cdot
        \vna\hat\Psi(\vx)
        +
        \frac{g}{2}
        \hat\Psi^\dagger(\vx)
        \hat\Psi^\dagger(\vx)
        \hat\Psi(\vx)
        \hat\Psi(\vx)
    } .
    \label{eq:HomogeneousGPHamiltonian}
\end{equation}
Using \eqref{eq:BoxFieldExpansion} we find
\begin{equation}
\begin{split}
    \hat H
    =&
    \frac{1}{V}
    \sum_{\vk,\vq}\lr{
    \frac{\hbar^2\vk\cdot\vq}{2m}\,
    \hat b^\dagger_{\vk}\hat b_{\vq}
    \int_V \dd^d\vx\,
    \ee^{\ci(\vq-\vk)\cdot\vx}}
    \\
    &+\frac{g}{2V^2}
    \sum_{\vk,\vk',\vq,\vq'}\lr{
    \hat b^\dagger_{\vk}
    \hat b^\dagger_{\vk'}
    \hat b_{\vq}
    \hat b_{\vq'}
    \int_V \dd^d\vx\,
    \ee^{\ci(\vq+\vq'-\vk-\vk')\cdot\vx}}\,.
\end{split}
\end{equation}
Hence, using mode orthogonality \eqref{eq:ModeOrthonormaityBox}, and before making any approximation, the Hamiltonian is
\begin{equation}
    \hat H
    =
    \sum_{\vk}
    \frac{\hbar^2\tk^2}{2m}
    \hat b^\dagger_{\vk}\hat b_{\vk}
    +
    \frac{g}{2V}
    \sum_{\vk,\vk',\vq}
    \hat b^\dagger_{\vk}
    \hat b^\dagger_{\vk'}
    \hat b_{\vq}
    \hat b_{\vk+\vk'-\vq}.
    \label{eq:ExactBoxHamiltonian}
\end{equation}
We now implement the Bogoliubov approximation. This consists on keeping only leading order terms in out-of-condensate operators, and replace condensate operators by its mean field value $\hat{b}_0\to\sqrt{N}\hat{\mathbb{I}}$ and $\hat{b}_0^\dagger\to\sqrt{N}\hat{\mathbb{I}}$. Since the gas is homogeneous, the condensate corresponds to the zero-momentum mode, which is macroscopically occupied (compared to the rest). Writing the total number operator as $\hat{N}=\hat{b}^\dagger_{0}\hat{b}_{0}+\sum_{\vk\neq0}\hat{b}^\dagger_{\vk}\hat{b}_{\vk}$, the Bogolyubov approximation at fixed total number of quanta in the system yields
\begin{equation}
    N_0^2\hat{\mathbb{I}}
    \approx
    N^2\hat{\mathbb{I}}
    -
    2N\sum_{\vk\neq0}
    \hat b^\dagger_{\vk}\hat b_{\vk}.
    \label{eq:N0Depletion}
\end{equation}
Taking this into account, we can apply the Bogolyubov approximation to \eqref{eq:ExactBoxHamiltonian} to find
\begin{equation}
    \hat H\approx
    \frac{gnN}{2}\hat{\mathbb{I}}
    +
    \sum_{\vk\neq0}
    \lrsq{
        \lr{ \frac{\hbar^2\tk^2}{2m}+gn}
        \hat b^\dagger_{\vk}\hat b_{\vk}
        +
        \frac{gn}{2}
        \lr{
            \hat b^\dagger_{\vk}
            \hat b^\dagger_{-\vk}
            +
            \hat b_{\vk}
            \hat b_{-\vk}
        }
    }\,,
    \label{eq:BogoliubovHamiltonianBox}
\end{equation}
where we have defined the atomic density $N/V$. To find \eqref{eq:QuantumLowEnergyHam}, we take the continuum limit at constant density through the replacements
\begin{equation}
    \sum_{\vk}
    \longrightarrow
    \frac{V}{(2\pi)^d}
    \int \dd^d\vk\,,
    \qquad\text{and}\qquad
    \hat b_{\vk}
    \to
    \lr{\frac{(2\pi)^d}{V}}^{1/2}
    \hat b_{\vk}.
    \label{eq:SumToIntegral}
\end{equation}
which take into account the proper normalization of the ladder operators in the continuum. As a remark, we note that the notation
$\int_{\vk\neq0}\dd^d\vk$ in \eqref{eq:QuantumLowEnergyHam} is merely a reminder that the condensate mode has already been separated before taking the continuum limit, as the point $\vk=0$ has zero measure in the continuum integral.
\end{problem}

\begin{problem}
    \label{prob:BdG}
\textbf{Bogoliubov-de Gennes problem}

\textit{3.1 Using the expansion $\eqref{eq:DefPertGPE}$, derive the coupled system of PDEs known as Bogoliubov-de Gennes problem for a stationary homogeneous condensate with chemical potential $\mu$ in the comoving frame of the condensate. Then obtain the Bogoliubov dispersion relation for plane waves of the form 
$$|\psi_{\omega\vk}\rangle=\ee^{-\ci(\omega t- \vk\cdot\vx)}|\psi_{\omega\vk}\rangle\,,$$
where $|\psi_{\omega\vk}\rangle$ are constant in space and time.}

\noindent \textit{3.2 verify that the ratio of the absolute value of the components of the eigenvectors corresponding to the positive eigenvalue branch is equal to the ratio $|u_\tk|/|v_\tk|$ found from \eqref{eq:BogoCoeffBECs}.}.

\noindent\textit{3.3 Repeat 3.1 using instead the expansion $\eqref{eq:DefPertGPELab}$.}

\noindent\textit{3.4 Prove that the two dispersion relations derived above, which should coincide with \eqref{eq:BogoliubovDispersionComoving} and \eqref{eq:doppler-shift}, are related by a Galilean transformation from comoving frame coordinates to lab frame coordinates.}

\medskip\noindent\textit{Solution to 3.1:}

Plugging $\Psi(\vx,t)=\ee^{-\ci \mu t/\hbar}\lrsq{\Psi_0(\vx)+{\psi}(t,\vx)}$ into the GPE we find
\begin{equation}
    \ci\hbar\partial_t\psi+\mu(\Psi_0+\psi)=\lrsq{-\frac{\hbar^2}{2m}\nabla^2+V_{\mathrm{ext}}}(\Psi_0+\psi) + gn_0(\Psi_0+2\psi) + g \Psi_0^2 \psi^* + \mathcal{O}(\psi^2).
\end{equation}
Taking into account that $\Psi_0$ satisfies the stationary GPE \eqref{eq:StaionaryGPE} with chemical potential $\mu$, the terms of order zero in the perturbations $\psi$ or $\psi^*$ cancel, which leads to 
\begin{equation}
    \ci\hbar\partial_t\psi=\lrsq{-\frac{\hbar^2}{2m}\nabla^2+V_{\mathrm{ext}}+2gn_0-\mu}\psi + g \Psi_0^2 \psi^* + \mathcal{O}(\psi^2).
    \label{eq:IntStep3}
\end{equation}
Using now $\eqref{eq:ChemPotHomog}$, the chemical potential of a homogeneous condensate in its comoving frame is $\mu=V_{\rm ext}+gn_0$, which leads to
\begin{equation}
    \ci\hbar\partial_t\psi=\lrsq{-\frac{\hbar^2}{2m}\nabla^2+gn_0}\psi + g \Psi_0^2 \psi^* + \mathcal{O}(\psi^2).
\end{equation}
Taking the complex conjugate equation, keeping only linear terms in $\psi$ and $\psi^*$, and defining $|\psi\rangle^\top=(\psi,\psi^*)$ we find
\begin{equation}
     \ci\hbar\partial_t\ket{\psi}=
     	\begin{pmatrix}
       		-\frac{\hbar^2}{2m}\nabla^2+gn_0 & g \Psi_0^2\\
      		- g \Psi_0^*{}^2 & \frac{\hbar^2}{2m}\nabla^2-gn_0
    	\end{pmatrix}\ket{\psi}\,,
\end{equation}
This is the BdG problem for a stationary homogeneous condensate in its comoving frame. We now assume plane waves of the form  $|\psi_{\omega\vk}\rangle=\ee^{-\ci(\omega t- \vk\cdot\vx)}|\psi_{\omega\vk}\rangle$, and the BdG operator turns into the matrix
\begin{equation}
     \hbar\omega\ket{\psi_{\omega\vk}}=
     	\begin{pmatrix}
       		\frac{\hbar^2\tk^2}{2m}+gn_0 & g \Psi_0^2\\
      		- g \Psi_0^*{}^2 & -\frac{\hbar^2\tk^2}{2m}-gn_0
    	\end{pmatrix}\ket{\psi_{\omega\vk}}\,,
\end{equation}
which means that $\ket{\psi_{\omega\vk}}$ is an eigenvector of the BdG operator with eigenvalue $\hbar\omega$. Now, we can compute the eigenvalues of the above BdG matrix in terms of $\tk$ and the quantities defining the background condensate. Using $|\Psi_0|^2=n_0$ these eigenvalues are
\begin{equation}
\pm\sqrt{\frac{\hbar^4\tk^4}{4m^2}+\frac{\hbar^2g n_0}{m} \tk^2}
\end{equation}
which implies
\begin{equation}
\omega^\pm_{\vk}=\pm\sqrt{\frac{\hbar^4\tk^4}{4m^2}+\frac{\hbar^2g n_0}{m} \tk^2}.
\end{equation}
This is the Bogoliubov dispersion relation \eqref{eq:BogoliubovDispersionComoving}.

\medskip\noindent\textit{Solution to 3.2:}

Take the BdG eigenvalue problem above with the eigenvectors of the form $\ket{\psi_{\omega\vk}}^\pm=(u^\pm_{\tk},v^\pm_{\tk})^\top$, so that
\begin{equation}
    \hbar\omega_{\tk}^\pm
    \begin{pmatrix}
        u^\pm_{\tk}\\
        v^\pm_{\tk}
    \end{pmatrix}
    =
    \begin{pmatrix}
        \frac{\hbar^2k^2}{2m}+gn_0 & gn_0\ee^{2\ci\theta_0}\\
        -gn_0\ee^{-2\ci\theta_0} & -\frac{\hbar^2k^2}{2m}-gn_0
    \end{pmatrix}
    \begin{pmatrix}
        u^\pm_{\tk}\\
        v^\pm_{\tk}
    \end{pmatrix},
\end{equation}
where $\Psi_0=\sqrt{n_0}\ee^{\ci\theta_0}$. The first row gives
\begin{equation}
    \lr{\frac{\hbar^2k^2}{2m}+gn_0}u^\pm_{\tk}
    +
    gn_0\ee^{2\ci\theta_0}v^\pm_{\tk}
    =
    \hbar\omega^\pm_\tk u^\pm_{\tk} .
\end{equation}
Hence
\begin{equation}
    \frac{v^\pm_{\tk}}{u^\pm_{\tk}}
    =
    \ee^{-2\ci\theta_0}
    \frac{\hbar\omega^\pm_\tk-\frac{\hbar^2k^2}{2m}-gn_0}{gn_0}.
\end{equation}
Using the definition of healing length \eqref{eq:HealingLength} we can rewrite this as
Hence
\begin{equation}
    \frac{v^\pm_{\tk}}{u^\pm_{\tk}}
    =
    \ee^{-2\ci\theta_0}D_\pm(\tk)
\end{equation}
where
\begin{equation}
    D_\pm(k)
    \coloneqq
    \frac{\hbar\omega^\pm_\tk-\frac{\hbar^2k^2}{2m}-gn_0}{gn_0}
    =
    -1-\frac{\xi^2k^2}{2}\pm\xi k \sqrt{1+\frac{\xi^2k^2}{4}}.
    \label{eq:DefinitionDEigenvectors}
\end{equation}
The eigenvectors can thus be written as
\begin{equation}
    |\psi_\pm(k)\rangle
    =
    \begin{pmatrix}
        \ee^{2\ci\theta_0}\\
        D_\pm(k)
    \end{pmatrix},
\end{equation}
in agreement with \eqref{eq:EigenstatesHomogBdG}. These eigenvectors satisfy $|u_\tk^\pm|^2-|v_\tk^\pm|^2=1-D_\pm(k)^2$. We now normalize the positive branch so that $|u_\tk^+|^2-|v_\tk^+|^2=1$, which shows that the ratio is the expected one. 

Multiplying $\ket{\Psi_+(\tk)}$ by $(1-D_+(\tk))^{-1/2}$  we find
\begin{equation}
   |u_\tk^+|=\frac{1}{\sqrt{1-D_+(k)^2}}\,,
   \qquad\qquad
   |v_\tk^+|=\frac{D_+(k)}{\sqrt{1-D_+(k)^2}}\,.
\end{equation}
Using the definition in \eqref{eq:DefinitionDEigenvectors}, squaring both equalities, multiplying by $g^2n_0^2$ and using 
\begin{equation}
    g^2n_0^2=\lr{\frac{\hbar^2\tk^2}{2m}+gn_0+\hbar\omega_\tk^+}\lr{\frac{\hbar^2\tk^2}{2m}+gn_0-\hbar\omega_\tk^+}
\end{equation}
we can write
\begin{equation}
    1-D^2_+(\tk)=\frac{2\hbar\omega^+_\tk}{\frac{\hbar^2\tk^2}{2m}+gn_0+\hbar\omega_\tk^+}
\end{equation}
wich, taking the inverse, after some algebraic manipulations, we find
\begin{equation}
\begin{split}
    &|u_\tk^+|=\sqrt{\frac{1}{2\hbar\w_\tk}\left(\frac{\hbar^2\tk^2}{2m}+gn\right)+\frac{1}{2}}
   \\
   &|v_\tk^+|=\sqrt{\frac{1}{2\hbar\w_\tk}\lr{\frac{\hbar^2k^2}{2m}+gn_0}-\frac{1}{2}}\,,
\end{split}
\end{equation}
 in agreement with \eqref{eq:BogoCoeffBECs}.

\medskip\noindent\textit{Solution to 3.3:}

We now repeat the process but now plugging $\Psi(\vx,t)=\ee^{-\ci\mu t/\hbar}\lrsq{\Psi_0+\ee^{\ci\vk_0\cdot\vx}\chi(t,\vx)}$ into the GPE. To do that efficiently, we start from \eqref{eq:IntStep3} with $\psi=\ee^{\ci \vk_0\cdot\vx}\chi$. We take into account that $\Psi_0=\ee^{\ci\vk_0\cdot\vx}\sqrt{n_0}$ now describes a condensate flowing at uniform velocity $\vtv=\hbar\vk_0/m$, the chemical potential now satisfies $\mu=V_{\rm ext}+gn_0+m \tv^2/2$, to arrive at
\begin{equation}
    \ci\hbar\partial_t\chi=\lrsq{-\frac{\hbar^2}{2m}\nabla^2-\ci\hbar\vtv\cdot\nabla+gn_0}\chi + g \Psi_0^2 \chi^* + \mathcal{O}(\chi^2).
\end{equation}
Again, taking the complex conjugate equation, keeping only linear terms in $\chi$ and $\chi^*$, and defining $|\chi\rangle^\top=(\chi,\chi^*)$ we find
\begin{equation}
     \ci\hbar\partial_t\ket{\chi}=
     	\begin{pmatrix}
       		-\frac{\hbar^2}{2m}\nabla^2-\ci\hbar\vtv\cdot\nabla+gn_0 & g n_0\\
      		- g n_0 & \frac{\hbar^2}{2m}\nabla^2-\ci\hbar\vtv\cdot\nabla-gn_0
    	\end{pmatrix}\ket{\chi}\,,
\end{equation}
which is the corresponding Bogoliubov problem for a stationary homogeneous condensate flowing at velocity $\vtv$ (i.e. in the lab frame). We now assume plane waves of the form  $|\chi_{\omega\vk}\rangle=\ee^{-\ci(\omega t- \vk\cdot\vx)}|\chi_{\omega\vk}\rangle$ we find
\begin{equation}
     \hbar\omega\ket{\chi_{\omega\vk}}=
     	\begin{pmatrix}
       		\frac{\hbar^2\tk^2}{2m}+\hbar\vtv\cdot\vk+gn_0 & gn_0\\
      		- g n_0 & -\frac{\hbar^2\tk^2}{2m}+\hbar\vtv\cdot\vk-gn_0
    	\end{pmatrix}\ket{\chi_{\omega\vk}}\,,
\end{equation}
which means that $\ket{\chi_{\omega\vk}}$ is an eigenvector of the BdG operator with eigenvalue $\hbar\omega$. Now, we can compute the eigenvalues of the above BdG matrix in terms of $\tk$ and the quantities defining the background condensate. Using $|\Psi_0|^2=n_0$ these eigenvalues are
\begin{equation}
\hbar \vtv\cdot\vk\pm\sqrt{\frac{\hbar^4\tk^4}{4m^2}+\frac{\hbar^2g n_0}{m} \tk^2}
\end{equation}
which implies
\begin{equation}
\omega^\pm_{\vk}=\vtv\cdot\vk\pm\sqrt{\frac{\hbar^4\tk^4}{4m^2}+\frac{\hbar^2g n_0}{m} \tk^2}.
\end{equation}
This is the Bogoliubov dispersion relation for a condensate moving at homogeneous velocity $\vtv$.

\medskip\noindent\textit{Solution to 3.4:}

For a general coordinate transformation defined by an invertible set of functions $\vx'(\vx,t)$ and $t'(\vx,t)$, the corresponding derivative operators associated to the two sets of coordinates are related by 
\begin{equation}
    \partial_t=\frac{\partial t'}{\partial t}\partial_t'+ \frac{\partial \tx'^j}{\partial t}\partial_{\tx'^j}\qquad\text{and}\qquad \partial_{\tx^i}=\frac{\partial t'}{\partial \tx^i}\partial_{t'} + \frac{\partial \tx'^j}{\partial\tx^i}\partial_{\tx'^j} \,.
\end{equation}
If a fluid moves with velocity $\vtv$ in the lab frame, the Galilean boost relating cartesian coordinates in the comoving frame $(\vx',t')$ to cartesian coordinates in the lab frame is given by
\begin{equation}
    \vx'(\vx,t)=\vx-\vtv t \qquad\text{and}\qquad t'(\vx,t)=t.
    \label{eq:GalileanBoost}
\end{equation}
the derivative operators related to the different coordinates are thus related by 
\begin{equation}    
\partial_{\tx'}=\partial_{\tx} \qquad\text{and}\qquad \partial_{t'}=\partial_t+\tv^i\cdot\partial_{\tx^i},
\end{equation}
where we have used homogeneity of the condensate. Therefore, the frequency associated to the lab coordinate frame $\omega_\mathrm{lab}$ by a Fourier expansion in plane waves is related to that of the comoving coordinate frame $\omega_\mathrm{com}$ by $-\ci\omega_{\mathrm{com}}=-\ci\omega_\mathrm{lab}+\ci\vtv\cdot\vk$, which leads to the standard Doppler shift between lab and comoving frame frequencies
\begin{equation}
    \omega_\mathrm{lab}=\vtv\cdot\vk+\omega_{\mathrm{com}}.
\end{equation}
This, together with the results above, indicates that different choices of the phase of the field variables describing linear perturbations are linked to describing sound waves in condensates in different frames.

\end{problem}

\begin{problem}
\noindent\textbf{Symplectic product: norm signs and norm conservation}
\label{prob:SymProd}

\noindent\textit{4.1 Prove that the symplectic product  \eqref{eq:SympProdDensPhase} is conserved in laboratory time for solutions to the density-phase perturbation equations \eqref{eq:DensPhasePerturbations}}.

\noindent{\textit{4.2 Prove that the sign of the norm of plane waves (which are solutions only for homogeneous condensates) is given by the sign of $\omega-\vtv\cdot\vk$}.}

\medskip\noindent\textit{Solution to 4.1:}

This can be done by directly computing $\partial_t (\varphi_1,\varphi_2)$ and using the linearized density-phase perturbation equations \eqref{eq:DensPhasePerturbations}, as well as the continuity equation satisfied by the background \eqref{eq:DensityPhaseGPE} together with Gauss' law.
\begin{equation*}
\begin{split}
    \partial_t(\varphi_1,\varphi_2)=\ci\int_\Sigma \dif V \partial_t (n_0\eta_1^* \varphi_2-\varphi_1^* n_0\eta_2).\\
\end{split}    
\end{equation*}
Using the continuity equation for $n_0$ as well as the perturbation equations for density and phase, we can substitute the time derivatives in terms of spatial derivatives. Regrouping the terms conveniently we find
\begin{equation*}
\begin{split}
    \partial_t(\varphi_1,\varphi_2)=\ci\int_\Sigma \dif V &\Bigg({\vna\cdot\lrsq{n_0(\varphi_1^*\eta_2-\eta_1^*\varphi_2)\vtv}+\frac{\hbar n_0}{m}\lrsq{\varphi_1^*\lr{\vna^2+\frac{\vna n_0}{n_0}\cdot\vna}\varphi_2-\varphi_2\lr{\vna^2+\frac{\vna n_0}{n_0}\cdot\vna}\varphi_1^*}}
    \\
    &+\frac{g n_0^2\xi^2}{4\hbar}\lrsq{\eta_1^*\lr{\vna^2+\frac{\vna n_0}{n_0}\cdot\vna}\eta_2-\eta_2\lr{\vna^2+\frac{\vna n_0}{n_0}\cdot\vna}\eta_1^*}\Bigg)
    \\
    =\ci\int_\Sigma \dif V&\vna\cdot\lrsq{n_0(\varphi_1^*\eta_2-\eta_1^*\varphi_2)\vtv+\frac{\hbar}{m}\lr{\varphi_1^* n_0\vna\varphi_2-\varphi_2 n_0\vna\varphi_1^*+\frac{\eta_1^* n_0\vna\eta_2-\eta_2 n_0\vna\eta_1^*}{4}}}
\end{split}    
\end{equation*}
where in the last equality we have used the definition of the healing length \eqref{eq:HealingLength} for the background condensate. Using Gauss law, the above integral can be turned into an integral in $\partial\Sigma$. Since the density of physically realistic BECs vanishes at the boundaries, the above integral vanishes and the product is conserved.

\medskip\noindent\textit{Solution to 4.2:}

In a homogeneous condensate we have $
    D_{n_0}^2=\nabla^2$. For plane wave solutions $\phi(t,\vx)=\phi_{\vk}\ee^{-\ci(\omega t-\vk\cdot\vx)}$ and $
    \eta(t,\vx)=\eta_{\vk}\ee^{-\ci(\omega t-\vk\cdot\vx)}$, equation \eqref{eq:EtaPhi} reads
\begin{equation}
    \eta_{\vk}=\frac{\ci\hbar \tk^2}{m(\omega-\vtv\cdot\vk)}\phi_{\vk}.
    \label{eq:EtaPhiPlaneWaveRelation}
\end{equation}

The norm of a plane wave as given by the density-phase inner product \eqref{eq:SympProdDensPhase} is then
\begin{equation}
    2n_0
    \frac{\hbar k^2}{m(\omega-\vtv\cdot\vk)}
    |\phi_{\vk}|^2
    \int_\Sigma \dd V .
\end{equation}
whose sign is given by the sign of 
\begin{equation}
    \omega-\vtv\cdot\vk.
\end{equation}
Note that this is the frequency of the plane wave as seen by observers comoving with the condensate.

\end{problem}

\begin{problem}
    \noindent\textbf{Relation between creation/annihilation operators and the sign of the norm of the modes}
\label{prob:LadderOpsFromProd}
    
\textit{Consider a basis of complex solutions to the classical density-phase perturbation equations \eqref{eq:DensPhasePerturbations} $\{\varphi_i,\varphi_i^*\}$ and $\{\eta_i,\eta_i^*\}$. Prove that operators $\hat{a}_i$ and $\hat{a}_i^\dagger$ defined by 
\begin{equation}
    \begin{split}
        \hat{a}_i\coloneqq(\varphi_i,\hat{\varphi})=\ci\int_\Sigma \dd V n_0(\eta_i^*\hat\varphi-\varphi_i^*\hat{\eta})
    \end{split}
\end{equation}
is an annihilation operator if $\varphi_i$ has positive norm and a creation  operator if it has negative norm. Here $\hat\varphi$ is the corresponding quantum field describing phase fluctuations and $\hat\eta$ the corresponding quantum field describing density fluctuations.} 

\medskip\noindent\textit{Solution:}

From the above definition, since $\hat\varphi$ and $\hat\eta$ are hermitian, we have
\begin{equation}
    \begin{split}
        \hat{a}_i^\dagger=-\ci\int_\Sigma \dd V n_0(\eta_i\hat\varphi-\varphi_i\hat{\eta}).
    \end{split}
\end{equation}
Therefore, the commutator of the two operators is given by
\begin{equation*}
    \begin{split}
        [\hat{a}_i,\hat{a}_j^\dagger]=\int_\Sigma \dd V_\vx \dd V_\vy n_0(t,\vx)n_0(t,\vy)\Big[\big(\eta_i^*(t,\vx)\hat\varphi(t,\vx)-\varphi_i^*(t,\vx)\hat{\eta}(t,\vx)\big),\big(\eta_j(t,\vy)\hat\varphi(t,\vy)-\varphi_j(t,\vy)\hat{\eta}(t,\vy)\big)\Big].
    \end{split}
\end{equation*}
Note that the times at which the integrands are evaluated in the same since they are integrated over the same spatial hypersurface of the foliation. The above can be rewritten as
\begin{equation}
    \begin{split}
        [\hat{a}_i,\hat{a}_j^\dagger]=\int_\Sigma &\dd V_\vx \dd V_\vy n_0(t,\vx)n_0(t,\vy)\Big(
        \eta_i^*(t,\vx)\eta_j(t,\vy)[\hat\varphi(t,\vx),\hat\varphi(t,\vy)]-\eta_i^*(t,\vx)\varphi_j(t,\vy)[\hat\varphi(t,\vx),\hat\eta(t,\vy)]
        \\
        &-\varphi_i^*(t,\vx)\eta_j(t,\vy)[\hat\eta(t,\vx),\hat\varphi(t,\vy)]+\varphi_i^*(t,\vx)\varphi_j(t,\vy)[\hat\varphi(t,\vx),\hat\varphi(t,\vy)]
        \Big).
    \end{split}
\end{equation}
Using the canonical commutation relations between density and phase fluctuation fields \eqref{eq:CCRDensPhase} we find
\begin{equation}
    \begin{split}
        [\hat{a}_i,\hat{a}_j^\dagger]=-\ci\int_\Sigma \dd V_\vx \dd V_\vy n_0(t,\vy)\Big(\varphi_i^*(t,\vx)\eta_j(t,\vy)
        -\eta_i^*(t,\vx)\varphi_j(t,\vy)
        \Big)\delta(\vx-\vy).
    \end{split}
\end{equation}
\begin{equation}
    [\hat\varphi(t,\vx),\hat\varphi(t,\vx')]=0,
    \qquad
    [\hat\eta(t,\vx),\hat\eta(t,\vx')]=0,
    \qquad
    [\hat\varphi(t,\vx),\hat\eta(t,\vx')]=-\frac{\ci}{n_0(t,\vx)} \delta(\vx-\vx'),
\end{equation}
Integrating the Dirac delta and using the definition of the Klein-Gordon symplectic product \eqref{eq:KGNorm} we find
\begin{equation}
    \begin{split}
        [\hat{a}_i,\hat{a}_j^\dagger]=-\ci\int_\Sigma \dd V_\vx \Big(\varphi_i^*(t,\vx)\eta_j(t,\vx)
        -\eta_i^*(t,\vx)\varphi_j(t,\vx)
        \Big)=(\varphi_i,\varphi_j).
    \end{split}
\end{equation}
Since $\{\varphi_i,\varphi_i^*\}$ is an orthonormal basis of solutions of the Klein-Gordon equation, we find $[\hat{a}_i,\hat{a}_j^\dagger]=N_{\varphi_i}\delta_{ij}=\pm\delta_{ij}$ with the plus sign if $\varphi_i$ has positive norm and the minus sign if it has negative norm. Hence, we find that $\hat{a}_i$ satisfies the canonical commutation algebra for an annihilation operator if $\varphi_i$ has positive norm and the canonical commutation algebra for a creation operator if it has negative norm. As a remark, note that the $\hbar^{-1}$ factor in the definition of the Klein-Gordon symplectic product \eqref{eq:SympProdDensPhase} is conveniently chosen so that when defining creation and annihilation operators in the quantum theory through the symplectic product they satisfy the standard algebra for ladder operators. Without this factor, keeping the same definitions one would find in the above derivation $[\hat{a}_i,\hat{a}^\dagger_j]=\pm\hbar\delta_{ij}$ instead.
\end{problem}

\begin{problem}
    \textbf{Characterization of superradiance}
    \label{prob:SRTheorem}

\textit{Prove theorem 6.1}

\medskip\noindent\textit{Solution:}

Since it is an if and only if statement, we will instead prove that $\bs{B}$ describes superradiant scattering if and only if it is a unitary matrix. To that end, consider the expression
\begin{equation}
\bs{B}^{\dagger}\cdot \bs{B}=
\begin{pmatrix}
|T|^2 +|R|^2 & T^* r +R^* t \\
T r^* +R t^* & |t|^2 +|r|^2
\end{pmatrix}.
\end{equation}
Therefore, $\bs{B}$ if and only if 
\begin{equation}
    |T|^2 +|R|^2=1, \quad |t|^2 +|r|^2=1, \quad \text{and} \quad T r^* +R t^*=0,
\end{equation} 
which requires that $|T|, |R|, |t|,$ and $|r|$ are all between 0 and 1. Therefore, a unitary $\bs{B}$ describes non-superradiant scattering. 

We will now show how non-superradiant scattering also implies the unitarity of $\bs{B}$. To that end, we use the linearity properties of the symplectic product \eqref{eq:KGNorm}, as well as orthonormality of left and right moving wavepackets, to find
\begin{align}
\begin{split}
    &N_{W^{\text{in}}_{r}}=|T|^2 N_{W^{\text{out}}_{r}} + |R|^2 N_{W^{\text{out}}_{l}},\\
    &N_{W^{\text{in}}_{l}}=|r|^2 N_{W^{\text{out}}_{r}} + |t|^2 N_{W^{\text{out}}_{l}},\\
    &0=T^* r\, N_{W^{\text{out}}_{r}} + R^* t\,,N_{W^{\text{out}}_{l}}.
    \label{eq:NormConsConstraints}
\end{split}
\end{align}
Since the wavepackets are normalized, for non-superradiant scattering, where $|T|, |R|, |t|,$ and $|r|$ are all less than unity, the two first relations imply that the symplectic norm of each of the four wave packets must have the same sign, either $+1$ or $-1$. Hence, the three relations will be satisfied only if
\begin{equation}
    |T|^2 +|R|^2=1,\quad |t|^2 +|r|^2=1, \quad \text{and} \quad T r^* +R t^*=0,
\end{equation} 
indicating that $\bs{B}$ is a unitary matrix.
\end{problem}

\begin{problem}
\label{prob:BogoShearLayer}
    Write a code that diagonalizes the Bogoliubov problem for the analysis of the planar ergosurface and reproduces Fig.\ref{fig:SSW-spectrum}. 

\medskip\noindent\textit{Solution:}

The solution is attached as a Jupyter notebook written in Julia \cite{Julia-2017}. It can be found as \texttt{7\_diag\_bogo\_ssw.ipynb} in the ancillary files on the arXiv page of these notes: \url{https://arxiv.org/abs/2512.14209}.

\end{problem}

\begin{problem}
\label{prob:wmax}
    \textbf{Threshold frequency in the SHE and show $\omax\to\infty$ the hydrodynamic limit}

\textit{Compute the threshold frequency in the SHE for an effectively 1-d condensate and show $\omax\to\infty$ the hydrodynamic limit.}

\medskip\noindent\textit{Solution:}

We need to find the maximum of the positive branch of the dispersion relation \eqref{eq:doppler-shift} for a transsonic effectively 1-dimensional configuration (or the minimum of the negative branch). Taking the derivative of \eqref{eq:doppler-shift} with respect to $\tk$ and equating it to $0$ we find the solutions

\begin{equation}
    k_{\rm max}=\mathrm{sign}(\tv-\cs)\frac{\sqrt{\tv^2+|\tv|\sqrt{8 \cs^2+\tv^2}-4 \cs^2}}{\sqrt{2} \cs \xi }.
\end{equation}
where $\tv$ is positive/negative for right/left moving flows. This solution is a real number only if $|\tv|>\cs$. The maximum frequency at which there are negative norm IN modes available to mix and produce the SHE is given by $\omax=\omega(\tk_{\rm max})$, which gives
\begin{equation}
\omax=\frac{\sqrt{\tv^4+|\tv|^3\sqrt{8 \cs^2+\tv^2}-4 \cs^2}-\sqrt{\cs^2\left(4 \cs^2+\tv^2-3|\tv|\sqrt{8 \cs^2+\tv^2}\right)+\tv^4+|\tv|^3\sqrt{8 \cs^2+\tv^2}}}{\sqrt{2} \cs \xi }\,.
\end{equation}
The hydrodynamic limit can be found by taking $\xi\to0$, in which case, there is no real value of $\tk$ at which $\dd\omega/\dd\tk=0$, so that there is no maximum in the dispersion relation. In fact, there are no negative norm IN modes at positive frequencies in the supersonic region for the SHE---in the hydrodynamic limit, no modes can travel against the flow in the supersonic region.
\end{problem}


\begin{problem}
\label{prob:HawkAmplitudes}

Write a code that solves for the scattering amplitudes of a one-dimensional step-interface BEC acoustic black holes, as discussed in Section \ref{sec:step-horizon}. Plot the scattering amplitudes as a function of the frequency to reproduce Figure \ref{fig:hawking-scattering}. 

\medskip\noindent\textit{Solution:}

A Jupyter notebook written in Julia \cite{Julia-2017} performing this is attached to these lecture notes. It can be found as \texttt{9\_matching\_BH.ipynb} in the ancillary files on the arXiv page: \url{https://arxiv.org/abs/2512.14209}.

\end{problem}


\begin{problem}
\label{prob:BHLasing}
Write a code that solves for the time evolution of the one-dimensional GPE, as outlined in Section \ref{sec:numerical-evolution-gpe}. Use it to observe the black hole instability and reproduce Figure \ref{fig:blackhole-lasing}.

\medskip\noindent\textit{Solution:}

A Jupyter notebook written in Julia \cite{Julia-2017} performing this is attached to these lecture notes. It can be found as \texttt{10\_gpe\_lasing.ipynb} in the ancillary files on the arXiv page: \url{https://arxiv.org/abs/2512.14209}.

\end{problem}

\end{appendices}


\bibliography{biblio}

\end{document}